\documentclass[envcountsame,envcountsect]{llncs}                         

\usepackage{amsmath,amssymb,enumerate,xspace,bbm,rotating,tabularx,url,times}
\usepackage{gastex,xcolor,framed}        

\newcommand{\dA}{\mathbb{A}}  
\newcommand{\dB}{\mathbb{B}}
\newcommand{\dC}{{\mathbb C}}
\newcommand{\dD}{\mathbb{D}}
                      
\newcommand{\mcA}{\mathcal{A}} 
\newcommand{\mcB}{\mathcal{B}} 
\newcommand{\mcT}{\mathcal{T}}

\newcommand{\alp}{\mathsf{alph}}
\newcommand{\AND}{\mathsf{and}}
\newcommand{\bad}{\mathsf{bad}} 
\newcommand{\bool}{\mathsf{bool}} 
\newcommand{\child}{\mathsf{child}} 
\newcommand{\cut}{\mathsf{cut}}
\newcommand{\EXPTIME}{\mathsf{EXPTIME}} 
\newcommand{\FALSE}{\mathsf{false}}
\newcommand{\good}{\mathsf{good}} 
\newcommand{\iso}{\mathsf{iso}} 
\newcommand{\leaf}{\mathsf{leaf}} 
\newcommand{\lex}{\mathsf{lex}} 
\newcommand{\lo}{\mathsf{lo}}
\newcommand{\Lo}{\mathsf{Lo}}
\newcommand{\NotNice}{\mathsf{Type\text{(3-5)}}}
\newcommand{\NSPACE}{\mathsf{NSPACE}} 
\newcommand{\N}{\mathbb{N}}
\newcommand{\omegaop}{\overline{\omega}}
\newcommand{\OR}{\mathsf{or}}    
\newcommand{\oT}{\mathsf{oT}} 
\newcommand{\out}{\mathsf{out}} 
\newcommand{\pref}{\mathsf{pref}} 
\newcommand{\Ptime}{\mathsf{P}} 
\newcommand{\PSPACE}{\mathsf{PSPACE}} 

\newcommand{\rest}{\mathord\restriction}
\newcommand{\rhs}{\mathsf{rhs}} 
\newcommand{\test}{\mathsf{test}} 
\newcommand{\TRUE}{\mathsf{true}} 

\newcommand{\unfold}{\mathsf{unfold}}
\newcommand{\uval}{\mathsf{uval}}
\newcommand{\Up}{\mathsf{Up}}
\newcommand{\up}{\mathsf{up}}
\newcommand{\val}{\mathsf{val}}

\DeclareMathOperator{\height}{\omega\eta-depth}

\DeclareMathOperator{\depth}{depth}

\DeclareMathOperator{\Bl}{Bl}
\DeclareMathOperator{\Reg}{Reg}

\renewcommand{\epsilon}{\varepsilon}
\renewcommand{\phi}{\varphi}

\begin{document}

\title{Isomorphism of regular trees and words}

\author{Markus Lohrey \and Christian Mathissen}

\institute{Institut f\"ur Informatik, Universit\"at Leipzig, Germany\\
  \texttt{\{lohrey,mathissen\}@informatik.uni-leipzig.de} }

\maketitle

\begin{abstract}
The computational complexity of the isomorphism problem 
for regular trees, regular linear orders, and regular words is analyzed.
A tree is regular if it is isomorphic to the prefix order on a regular language. 
In case regular languages are represented
by NFAs (DFAs), the isomorphism problem for regular trees
turns out to be $\EXPTIME$-complete (resp. $\Ptime$-complete).
In case the input automata are acyclic NFAs (acyclic DFAs), 
the corresponding trees are (succinctly represented) finite trees,
and the isomorphism problem turns out to be 
$\PSPACE$-complete (resp. $\Ptime$-complete).
A linear order is regular if it is isomorphic to the lexicographic 
order on a regular language. A polynomial time algorithm
for the isomorphism problem for regular linear orders (and even 
regular words, which generalize the latter)
given by DFAs is presented. This solves an open problem by {\'E}sik and Bloom.
\end{abstract}

\section{Introduction}

Isomorphism problems for infinite but finitely presented structures
are an active research topic in algorithmic model theory \cite{BaGrRu10}. It is a
folklore result in computable model theory that the isomorphism
problem for computable structures (i.e., structures, where the domain
is a computable set of natural numbers and all relations are
computable too) is highly undecidable --- more precisely, it is
$\Sigma^1_1$-complete, i.e., complete for the first existential level
of the analytical hierarchy. Khoussainov et al.~proved in \cite{KhoNRS07}
that even for automatic structures (i.e., structures, where the domain
is a regular set of words and all relations can be recognized by
synchronous multitape automata), the isomorphism problem is
$\Sigma^1_1$-complete. In \cite{KuLiLo11}, this result was further improved to
automatic order trees and automatic linear orders.
On the decidability side, Courcelle proved that the isomorphism 
problem for equational graphs is decidable \cite{Cour89}. Recall that 
a graph is equational if it is the least solution of a system of 
equations over the HR graph operations. 
We remark that Courcelle's algorithm for the isomorphism problem
for equational graphs has very high complexity (it is not elementary),
since it uses the decidability of monadic second-order logic on
equational graphs.

In this paper, we continue the investigation of 
isomorphism problems for infinite but finitely presented structures
at the lower end of the spectra. We focus on two very simple 
classes of infinite structures: {\em regular trees} and {\em regular
  words}.  Both are particular automatic structures.
Recall that a countable tree is regular if it has only finitely
many subtrees up to isomorphism. This definition works for 
ordered trees (where the children of a node are linearly ordered)
and unordered trees. An equivalent characterization in the unordered
case uses regular
languages: An unordered (countable) tree $T$ is regular if and only if 
there is a regular language $L \subseteq \Sigma^*$ which contains 
the empty word and such that $T$ is isomorphic to the tree obtained by
taking the prefix order on $L$ (the empty word word is the root of the
tree). Hence, a regular tree can be
represented by a finite deterministic or nondeterministic
automaton (DFA or NFA), and the isomorphism problem for regular trees becomes
the following computational problem: Given two DFAs (resp., NFAs)
accepting both the empty word, are the corresponding
regular trees isomorphic? It is is not difficult to prove that this
problem can be solved in polynomial time if the two input automata
are assumed to be DFAs; the algorithm is very similar to the 
well-known partition refinement algorithm for 
checking bisimilarity of finite state systems \cite{KanSmoS90}, see
Section~\ref{sec:upper-bounds-trees}.
Hence, the isomorphism problem for regular trees that are represented
by NFAs can be solved in exponential time. 
Our first main result states that this problem is in fact
$\EXPTIME$-complete, see Section~\ref{sec:lower-bound-exptime-tree}.
The proof of the $\EXPTIME$ lower bound uses three main 
ingredients: (i) 
$\EXPTIME$  coincides with alternating polynomial space \cite{CKS81},
(ii) a construction from \cite{JKMT03}, which reduces the evaluation problem
for Boolean expressions to the isomorphism problem for (finite) trees,
and (iii) a small NFA accepting all words that
do {\em not} represent an accepting computation of a polynomial space
machine \cite{StMe73}.\footnote{This construction is used in 
\cite{StMe73} to prove that the universality problem for 
NFAs is $\PSPACE$-complete.}.
Our proof technique yields another result too: 
It is $\PSPACE$-complete to check for 
two given {\em acyclic} NFAs $\mcA_1$, $\mcA_2$ 
(both accepting  the empty word), whether the trees that result 
from the prefix orders on $L(\mcA_1)$ and $L(\mcA_2)$, respectively,
are isomorphic. 
Note that these two trees are clearly finite (since
the automata are acyclic), but the size of $L(\mcA_i)$
can be exponential in the number of states of $\mcA_i$. 
In this sense, acyclic NFAs can be seen as a succinct 
representation of finite trees. 
The $\PSPACE$-upper bound for acyclic NFAs
follows easily from Lindell's result \cite{Lindell92}
that isomorphism of explicitly given trees can be checked 
in logarithmic space.

The second part of this paper studies the isomorphism problem for
{\em regular words}, which were introduced in \cite{Cou78ITA}.
A {\em generalized word} over an alphabet $\Sigma$ 
is a countable linear order together with a $\Sigma$-coloring of 
the elements. A generalized word is regular if it can be obtained
as the least solution (in a certain sense made precise 
in \cite{Cou78ITA}) of a system $X_1 = t_1, \ldots, X_n = t_n$.
Here, every $t_i$ is a finite word over the alphabet
$\Sigma \cup \{X_1, \ldots, X_n\}$. 
For instance, the system $X = abX$ defines the regular word $(ab)^\omega$. 
Courcelle \cite{Cou78ITA} gave an alternative characterization of
regular words: A generalized word is regular if and only if it is equal to the 
frontier word of a finitely-branching ordered regular tree, where 
the leaves are colored by symbols from $\Sigma$. Here, the frontier
word is obtained by ordering the leaves in the usual left-to-right
order (note that the tree is ordered). Alternatively, 
a regular word can be represented by a DFA $\mcA$, where the set of final
states is partitioned into sets $F_a$ ($a \in \Sigma$);
we call such a DFA a {\em partitioned DFA}. The
corresponding regular word is obtained by ordering the language 
of $\mcA$ lexicographical and coloring a word $w \in L(\mcA)$ with
$a$ if $w$ leads from the initial state to a state from $F_a$.
A third characterization of regular words 
was provided by Heilbrunner \cite{Heilbrunner80}:
A generalized word is regular if it can be obtained from 
singleton words (i.e., symbols from $\Sigma$) using 
the operations of concatenation, $\omega$-power,
$\overline{\omega}$-power and dense shuffle. 
For a generalized word $u$, its $\omega$-power (resp. 
$\overline{\omega}$-power) is the generalized word 
$u u u \cdots$ (resp. $\cdots u u u$). Moreover, the 
shuffle of generalized words $u_1, \ldots, u_n$ is obtained
by choosing a dense coloring of the rationals with colors
$\{1,\ldots,n\}$ (up to isomorphism, there is only a single
such coloring \cite{Ros82}) and then replacing every $i$-colored
rational by $u_i$. In fact, Heilbrunner presents an algorithm
which computes from a given system of equations (or, alternatively,
a partitioned DFA) an expression 
over the above set of operations (called a {\em regular expression}
in the following) which defines the least solution of the system of 
equations. A simple analysis of Heilbrunner's algorithm shows
that the computed regular expression in general has exponential
size with respect to the input system of equations and it is easy
to see that this cannot be avoided.\footnote{Take for instance
the system $X_i = X_{i+1} X_{i+1}$ ($1 \leq i \leq n$), $X_n = a$,
which defines the finite word $a^{2^n}$.}
The next step was taken by Thomas in \cite{Thomas86}, where he
proved that the isomorphism problem for regular words is decidable.
For his proof, he uses the decidability of the monadic second-order
theory of linear orders; hence his proof does not yield an 
elementary upper bound for the isomorphism problem for regular words.
Such an algorithm was presented later by Bloom and {\'E}sik in \cite{BloomE05},
where the authors present a polynomial time algorithm for checking
whether two given regular expressions define isomorphic regular words.
Together with Heilbrunner's algorithm, this yields an exponential
time algorithm for checking whether the least solutions of two 
given systems of equations (or, alternatively, the regular words
defined by two partitioned DFAs)
are isomorphic. It was asked in 
\cite{BloomE05}, whether a polynomial time algorithm for this problem 
exists. Our second main result answers this question 
affirmatively. In fact, we prove that the problem,
whether two given partitioned DFAs define isomorphic regular words,
is $\Ptime$-complete. A large part of this paper deals with the 
polynomial time upper bound. The first step is simple. By reanalyzing
Heilbrunner's algorithm, it is easily seen that from a given
partitioned DFA (defining a regular word $u$) one can compute 
in {\em polynomial time} a {\em succinct representation} of a regular
expression for $u$. This succinct representation consists of a DAG
(directed acyclic graph), whose unfolding is a regular expression for
$u$. The second and main step of the proof shows that the polynomial 
time algorithm of Bloom and {\'E}sik for regular expressions can be 
refined in such a way that it works (in polynomial time) for succinct
regular expressions too. The main tool in our proof is (besides the 
machinery from   \cite{BloomE05}) algorithmics on compressed 
strings (see \cite{Ryt04} for a survey), in particular Plandowski's result that equality of strings
that are represented by {\em straight-line programs} (i.e., context
free grammars that only generate  a single word) can be checked in
polynomial time \cite{Pla94}. It is a simple observation that an {\em acyclic}
partitioned DFA is basically a straight-line program. 
Hence, we show how to extend Plandowski's polynomial time algorithm
from acyclic partitioned DFAs to general partitioned DFAs.

An immediate corollary of our result is that it can be checked
in polynomial time whether the lexicographic orderings on the 
languages defined by two given DFAs (so called regular linear orderings)
are isomorphic.
For the special case that the two input DFAs accept well-ordered
languages, this was shown in \cite{Esik10}.
Let us mention that it is highly undecidable ($\Sigma^1_1$-complete) to check, whether the
lexicographic orderings on the 
languages defined by two given deterministic pushdown automata
(these are the algebraic linear orderings \cite{BlEsik10}) 
are isomorphic \cite{KuLiLo11}.

\section{Preliminaries}

For an equivalence relation $R$ on a set $A$ and 
$a \in A$ we denote with $[a]_R$ the equivalence 
class containing $R$. Moreover, $[A]_R = \{ [a]_R \mid a \in A \}$.
Let us take a finite alphabet $\Sigma$. 
The length of a finite words $u \in \Sigma^*$ is denoted by $|u|$.
Let $\Sigma^+ = \{ u \in \Sigma^* \mid |u|>0\}$,
$\Sigma^k = \{ u \in \Sigma^* \mid |u|=k\}$,
$\Sigma^{\leq k} = \{ u \in \Sigma^* \mid |u|\leq k\}$,
and $\Sigma^{\geq k} = \{ u \in \Sigma^* \mid |u|\geq k\}$.
For $u, v \in \Sigma^*$, we write $u \leq_{\pref} v$ if 
there exists $w \in \Sigma^*$ with $v = u w$, i.e., $u$ is a {\em prefix}
of $v$.  We write $u <_{\pref} v$ if $u \leq_{\pref} v$ and $u \neq v$.
For a language $L \subseteq \Sigma^*$ let 
$\pref(L) = \{ u \in \Sigma^* \mid \exists v \in L : u \leq_{\pref} v \}$.
For a fixed linear order $\leq$ on the alphabet $\Sigma$ we define
the {\em lexicographic order} $\leq_{\lex}$ on $\Sigma^*$ as follows:
$u \leq_{\lex} v$ if $u \leq_{\pref} v$ or there exist words $w,x,y$ and 
$a, b \in \Sigma$ such that $a < b$, $u = wax$, and $v = wby$.

\subsection{Complexity theory}

We assume that the reader has some basic 
background in complexity theory, in particular concerning the complexity classes $\mathsf{NL}$,
$\Ptime$, $\PSPACE$, and $\EXPTIME$, see e.g.~\cite{Papa94}.
All completeness results in this paper refer to logspace reductions.

A $\PSPACE$-transducer is a deterministic Turing machine with a 
read-only input tape, a write-only output tape 
and a work tape, whose length is bounded by $n^{O(1)}$,
where $n$ is the input length. The output is written
from left to right on the output tape, i.e., 
in each step the transducer either outputs a new symbol
on the output tape, in which case the output head moves
one cell to the right, or the transducer does not output
a new symbol in which case the output head does not move.
Moreover, we assume that the transducer terminates for every
input. This implies that a $\PSPACE$-transducer computes a 
mapping $f : \Sigma^* \to \Theta^*$, where $|f(w)|$ is bounded
by $2^{|w|^{O(1)}}$. We need the following simple lemma:

\begin{lemma} \label{PSPACE}
Assume that the mapping $f : \Sigma^* \to \Theta^*$ 
can be computed by a $\PSPACE$-transducer and let
$L \subseteq \Theta^*$ be a language in $\NSPACE(\log^k(n))$
for some constant $k$. Then $f^{-1}(L)$ belongs to $\PSPACE$.
\end{lemma}

\begin{proof}
The proof uses the same idea that shows that the composition
of two logspace computable mappings is again logspace computable.
Let $w \in \Sigma^*$ be an input.
Basically, we run the $\NSPACE(\log^k(n))$-algorithm for $L$
on the input $f(w)$. But since $f$ can be computed by a 
$\PSPACE$-transducer (which can generate an exponentially long 
output) the length of $f(w)$ can be only bounded by $2^{|w|^{O(1)}}$.
Hence, we cannot construct $f(w)$ explicitly. But this is not 
necessary. We only store  a pointer to some position $f(w)$ 
(this pointer needs space $|w|^{O(1)}$) while
running the $\NSPACE(\log^k(n))$-algorithm for $L$. Each 
time, this algorithm needs the $i^{th}$ letter of $f(w)$, we 
run the PSPACE-transducer for $L$ until the $i^{th}$ output symbol
is generated. The first $i-1$ symbols of $f(w)$ are not written
on the output tape.
Note that the $\NSPACE(\log^k(n))$-algorithm for $L$
needs space $\log^k(2^{|w|^{O(1)}}) = |w|^{O(1)}$ while running on 
$f(w)$. Hence, the total space requirement is bounded by $|w|^{O(1)}$.
\qed
\end{proof}
An {\em alternating Turing machine} is an ordinary nondeterministic
Turing machine, where in addition the set of states $Q$ is partitioned
into existential states ($Q_\exists$) and universal states ($Q_\forall$).
A configuration, where the current state is existential (resp.,
universal) is called an existential (resp., universal) configuration.
Let us assume that $M$ is an alternating Turing machine without
infinite computation paths. Then, we define inductively the notion 
of an {\em accepting configuration} as follows: If $c$ is an
existential configuration, then $c$ is accepting if and only if 
$c$ has an accepting successor configuration. If  $c$ is a
universal configuration, then $c$ is accepting if and only if 
all successor configurations of $c$ are accepting. Note that
a universal configuration without successor configurations is
accepting, whereas an existential configuration without successor 
configurations is not accepting. An input $x$ is accepted by $M$
(briefly, $x \in L(M)$) if and only if the initial configuration 
with input $x$ is accepting.

The complexity class $\mathsf{C}_=\mathsf{P}$ consists of all 
languages $L \subseteq \Sigma^*$ such that there exist 
nondeterministic polynomial time Turing machines $M_1$ and $M_2$
with input alphabet $\Sigma$ such that for every input $w \in \Sigma^*$:
$w \in L$ if and only if the number of accepting computations of $M_1$ on input 
$w$ equals  the number of accepting computations of $M_2$ on input 
$w$. If we replace in this definition 
nondeterministic polynomial time Turing machines
by nondeterministic logspace Turing machines, we obtain the class
 $\mathsf{C}_=\mathsf{L}$.

\subsection{Finite automata and transducer}

Let $\mcA = (Q,\Sigma,\delta,q_0,F)$ be a
nondeterministic finite automaton, briefly
{\em NFA}, where $Q$ is the set of states, $\Sigma$ is the input alphabet,
$\delta \subseteq Q \times \Sigma \times Q$ is the transition
relation, $q_0 \in Q$ is the initial state, and $F \subseteq Q$ is 
the set of final states.
A state  $q\in Q$ is {\em accessible}
(resp. {\em coaccessible}), if $q$ can be reached from the initial
state $q_0$ (resp., if a final state from $F$ can be reached from $q$).
We say that $\mcA$ is accessible 
(resp., coaccessible), if every state of $\mcA$ is accessible  (resp, coaccessible). 
An NFA $\mcA$ is called {\em prefix-closed} if every state of 
$\mcA$ is a final state. In that 
case, the language $L(\mcA)$ is prefix-closed.
Moreover, if $\mcA$ is coaccessible and the prefix-closed NFA 
$\mcB$ results from $\mcA$ by
making every state final, then clearly $L(\mcB) = \pref(L(\mcA))$.
For a DFA (deterministic finite automaton), $\delta$ is a partial
map from $Q \times \Sigma$ to $Q$.
Sometimes, we will also
deal with NFAs (DFAs) without an initial state. If $\mcA$ is 
an NFA without an initial state and $q$ is a state of $\mcA$,
then $L(\mcA,q)$ is the language accepted by $\mcA$, when
$q$ is declared to be the initial state.
We will need the following simple lemma, which is probably folklore:

\begin{lemma} \label{lemma:counting}
For a given a DFA $\mcA = (Q,\Sigma,\delta,q_0,F)$, we can compute
the cardinality $|L(\mcA)| \in \mathbb{N} \cup \{\infty\}$ in polynomial time.
\end{lemma}

\begin{proof}
W.l.o.g we can assume that $\mcA$ is accessible and coaccessible.
Then $L(\mcA)$ is finite if and only if $\mcA$ is acyclic. So assume that
$\mcA$ is acyclic. Since $\mcA$ is deterministic, the size of
$L(\mcA)$ equals the number of paths from $q_0$ to $F$. Now,
in a directed acyclic graph, the number of paths from a source
node to all other nodes can be easily computed 
by dynamic programming  in polynomial time.
\qed
\end{proof}
A {\em partitioned DFA} is a tuple $\mcA = (Q,\Sigma,\delta,q_0,(F_a)_{a \in\Gamma})$, where 
$\Gamma$ is a finite alphabet,  
$\mcB = (Q,\Sigma,\delta,q_0, \bigcup_{a \in \Gamma} F_a)$ 
is an ordinary DFA and $F_a \cap F_b = \emptyset$ for $a \neq b$. 
Since $\mcB$ is a DFA, it follows that
the language $L(\mcB)$ is partitioned by the languages
$L(\mcA_a)$, where  $\mcA_a = (Q,\Sigma,\delta,q_0,F_a)$
($a  \in \Gamma$).
We use partitioned DFAs to label elements of a structure
with symbols from $\Gamma$. The language
$L(\mcA_a)$  will be 
the set of $a$-labelled elements. We do not introduce
partitioned NFAs, since for NFAs the languages 
$L(\mcA_a)$ ($a  \in \Gamma$)
would not partition $L(\mcB)$ (thus, a point could
get several labels). 

A ($\varepsilon$-free) {\em rational transducer} is a tuple $\mcT = (Q,\Sigma,\Gamma,\delta,
q_0,F)$, where $Q$ (the set of states), $\Sigma$ (the input alphabet),
and $\Gamma$ (the output alphabet) are finite sets, $q_0 \in Q$ is the
initial state, $F \subseteq Q$ is the set of final states, and 
$\delta \subseteq Q \times \Sigma \times \Gamma^+ \times Q$ is the
transition relation. A transition $(q,a,w,p) \in \delta$ is also
written as $q \xrightarrow{a|w} p$.
The rational transducer $\mcT$ defines a binary relation
$[\![ \mcT ]\!] \subseteq \Sigma^* \times \Gamma^*$ in the usual
way. For a language $L \subseteq \Sigma^*$ let
$\mcT(L) = \{ v \in \Gamma^* \mid \exists u \in L : (u,v) \in [\![ \mcT ]\!]\}$.

\subsection{Trees} \label{sec:trees}

A {\em tree} is a partial order $T = (A; \leq)$, where 
$\leq$ has a smallest
element (the root of the tree; in particular $A \neq \emptyset$) 
and for every $a \in A$, the set
$\{b \in A \mid b \leq a\}$ is finite and linearly ordered by $\leq$.
We write $a \lessdot b$ if $a < b$ and there does not exist
$c \in A$ with $a < c < b$.
For $a \in A$, let $\child(a,T)$ (the set of children of $a$) be the set of 
all $b \in A$ such that $a \lessdot b$. The set of leaves of $T$ is 
$\leaf(T) = \{ a \in A \mid \child(a,T) = \emptyset\}$.
For $a \in A$ let $T\rest_a$ be the subtree of $T$ rooted at $a$,
i.e., the set of nodes of $T\rest_a$ is $\{ b \in A \mid a \leq b \}$.
The tree $T$ is {\em finitely branching} if $\child(a,T)$ is finite
for all $a \in A$.
An {\em infinite path} of $T$ is an infinite chain $a_0 \lessdot a_1 \lessdot a_2 \lessdot \ \cdots$;
{\em finite paths} are defined analogously. 
If $T$ is finite and $a \in A$, then the {\em height} of $a$ in $T$ is the maximal length of a path 
that starts in $a$.
For trees $T_1$ and $T_2$ we write $T_1 \cong T_2$ in case 
$T_1$ and $T_2$ are isomorphic.

A {\em tree over the finite alphabet} $\Sigma$ is a 
pair $T = (L; \leq_{\pref})$, where $L \subseteq \Sigma^*$ 
is a language with $\varepsilon \in L$.
Note that $T$ is indeed a tree in the above sense.
Most of the time, we will identify the language $L$ with the
tree $(L; \leq_{\pref})$.
Moreover, if $L = \pref(L)$ (i.e., $L$ is prefix-closed), then
$T$ is a finitely branching tree.

A countable tree $T$ is called {\em regular} if $T$ has only finitely many subtrees
up to isomorphism. Equivalently, a countable tree is regular if it is 
isomorphic to a tree of the form $(L;\leq_{\pref})$, where
$L$ is a regular language with $\varepsilon \in L$.
We require that the empty word $\varepsilon$ belongs to $L$ 
in order to ensure the existence of a root (otherwiese
$(L;\leq_{\pref})$ would be only a forest).
If $L$ is accepted by the accessible DFA $\mcA$, 
then the subtrees of $(L;\leq_{\pref})$ correspond to the final states of $\mcA$.  
Note that by our definition, a regular tree need not be 
finitely branching. 


Our definition of a regular tree 
(having only finitely many 
subtrees up to isomorphism) makes sense for other types 
of trees as well, e.g. for node-labeled trees or 
ordered trees (where the children of a node are linearly ordered). 
These variants of regular trees can be generated by
finite automata as well. For instance, a node-labeled
regular tree $(L; \leq_{\pref}, (L_a)_{a\in\Gamma})$, where
$\Gamma$ is the finite labeling alphabet and $L_a$ is the set
of $a$-labeled nodes 
can be specified by a partitioned DFA $(Q,\Sigma,\delta,q_0,(F_a)_{a \in \Gamma})$
with  $L_a = L(Q,\Sigma,\delta,q_0,F_a)$ and $L = \bigcup_{a\in\Gamma} L_a$.
We do not consider node labels in this paper, since
it makes no difference for the isomorphism problem
(node labels can be eliminated by adding additional
children to nodes).
Ordered regular trees will be briefly considered in 
Section~\ref{sec:order-trees}.

\subsection{Linear orders}

See \cite{Ros82} for a thorough introduction into linear orders.
Let $\eta$ be the order type of the rational numbers, $\omega$
the order type of the natural number, and $\omegaop$ be the order type
of the negative integers. With $\mathbf{n}$ we denote a
finite linear order with $n$ elements.
Let $\Lambda = (L; \leq)$ be a linear order.
$\Lambda$ is {\em dense} if $L$ consists of at least two elements,
and for all $x < y$ there exists $z$ with
$x<z<y$.
By Cantor's theorem,  every countable dense linear
order, which neither has a smallest nor largest
element is isomorphic to $\eta$.
Hence, if we take symbols $0$ and $1$ with $0 < 1$, 
then $(\{0,1\}^*1; \leq_{\lex}) \cong \eta$.
The linear order $\Lambda$ is {\em scattered}
if there does not exist an injective order morphism $\phi: \eta \to \Lambda$.
Clearly, $\omega$, $\omegaop$, as well as every finite linear order are scattered.
A linear order is {\em regular} if it is isomorphic to 
a linear order $(L; \leq_{\lex})$ for a regular language $L$.
Hence, for instance, $\eta$, $\omega$, $\omegaop$, and every
finite linear order are regular linear orders.

For two linear orders $\Lambda_1 = (L_1; \leq_1)$ and $\Lambda_1 =
(L_2; \leq_2)$ with $L_1 \cap L_2 = \emptyset$ we define the 
sum $\Lambda_1 + \Lambda_2 = (L_1 \cup L_2; \leq)$, where
$x \leq y$ if and only if either $x,y \in L_1$ and 
$x \leq_1 y$, or $x,y \in L_2$ and 
$x \leq_2 y$, or $x \in L_1$ and $y \in L_2$.
We define the product $\Lambda_1 \cdot \Lambda_2 = (L_1 \times L_2;
\leq)$ where $(x_1,x_2)  \leq (y_1,y_2)$ if and only if either
$x_2 <_2 y_2$ or ($x_2 = y_2$ and $x_1 \leq_1 y_1$).

An \emph{interval} of $\Lambda$ is a subset $I\subseteq L$ such that $x<z<y$
and $x,y\in I$ implies $z\in I$. 
An interval is {\em right-closed} (resp.
{\em left-closed}) if it has a greatest (resp. smallest) element and 
it is {\em closed} if it is both  right-closed and  left-closed.
An interval $I$ is \emph{dense} (resp., \emph{scattered}) if the linear order $\leq$ restricted
to $I$ is dense (resp., \emph{scattered}). 
A predecessor (resp., successor) of $x \in L$ is a largest (resp., smallest)
element of $\{ y \in L \mid y < x \}$ (resp., $\{ y \in L \mid x < y \}$). Of course,
a {\em predecessor} (resp., {\em successor}) of $x$ need not exist, but if it exists then it 
is unique.

\subsection{Generalized words}

Generalized words are countable colored linear orders.
Let $\Sigma$ be a (possibly infinite) alphabet. A \emph{generalized
 word} (or simply word) $u$ over $\Sigma$ is a triple $(L;\leq,\tau)$
such that $L$ is a finite or countably infinite set, $\leq$ is a
linear order on $L$ and $\tau:L\to \Sigma$ is a coloring of $L$.  The
alphabet $\alp(u)$ equals the image of $\tau$.  If $L$ is finite, we
obtain a finite word in the usual sense.
As for trees, we write $u \cong v$ for generalized words $u$ and $v$
in case $u$ and $v$ are isomorphic.

Let $u=(L;\leq,\tau)$ be a generalized word over $\Sigma$ with
$\Gamma = \alp(u)$. Let $v_a=(L_a;\leq_a,\tau_a)$ be a generalized word
for each $a\in \Gamma$. We define the generalized word 
$u[ (a/v_a)_{a \in \Gamma}]=(L';\leq,\tau')$ as follows:
\begin{itemize}
\item $L'=\{(x,y) \mid y \in L, x \in L_{\tau(y)}\}$,
\item $(x,y)\leq (x',y')$ if and only if either $y<y'$ or ($y=y'$ and $x\leq_{\tau(y)} x'$), and
\item $\tau'(x,y)=\tau_{\tau(y)}(x)$.
\end{itemize} 
Thus, $u[ (a/v_a)_{a \in \Gamma}]$ is obtained from $u$ by 
replacing every $a$-labelled point by $v_a$ (for all $a \in \Sigma$).
Now we can define the regular operations on words. In order to do so
we need the following words. The words $ab$ and $a^\omega$ for $a,b\in
\Sigma$ are as usual. The generalized word $a^{\omegaop}$ has 
$\omegaop$ as underlying order and every element is colored with
$a$. Finally, we let $[a_1,\ldots,a_n]^\eta$ be the generalized word with
underlying order $\eta$ where the coloring is such that any point
is labeled by some $a_i$ $(1\leq i\leq n)$ and, moreover, for any two
points $x<y$ and any $1\leq i\leq n$ we find a point $z$ with $x< z<
y$ colored by $a_i$. It can be shown that this describes a unique word
up to isomorphism \cite{Ros82}.
\begin{definition}[Regular Operations]
  Let $u,v,u_1,\ldots,u_n$ be words over $\Sigma$. We let:
  \begin{alignat*}{2}
    uv&=(ab)[a/u,b/v]   &  \qquad    u^\omega&=a^\omega[a/u]\\
    [u_1,\ldots,u_n]^\eta&=[a_1,\ldots,a_n]^\eta[a_1/u_1,\ldots,a_n/u_n]
    & u^{\omegaop}&=a^{\omegaop}[a/u] .
  \end{alignat*}
\end{definition}
Thus, the underlying linear order of $uv$ 
is the sum of the underlying linear orders of $u$ and $v$.
Intuitively, we have $u^\omega = uuu\cdots$ and 
$u^\omega = \cdots uuu$.
Note that since $[u_1,\ldots,u_n]^\eta$ is invariant under
permutations of the $u_i$ we also sometimes use the notation $X^\eta$
for a finite set $X$.  The least set of words which is closed under
the regular operations and contains the singleton words $a$ for $a\in
\Sigma$ is called the set of {\em regular words} over $\Sigma$, denoted $\Reg(\Sigma)$.
Note that this implies that every regular word is non-empty, i.e., its
domain is a non-empty set. Moreover, although we allow $\Sigma$ to be infinite (this will be useful
later), the alphabet $\alp(u)$ of 
a regular word $u$ must be finite. 
Clearly, every regular word can be described by a {\em regular expression} over 
the above operations, but this regular expression is in general not unique.
\begin{example}
Here are some typical identities between regular words, where
$X$ is a finite set of regular words,  $n \geq 0$, $m \geq 1$,
$u,u_1,\ldots,u_n \in X$, every $v_i$ ($1 \leq i \leq m$) has one of the forms $X^\eta$, $y X^\eta$, $X^\eta z$, $y X^\eta z$
  with $y,z \in X$, 
 and $v, w$ are regular words:
\begin{gather*}
X^\eta X^\eta \cong  X^\eta u X^\eta \cong (X^\eta)^\omega \cong  (X^\eta u)^\omega  \cong (X^\eta)^{\omegaop} \cong  (u X^\eta)^{\omegaop}  \cong X^\eta, \\
[u_1, \ldots, u_n, v_1, \ldots, v_m]^\eta \cong X^\eta, \\
 (vw)^\omega = v (wv)^\omega, \quad (vw)^{\omegaop} = (wv)^{\omegaop} w .
\end{gather*}
See \cite{BloomE05} for a complete axiomatization of the equational theory of regular words.
\end{example}
By a result of Heilbrunner \cite{Heilbrunner80},
regular words can be characterized by 
partitioned DFAs as follows:
Let $\mcA = (Q,\Gamma,\delta,q_0,(F_a)_{a \in\Sigma})$
be a partitioned DFA, and let 
$\mcB = (Q,\Gamma,\delta,q_0,\bigcup_{a \in \Sigma}F_a)$.
Let us fix a linear order on the alphabet $\Gamma$, so 
that the lexicographic order $\leq_{\lex}$ is defined
on $\Gamma^*$.
Then we denote with $w(\mcA)$ the 
generalized word
$$
w(\mcA) = (L(\mcB); \leq_{\lex},\tau),
$$
where $\tau(u) = a$ ($a \in \Sigma$, $u \in L(\mcB)$) 
if and only if $u \in L(Q,\Gamma,\delta,q_0,F_a)$.
It is easy to construct from a given regular expression
(describing the regular word $u$)
a partitioned DFA $\mcA$ with $u \cong w(\mcA)$, see e.g.
\cite[proof of Proposition~2]{Thomas86} for a simple construction.
The other direction is more difficult.
Heilbrunner has shown in \cite{Heilbrunner80} how to compute
from a given partitioned DFA $\mcA$ (such that
$w(\mcA)$ is non-empty) a regular expression
for the word $w(\mcA)$, which is therefore regular.\footnote{In
fact, Heilbrunner speaks about systems of equations and their 
least solutions instead of partitioned DFAs. But these two formalisms can be easily
(and efficiently) transformed into each other.}
Unfortunately, the size of the regular expression produced by
Heilbrunner's algorithm is exponential in the size of 
$\mcA$. In Section~\ref{sec:heilbrunner}, we will see that
a succinct representation of a regular expression for 
$w(\mcA)$ can be produced in polynomial time.

One can show that the isomorphism problem for 
regular words (given by partitioned DFAs) can be reduced (in logspace)
to the isomorphism problem for regular linear orders (given by DFAs).
In other words, node labels can be eliminated as for regular trees 
(as remarked at the end of Section~\ref{sec:trees}). So, the reader might ask, why we consider
the isomorphism problem for regular words and do not restrict
to regular linear orders. The point is that even if we start with
regular linear orders, in the course of our polynomial 
isomorphism check regular words will naturally arise.

\section{Isomorphism problem for regular trees} \label{sec:reg-trees}

In this section, we investigate the isomorphism problem
for (unordered) regular trees. We consider two input representations for 
regular trees: DFAs and NFAs. It turns out that 
while the isomorphism problem for DFA-represented regular trees
is $\Ptime$-complete, the same problem becomes $\EXPTIME$-complete
for NFA-represented regular trees.
Moreover, we show that for {\em finite} trees that are 
succinctly represented by {\em acyclic} NFAs, isomorphism is $\PSPACE$-complete.

\subsection{Upper bounds} \label{sec:upper-bounds-trees}

\begin{theorem} \label{thm:upper-bound-P}
The following problem can be solved in polynomial time:

\medskip
\noindent
INPUT: Two DFAs $\mcA_1$ and $\mcA_2$ such that
$\varepsilon \in L(\mcA_1) \cap L(\mcA_2)$.

\noindent
QUESTION: $(L(\mcA_1); \leq_{\pref}) \cong (L(\mcA_2); \leq_{\pref})$?
\end{theorem}

\begin{proof}
By taking the disjoint union of $\mcA_1$ and $\mcA_2$, 
it suffices to solve the following problem in polynomial time:

\medskip
\noindent
INPUT: A DFA $\mcA$ without initial state and two final states
$p,q$ of $\mcA$.

\noindent
QUESTION: $(L(\mcA,p); \leq_{\pref}) \cong (L(\mcA,q); \leq_{\pref})$?

\medskip
\noindent
Note that $\varepsilon \in L(\mcA,p) \cap L(\mcA,q)$ since
$p$ and $q$ are final.
Let $\mcA = (Q,\Sigma, \delta,F)$.
In fact, we will compute in polynomial time the equivalence relation
$$
\iso = \{(p,q) \in F \times F \mid (L(\mcA,p); \leq_{\pref}) \cong (L(\mcA,q); \leq_{\pref}) \}.
$$
This will be done
similarly to the classical partition refinement algorithm for 
checking bisimilarity of finite state systems \cite{KanSmoS90}.

For $p \in F$ and $C\subseteq F$ let $L(\mcA, p,C)$ be the set of all
words accepted by the DFA $(Q,\Sigma, \delta, p, C)$. Hence,
the sets $L(\mcA, p,\{q\})$ ($q \in F$) partition $L(\mcA,p)$.
Let us say that a node $u \in L(\mcA,p)$ is of type $q$ if 
$u \in L(\mcA, p,\{q\})$. 
For $p \in F$ and $C \subseteq F$ let us define the subset
$K(\mcA,p,C) \subseteq L(\mcA,p,C)$ as the set of all words over $\Sigma$ 
labeling a path from $p$ to a state from $C$ without
intermediate final states; this is clearly a regular language and a DFA
for $K(\mcA,p,C)$ can be easily computed in polynomial time from $\mcA$, $p$, and $C$:
We take the DFA $\mcA$ and remove every transition leaving a final
state from $F$. Moreover, we introduce a copy $p'$ of $p$,
which will be the new initial state and there is an $a$-labeled transition from
$p'$ to $q$ if and only if there is an $a$-labeled transition from
$p$ to $q$ in $\mcA$. Finally, $C$ is the set of final states.

Note that if $u \in L(\mcA,p)$ is of type $q$, then
the nodes $u v$ with $v \in K(\mcA,q,F)$ are exactly
the children of $u$ in the tree $(L(\mcA,p); \leq_{\pref})$.
Let $n(p,q) \in \mathbb{N} \cup \{\infty\}$  be the cardinality of the language $K(p,\{q\})$.
By Lemma~\ref{lemma:counting},  each of these 
numbers $n(p,q)$ can be computed in polynomial time.
For $C\subseteq F$ let $n(p,C) = \sum_{q \in F} n(p,q)$.
Thus $n(p,C)$ is the cardinality of the language $K(p,C)$.

Let us now compute the equivalence relation $\iso$.
As already remarked, this will be done by a partition refinement algorithm.
Assume that $R$ is an equivalence relation on $F$.
We define the new equivalence relation $\widetilde{R}$ on $F$
as follows:
$$
\widetilde{R} = \{ (p,q) \in R \mid n(p,C) = n(q,C) \text{ for every equivalence class $C$ of $R$}\} .
$$
Thus, $\widetilde{R}$ is a refinement of $R$ which can be computed
in polynomial time from $R$.
Let us define a sequence of equivalence relations $R_0, R_1,
\ldots$ on $F$ as follows: $R_0 = F \times F$, 
$R_{i+1} = \widetilde{R}_i$.
Then, there exists $k < |F|$ such that $R_k = R_{k+1}$.
We claim that  $R_k = \iso$.
A simple argument shows that for every equivalence relation
$R$ on $F$ with $\iso \subseteq R$, one has $\iso \subseteq
\widetilde{R}$ as well. Hence, by induction over $i \geq 0$, one gets
$\iso \subseteq R_i$ for all $i \geq 0$.

For the other direction, we show that if $R$ is an equivalence
relation on $F$ such that $R = \widetilde{R}$ (this holds for $R_k$),
then $R \subseteq \iso$. 
So, assume that $(p_1,p_2) \in R =  \widetilde{R}$.
We will define an isomorphism 
 $f:   (L(\mcA,p_1); \leq_{\pref}) \to (L(\mcA,p_2); \leq_{\pref})$ as the
limit of isomorphisms $f_n$, $n \geq 1$. Here, $f_n$ is an isomorphism between the 
trees that result from $(L(\mcA,p_1); \leq_{\pref})$ and $(L(\mcA,p_2); \leq_{\pref})$ by cutting off
all nodes below level $n$ (the roots are one level 1). Let us call these trees 
$(L(\mcA,p_i); \leq_{\pref}) \rest_n$ ($i \in \{1,2\}$).
Moreover, $f_n$ has the additional property that if $f_n$ maps a node 
$u_1$ of type $q_1$ to a node $u_2$ of type $q_2$, then 
we will have $(q_1, q_2) \in R$. 
Assume that $f_n$ is already constructed and let $u_1$ of type $q_1$ be a 
leaf of $(L(\mcA,p_1); \leq_{\pref}) \rest_n$. Let $u_2 = f(u_1)$ be of type $q_2$; it is
a leaf of $(L(\mcA,p_2); \leq_{\pref}) \rest_n$. 
Then we have $(q_1, q_2) \in R = \widetilde{R}$ and hence for every equivalence 
class $C$ of $R$ we have $n(q_1,C) = n(q_2,C)$.
We can therefore find a bijection $g$ between the languages
$K(q_1,F)$ and $K(q_2,F)$ such that $(u,g(u)) \in R$ for all 
$u \in K(q_1,F)$. Note that the nodes $u_i v$ with $v \in K(q_i,F)$
are the children of $u_i$ in the tree $(L(\mcA,p_1); \leq_{\pref})$.
We now extend the isomorphism $f_n$ by $g$ and do this for all leaves
$u_1$ of $(L(\mcA,p_1); \leq_{\pref}) \rest_n$. This gives us the isomorphism
$f_{n+1}$.
\qed
\end{proof}

\begin{corollary} \label{coro-NFA-prefix}
The following problem belongs to $\EXPTIME$:

\medskip
\noindent
INPUT: Two NFAs $\mcA_1$ and $\mcA_2$ such that
$\varepsilon \in L(\mcA_1) \cap L(\mcA_2)$.

\noindent
QUESTION: $(L(\mcA_1); \leq_{\pref}) \cong (L(\mcA_2); \leq_{\pref})$?
\end{corollary}

\begin{proof}
In exponential time, we can transform $\mcA_1$ and $\mcA_2$
into DFAs using the powerset construction.
Then we can apply Theorem~\ref{thm:upper-bound-P}.
\qed
\end{proof}

\begin{theorem} \label{thm:PSPACE-acyclic}
The following problem belongs to $\PSPACE$:

\medskip
\noindent
INPUT: Two acyclic NFAs $\mcA_1$ and $\mcA_2$ such that
$\varepsilon \in L(\mcA_1) \cap L(\mcA_2)$.

\noindent
QUESTION: $(L(\mcA_1); \leq_{\pref}) \cong (L(\mcA_2); \leq_{\pref})$?
\end{theorem}

\begin{proof}
By \cite{Lindell92}, isomorphism for finite trees, given explicitly by
adjacency lists, can be decided in deterministic logspace. Hence, by
Lemma~\ref{PSPACE} it suffices to show that for a given acyclic
NFA, the adjacency list representation for the tree  
$(L(\mcA); \leq_{\pref})$ can be computed by a $\PSPACE$-transducer.
This is straightforward. Assume that $\Sigma$ is the alphabet
of $\mcA$ and that $n$ is the number of states of $\mcA$.
Let us fix an arbitrary order on $\Sigma$ and let $z$ be the largest
symbol in $\Sigma$. 
 
The language $L(\mcA)$ only contains words of length at most $n-1$.
In an outer loop we generate the language
$L(\mcA)$. For this, we enumerate all words (e.g. in lexicographic
order) of length at most $n-1$ and test whether the current word
is accepted by $\mcA$. For each enumerated word $u \in L(\mcA)$,
we have to output a list of all children of $u$ in the tree
$(L(\mcA); \leq_{\pref})$. In an inner loop,
we enumerate (again in lexicographic order) all
words $uv$ ($v \in \Sigma^+$) of length at most $n-1$  
and check whether $uv \in L(\mcA)$.
In case, we find such a word $uv \in L(\mcA)$, we 
output $uv$ and do the following: If $v \in \{z\}^+$, then the inner
loop terminates. On the other hand, if $v = v' a z^k$, where $a \neq z$,
then we jump in the inner loop to the word $u v' b$, where $b$ is the symbol following $a$
in our order.
\qed
\end{proof}

\subsection{Lower bounds}

The main result of this section states that the isomorphism problem for regular
trees that are represented by NFAs is $\EXPTIME$-hard, which matches the
upper bound from the  previous section. It is straightforward to prove
$\PSPACE$-hardness. If $\Sigma$ is the underlying alphabet of a given
NFA $\mcA$, then $(L(\mcA); \leq_{\pref})$ is a full $|\Sigma|$-ary tree 
if and only if $L(\mcA) = \Sigma^*$. But universality for NFAs is 
$\PSPACE$-complete \cite{StMe73}. The proof for the $\EXPTIME$ lower bound
is more involved. Here is a rough outline: $\EXPTIME$  coincides 
with alternating polynomial space \cite{CKS81}. Checking whether a given input
is accepted by a polynomial space bounded alternating Turing machine $M$
amounts to evaluate a Boolean expression whose gates correspond to 
configurations of $M$. Using a construction from \cite{JKMT03}, the evaluation problem
for (finite) Boolean expressions can be reduced to the isomorphism problem for (finite) trees.
In our case, the Boolean expression will be infinite. Nevertheless, the infinite
Boolean expressions we have to deal with can be evaluated because on every
infinite path that starts in the root (the output gate) there will be either an $\AND$-gate,
where one of the inputs is a $\FALSE$-gate, or an $\OR$-gate,
where one of the inputs is a $\TRUE$-gate. Applying the construction from
\cite{JKMT03} to an infinite Boolean expression (that arises from our construction) will 
yield two infinite trees, which are isomorphic if and only if our Boolean expression
evaluates to $\TRUE$. Luckily, these two trees turn out to be regular, and they
can be represented by small NFAs.

\subsubsection{Infinite Boolean formulas.} \label{sec:inf-formulas}

Let us fix the alphabet
\begin{equation}\label{eq:Omega}
\Omega = \{a,\ell_\wedge,\ell'_\wedge,r_\wedge,\ell_\vee,\ell'_\vee,r_\vee\} .
\end{equation}
In the following, we will only consider {\em prefix-closed} trees over the alphabet
$\Omega$ (we will not mention this explicitly all the time).
Moreover, we will identify the tree
$(L;\leq_{\pref})$ with the language $L$.
Now, consider such a tree $T \subseteq \Omega^*$.
Then, $T$ is {\em well-formed}, if the 
following conditions hold:
\begin{enumerate}[(a)]
\item
If $u = \varepsilon$ or $u \in T$ ends with 
$\ell_\vee$, $\ell_\wedge$, $r_\vee$, or $r_\wedge$,
then $\child(u,T)$ is one of the following sets, where $\circ \in \{\vee,\wedge\}$:
$\{u\,\ell_\circ,  u\,r_\circ \}$,
$\{u\,\ell'_\circ, u\,r_\circ\}$,
$\{ua, u\,\ell'_\circ, u \, r_\circ\}$.
\item If $u \in T$ ends with $a$, $\ell'_\vee$, or $\ell'_\wedge$, then $u$ is a leaf of $T$.
\item
For every infinite path $P$ in $T$, there exists $u \in P$ 
with $ua \in T$.
\end{enumerate}
Note that a well-formed tree $T$ is always infinite; it contains
an infinite path of the form $r_1 r_2 r_3 \cdots$, where 
$r_i \in \{r_\wedge, r_\vee\}$ for all $i \geq 1$. 
Let us define the set
\begin{equation} \label{eq:def_C(T)}
\cut(T) = \{ u \in T \mid  ua \in T, \
 \forall v <_{\pref} u :  va \not\in  T \} .
\end{equation}
Hence, on every infinite path in $T$ there is a unique node from $\cut(T)$.

With a well-formed tree $T$ we associate 
an infinite Boolean expression
$\bool(T)$ as follows:
The gates of $\bool(T)$ are the nodes of $T$ that do not end with $a$.
\begin{itemize}
\item 
The set of input gates for $u \in T$ is $\child(u,T) \setminus \{ua\}$.
\item If $u r_\vee \in T$ (resp. $u r_\wedge \in T$), then $u$ is an $\OR$-gate
(resp. $\AND$-gate).
\item If $u \ell'_\wedge \in T$ and 
$u a \not\in T$, then $u \ell'_\wedge$ is a $\TRUE$-gate.
\item If $u \ell'_\wedge \in T$ and 
$u a \in T$, then $u \ell'_\wedge$ is a $\FALSE$-gate.
\item If $u \ell'_\vee \in T$ and 
$u a \not\in T$, then $u \ell'_\vee$ is a $\FALSE$-gate.
\item If $u \ell'_\vee \in T$ and 
$u a \in T$, then $u \ell'_\vee$ is a $\TRUE$-gate.
\end{itemize}
Although $\bool(T)$ is an infinite Boolean formula,
the fact that $T$ is well-formed ensures that 
the root of $\bool(T)$ can be evaluated:
We simply remove from $T$ all nodes that have a proper
prefix from  $\cut(T)$. The resulting
tree has no infinite path and since it is finitely branching
it is finite by K\"onig's lemma. If $u \in \cut(T)$ is such that
$u \ell'_\wedge \in T$ (resp., $u \ell'_\vee \in T$), then
$u$ can be transformed into a $\FALSE$-gate (resp., $\TRUE$-gate).
Then, one has to evaluate the resulting finite Boolean expression.

\begin{figure}[t]                                                   
    \centering                                                        
    \begin{picture}(100,65)(0,-20)                                           
    \unitlength=1mm 
    \gasset{Nframe=n,loopdiam=9,ELdist=.7}                                                        
    \gasset{Nadjust=wh,Nadjustdist=1}                    
    \node(1)(35,25){$q_1$} 
    \node(2)(65,25){$q_2$}                                
   \node(s)(50,-15){$s$} 
    \drawloop[loopangle=160,ELdist=-2](1){$\begin{array}{rl}\ell_\wedge |& \!\ell \\[1mm] r_\wedge| &
        \!r\ell \\[1mm] \ell_\vee| & \!\ell^2 \\[1mm]  r_\vee| & \!r^2\ell \end{array}$}
    \drawloop[loopangle=20,ELdist=-2](2){$\begin{array}{rl}\ell_\wedge |& \!\ell \\[1mm] r_\wedge| &
        \!r\ell \\[1mm] \ell_\vee| & \!r\ell \\[1mm]  r_\vee| &
        \!r^2\ell \end{array}$}
    \drawedge[curvedepth=5](1,2){$\begin{array}{rl}\ell_\vee| & \!r\ell
        \\[1mm]  r_\vee| & \!\ell r\ell \end{array}$}
    \drawedge[curvedepth=5](2,1){$\begin{array}{rl}\ell_\vee| &
        \!\ell^2 \\[1mm]  r_\vee| & \!\ell r\ell \end{array}$} 
    \drawedge[ELside=r,curvedepth=-10,ELdist=0,ELpos=45](1,s){$\begin{array}{rl} \ell'_\vee |& \!\ell^2
        \\[1mm]  \ell'_\vee |& \! r\ell^2 \\[1mm]  \ell'_\wedge |& \! \ell \\[1mm]  a | & \! \ell^3 \end{array}$}
    \drawedge[ELside=l,curvedepth=10,ELdist=0,ELpos=45](2,s){$\begin{array}{rl} \ell'_\vee |& \!\ell^2
        \\[1mm]  \ell'_\vee |& \! r\ell^2 \\[1mm]  \ell'_\wedge |& \! \ell \\[1mm]  a | & \! \ell^3 \\[1mm]  a | & \! \ell r \end{array}$}
    \end{picture}                                                     
    \caption{The transducer}                                    
    \label{fig:trans}                                                   
  \end{figure}
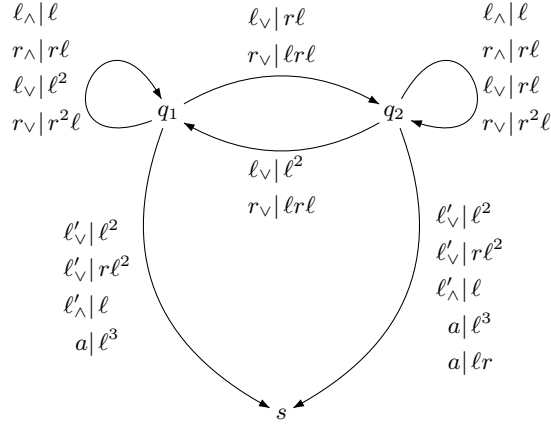

We next transform a tree $T \subseteq \Omega^*$
into trees $[T]_1, [T]_2 \subseteq \{\ell,r\}^*$
using two rational transducers. 
These two transducers only
differ in their initial state.
For $i \in \{1,2\}$, let $\mcT_i$ be the transducer from 
Figure~\ref{fig:trans}, where the initial state is $q_i$
and all states are final. 
Then, for a tree $T \subseteq \Omega^*$ and 
$i \in \{1,2\}$ let $[T]_i = \pref(\mcT_i(T))$. 
We will show that for every well-formed tree $T \subseteq \Omega^*$:
$\bool(T)$ evaluates to true if and only if 
$[T]_1 \cong [T]_2$.
(Lemma~\ref{lemma_inf:toran})
For this, we first have to show a few lemmas.

\begin{lemma}\label{lemma_inf:1}
Let $T = \{\varepsilon, \ell'_\vee\} \cup r_\vee U$ 
or $T = \{\varepsilon,\ell'_\wedge\} \cup r_\wedge U$ 
for a tree $U$ (hence, also $T$ is a tree).
Then $[T]_1 \cong [T]_2$ if and only if $[U]_1 \cong [U]_2$.
\end{lemma}

\begin{proof}
We only prove the lemma for 
$T = \{\varepsilon, \ell'_\vee\} \cup r_\vee U$; the 
statement for $T = \{\varepsilon, \ell'_\wedge\} \cup r_\wedge U$ 
can be shown analogously.
Let us compute compute  $\mcT_1(T)$ and $\mcT_2(T)$. 
We  have 
\begin{equation} \label{eq:mcT_1:1}
\mcT_1(\ell'_\vee) =  \mcT_2(\ell'_\vee) = \{\ell^2, r\ell^2\}.
\end{equation}
Next, we have to compute $\mcT_1(r_\vee U)$.
There are two transitions starting in $q_1$, where $r_\vee$ can
be read, namely 
$$q_1 \xrightarrow{r_\vee|\ell r \ell} q_2 \quad\text{and}\quad
  q_1 \xrightarrow{r_\vee|r^2 \ell} q_1 .
$$
Hence, we get 
\begin{equation} \label{eq:mcT_1:2}
\mcT_1(r_\vee U) = 
r^2\ell\, \mcT_1(U) \cup \ell r \ell\, \mcT_2(U) .
\end{equation}
Similarly, we get
\begin{equation} \label{eq:mcT_2:1}
\mcT_2(r_\vee U) = \
r^2\ell\, \mcT_2(U) \cup \ell r\ell\, \mcT_1(U) .
\end{equation}
From \eqref{eq:mcT_1:1}, \eqref{eq:mcT_1:2}, and \eqref{eq:mcT_2:1}
it follows that the trees 
$[T]_i = \pref(\mcT_i(\{\varepsilon, \ell'_\vee\} \cup r_\vee U))$ ($i \in \{1,2\}$) are the ones
shown in Figure~\ref{fig:lemma_inf:1}.
The equivalence of $[T]_1 \cong [T]_2$ and $[U]_1 \cong [U]_2$
is obvious from these diagrams.
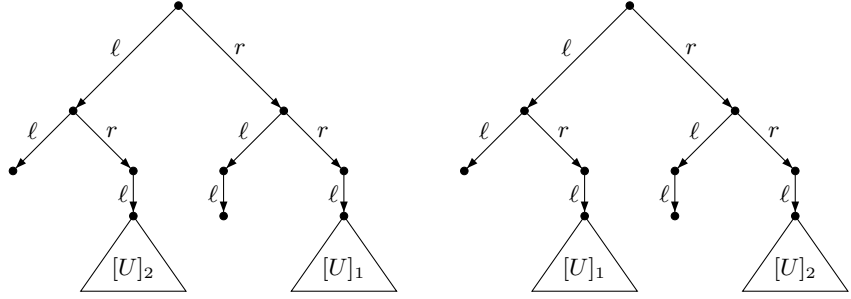
\begin{figure}[t]
\begin{center}                                                        
\begin{picture}(111,40)(0,15)                                           
\unitlength=1mm 
\gasset{Nframe=y,Nfill=y,Nw=1,Nh=1,ELdist=.7,ExtNL=y,NLangle=270,NLdist=5}
\node(root)(22,50){}
\node(l)(8,36){}
\node(r)(36,36){}
\drawedge[ELside=r](root,l){$\ell$}
\drawedge(root,r){$r$}
\node(ll)(0,28){}
\node(lr)(16,28){}
\node(rl)(28,28){}
\node(rr)(44,28){}
\drawedge[ELside=r](l,ll){$\ell$}
\drawedge[ELside=l](l,lr){$r$}
\drawedge[ELside=r](r,rl){$\ell$}
\drawedge[ELside=l](r,rr){$r$}
\node(lrl)(16,22){$[U]_2$}
\node(rrl)(44,22){$[U]_1$}
\drawedge[ELside=r](lr,lrl){$\ell$}
\drawedge[ELside=r](rr,rrl){$\ell$}
\drawpolygon[Nframe=y,Nfill=n](16,22)(9,12)(23,12)
\drawpolygon[Nframe=y,Nfill=n](44,22)(37,12)(51,12)
\node(rll)(28,22){}
\drawedge[ELside=r](rl,rll){$\ell$}
\node(root')(82,50){}
\node(l')(68,36){}
\node(r')(96,36){}
\drawedge[ELside=r](root',l'){$\ell$}
\drawedge(root',r'){$r$}
\node(ll')(60,28){}
\node(lr')(76,28){}
\node(rl')(88,28){}
\node(rr')(104,28){}
\drawedge[ELside=r](l',ll'){$\ell$}
\drawedge[ELside=l](l',lr'){$r$}
\drawedge[ELside=r](r',rl'){$\ell$}
\drawedge[ELside=l](r',rr'){$r$}
\node(lrl')(76,22){$[U]_1$}
\node(rrl')(104,22){$[U]_2$}
\drawedge[ELside=r](lr',lrl'){$\ell$}
\drawedge[ELside=r](rr',rrl'){$\ell$}
\drawpolygon[Nframe=y,Nfill=n](76,22)(69,12)(83,12)
\drawpolygon[Nframe=y,Nfill=n](104,22)(97,12)(111,12)
\node(rll')(88,22){}
\drawedge[ELside=r](rl',rll'){$\ell$}
\end{picture}
\end{center}
\caption{\label{fig:lemma_inf:1} $[T]_1$ (left)  and $[T]_2$ (right) from Lemma~\ref{lemma_inf:1}}
\end{figure}
\qed
\end{proof}
The following three lemmas can be shown with the same kinds of arguments
as for Lemma~\ref{lemma_inf:1}. We therefore only sketch the proofs.

\begin{lemma}\label{lemma_inf:2}
Let $T = \{\varepsilon, \ell'_\vee, a\} \cup r_\vee U$
for a tree $U$ (hence, also $T$ is a tree).
Then $[T]_1 \cong [T]_2$.
\end{lemma}

\begin{proof}
We have 
$\mcT_1(a) = \{\ell^3\}$ and 
$\mcT_2(a) = \{\ell^3, \ell r\}$.
It follows, that the trees $[T]_1$ and $[T]_2$ are as shown in
Figure~\ref{fig:lemma_inf:2}. Clearly, we have $[T]_1 \cong [T]_2$.
\begin{figure}[t]
\begin{center}                                                        
\begin{picture}(111,40)(0,15)                                           
\unitlength=1mm 
\gasset{Nframe=y,Nfill=y,Nw=1,Nh=1,ELdist=.7,ExtNL=y,NLangle=270,NLdist=5}
\node(root)(22,50){}
\node(l)(8,36){}
\node(r)(36,36){}
\drawedge[ELside=r](root,l){$\ell$}
\drawedge(root,r){$r$}
\node(ll)(0,28){}
\node(lr)(16,28){}
\node(rl)(28,28){}
\node(rr)(44,28){}
\drawedge[ELside=r](l,ll){$\ell$}
\drawedge[ELside=l](l,lr){$r$}
\drawedge[ELside=r](r,rl){$\ell$}
\drawedge[ELside=l](r,rr){$r$}
\node(lll)(0,22){}
\node(lrl)(16,22){$[U]_2$}
\node(rrl)(44,22){$[U]_1$}
\drawedge[ELside=r](ll,lll){$\ell$}
\drawedge[ELside=r](lr,lrl){$\ell$}
\drawedge[ELside=r](rr,rrl){$\ell$}
\drawpolygon[Nframe=y,Nfill=n](16,22)(9,12)(23,12)
\drawpolygon[Nframe=y,Nfill=n](44,22)(37,12)(51,12)
\node(rll)(28,22){}
\drawedge[ELside=r](rl,rll){$\ell$}
\node(root')(82,50){}
\node(l')(68,36){}
\node(r')(96,36){}
\drawedge[ELside=r](root',l'){$\ell$}
\drawedge(root',r'){$r$}
\node(ll')(60,28){}
\node(lr')(76,28){}
\node(rl')(88,28){}
\node(rr')(104,28){}
\drawedge[ELside=r](l',ll'){$\ell$}
\drawedge[ELside=l](l',lr'){$r$}
\drawedge[ELside=r](r',rl'){$\ell$}
\drawedge[ELside=l](r',rr'){$r$}
\node(lll')(60,22){}
\node(lrl')(76,22){$[U]_1$}
\node(rrl')(104,22){$[U]_2$}
\drawedge[ELside=r](ll',lll'){$\ell$}
\drawedge[ELside=r](lr',lrl'){$\ell$}
\drawedge[ELside=r](rr',rrl'){$\ell$}
\drawpolygon[Nframe=y,Nfill=n](76,22)(69,12)(83,12)
\drawpolygon[Nframe=y,Nfill=n](104,22)(97,12)(111,12)
\node(rll')(88,22){}
\drawedge[ELside=r](rl',rll'){$\ell$}
\end{picture}
\end{center}
\caption{\label{fig:lemma_inf:2} $[T]_1$ (left)  and $[T]_2$ (right) from Lemma~\ref{lemma_inf:2}}
\end{figure}
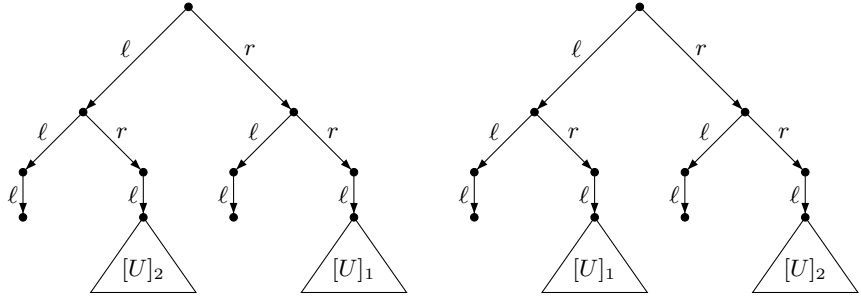
\qed
\end{proof}

\begin{lemma}\label{lemma_inf:3}
Let $T = \{\varepsilon, \ell'_\wedge, a\} \cup r_\wedge U$
for a tree $U$ (hence, also $T$ is a tree).
Then $[T]_1 \not\cong [T]_2$.
\end{lemma}

\begin{proof}
The trees $[T]_1$ and $[T]_2$ are shown in
Figure~\ref{fig:lemma_inf:3}. Clearly, we have $[T]_1 \not\cong [T]_2$.
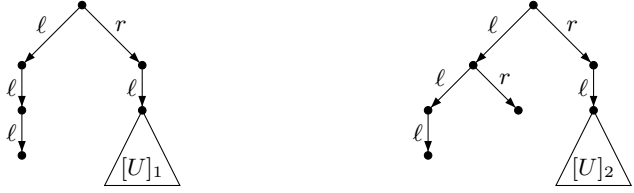
\begin{figure}[t]
\begin{center}                                                        
\begin{picture}(115,26)(-5,15)                                           
\unitlength=1mm 
\gasset{Nframe=y,Nfill=y,Nw=1,Nh=1,ELdist=.7,ExtNL=y,NLangle=270,NLdist=5.7}
\node(root)(22,36){}
\node(l)(14,28){}
\node(ll)(14,22){}
\node(lll)(14,16){}
\node(r)(30,28){}
\drawedge[ELside=r](root,l){$\ell$}
\drawedge[ELside=r](l,ll){$\ell$}
\drawedge[ELside=r](ll,lll){$\ell$}
\drawedge[ELside=l](root,r){$r$}
\node(rl)(30,22){$[U]_1$}
\drawedge[ELside=r](r,rl){$\ell$}
\drawpolygon[Nframe=y,Nfill=n](30,22)(25,12)(35,12)
\node(root')(82,36){}
\node(l')(74,28){}
\node(r')(90,28){}
\node(ll')(68,22){}
\node(lr')(80,22){}
\node(lll')(68,16){}
\drawedge[ELside=r](root',l'){$\ell$}
\drawedge[ELside=r](l',ll'){$\ell$}
\drawedge[ELside=l](l',lr'){$r$}
\drawedge[ELside=r](ll',lll'){$\ell$}
\drawedge[ELside=l](root',r'){$r$}
\node(rl')(90,22){$[U]_2$}
\drawedge[ELside=r](r',rl'){$\ell$}
\drawpolygon[Nframe=y,Nfill=n](90,22)(85,12)(95,12)
\end{picture}
\end{center}
\caption{\label{fig:lemma_inf:3} $[T]_1$ (left)  and $[T]_2$ (right) from Lemma~\ref{lemma_inf:3}}
\end{figure}
\qed
\end{proof}

\begin{lemma}\label{lemma_inf:4}
Let $T = \{\varepsilon\} \cup \ell_\vee U \cup r_\vee V$
for well-formed trees $U,V$ (hence, also $T$ is well-formed).
Then $[T]_1 \cong [T]_2$ if and only if 
$([U]_1 \cong [U]_2$ or  $[V]_1 \cong [V]_2)$.
\end{lemma}

\begin{proof}
The trees $[T]_1$ and $[T]_2$ are shown in
Figure~\ref{fig:lemma_inf:4}. Since $U$ and $V$ 
are well-formed, in each of the trees $[U]_1$, $[U]_2$, 
$[V]_1$, and $[V]_2$, the root has two children.
It follows easily that $[T]_1 \cong [T]_2$ if and only if 
($[U]_1 \cong [U]_2$ or  $[V]_1 \cong [V]_2$).
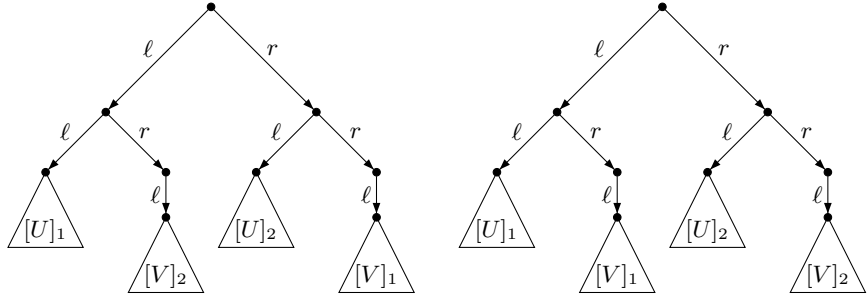
\begin{figure}[t]
\begin{center}                                                        
\begin{picture}(115,40)(-5,15)                                           
\unitlength=1mm 
\gasset{Nframe=y,Nfill=y,Nw=1,Nh=1,ELdist=.7,ExtNL=y,NLangle=270,NLdist=5.7}
\node(root)(22,50){}
\node(l)(8,36){}
\node(r)(36,36){}
\drawedge[ELside=r](root,l){$\ell$}
\drawedge(root,r){$r$}
\node(ll)(0,28){$[U]_1$}
\node(lr)(16,28){}
\node(rl)(28,28){$[U]_2$}
\node(rr)(44,28){}
\drawedge[ELside=r](l,ll){$\ell$}
\drawedge[ELside=l](l,lr){$r$}
\drawedge[ELside=r](r,rl){$\ell$}
\drawedge[ELside=l](r,rr){$r$}
\node(lrl)(16,22){$[V]_2$}
\node(rrl)(44,22){$[V]_1$}
\drawedge[ELside=r](lr,lrl){$\ell$}
\drawedge[ELside=r](rr,rrl){$\ell$}
\drawpolygon[Nframe=y,Nfill=n](0,28)(-5,18)(5,18)
\drawpolygon[Nframe=y,Nfill=n](28,28)(23,18)(33,18)
\drawpolygon[Nframe=y,Nfill=n](16,22)(11,12)(21,12)
\drawpolygon[Nframe=y,Nfill=n](44,22)(39,12)(49,12)
\node(root')(82,50){}
\node(l')(68,36){}
\node(r')(96,36){}
\drawedge[ELside=r](root',l'){$\ell$}
\drawedge(root',r'){$r$}
\node(ll')(60,28){$[U]_1$}
\node(lr')(76,28){}
\node(rl')(88,28){$[U]_2$}
\node(rr')(104,28){}
\drawedge[ELside=r](l',ll'){$\ell$}
\drawedge[ELside=l](l',lr'){$r$}
\drawedge[ELside=r](r',rl'){$\ell$}
\drawedge[ELside=l](r',rr'){$r$}
\node(lrl')(76,22){$[V]_1$}
\node(rrl')(104,22){$[V]_2$}
\drawedge[ELside=r](lr',lrl'){$\ell$}
\drawedge[ELside=r](rr',rrl'){$\ell$}
\drawpolygon[Nframe=y,Nfill=n](60,28)(55,18)(65,18)
\drawpolygon[Nframe=y,Nfill=n](88,28)(83,18)(93,18)
\drawpolygon[Nframe=y,Nfill=n](76,22)(71,12)(81,12)
\drawpolygon[Nframe=y,Nfill=n](104,22)(99,12)(109,12)
\end{picture}
\end{center}
\caption{\label{fig:lemma_inf:4} $[T]_1$ (left)  and $[T]_2$ (right) from Lemma~\ref{lemma_inf:4}}
\end{figure}
\qed
\end{proof}

\begin{lemma}\label{lemma_inf:5}
Let $T = \{\varepsilon\} \cup \ell_\wedge U \cup r_\wedge V$
for well-formed trees $U,V$ (hence, also $T$ is well-formed).
Then $[T]_1 \cong [T]_2$ if and only if 
$([U]_1 \cong [U]_2$ and  $[V]_1 \cong [V]_2)$.
\end{lemma}

\begin{proof}
The trees $[T]_1$ and $[T]_2$ are as shown in
Figure~\ref{fig:lemma_inf:5}. Since $U$ and $V$ 
are well-formed, in each of the trees $[U]_1$, $[U]_2$, 
$[V]_1$, and $[V]_2$, the root has two children.
It follows easily that $[T]_1 \cong [T]_2$ if and only if 
($[U]_1 \cong [U]_2$ and  $[V]_1 \cong [V]_2$).
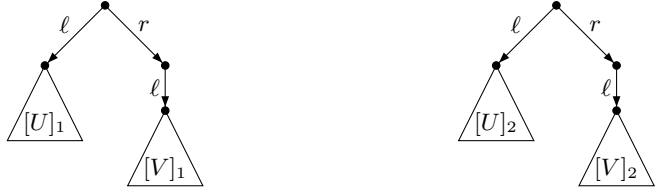
\begin{figure}[t]
\begin{center}                                                        
\begin{picture}(115,26)(-5,15)                                           
\unitlength=1mm 
\gasset{Nframe=y,Nfill=y,Nw=1,Nh=1,ELdist=.7,ExtNL=y,NLangle=270,NLdist=5.7}
\node(l)(22,36){}
\node(ll)(14,28){$[U]_1$}
\node(lr)(30,28){}
\drawedge[ELside=r](l,ll){$\ell$}
\drawedge[ELside=l](l,lr){$r$}
\node(lrl)(30,22){$[V]_1$}
\drawedge[ELside=r](lr,lrl){$\ell$}
\drawpolygon[Nframe=y,Nfill=n](14,28)(9,18)(19,18)
\drawpolygon[Nframe=y,Nfill=n](30,22)(25,12)(35,12)
\node(l')(82,36){}
\node(ll')(74,28){$[U]_2$}
\node(lr')(90,28){}
\drawedge[ELside=r](l',ll'){$\ell$}
\drawedge[ELside=l](l',lr'){$r$}
\node(lrl')(90,22){$[V]_2$}
\drawedge[ELside=r](lr',lrl'){$\ell$}
\drawpolygon[Nframe=y,Nfill=n](74,28)(69,18)(79,18)
\drawpolygon[Nframe=y,Nfill=n](90,22)(85,12)(95,12)
\end{picture}
\end{center}
\caption{\label{fig:lemma_inf:5} $[T]_1$ (left)  and $[T]_2$ (right) from Lemma~\ref{lemma_inf:5}}
\end{figure}
\qed
\end{proof}

\begin{lemma}\label{lemma_inf:toran}
For every well-formed tree $T \subseteq \Omega^*$, we have:
$\bool(T)$ evaluates to true if and only if 
$[T]_1 \cong [T]_2$.
\end{lemma}

\begin{proof}
Recall the definition of the set $\cut(T)$ from \eqref{eq:def_C(T)}.
From the definition it follows that
$\pref(\cut(T))$ is a finitely branching tree without
infinite paths. Hence, by K\"onig's lemma it is finite.
Moreover, for every $u \in \pref(\cut(T))$, the subtree $T\rest_u$ 
is well-formed as well (since 
$\pref(\cut(T)) \subseteq \{\varepsilon\} \cup \Omega^*\{\ell_\vee,\ell_\wedge,r_\vee,r_\wedge\}$).
Inductively over the height of $u \in \pref(\cut(T))$
in the finite tree $\pref(\cut(T))$, we will prove 
for every $u \in \pref(\cut(T))$:  $[T\rest_u]_1 \cong [T\rest_u]_2$
if and only if  $\bool(T\rest_u)$ evaluates to $\TRUE$.

For the induction base, let $u \in \cut(T)$ be a leaf of $\pref(\cut(T))$.
Hence, we have $ua \in T$.
If $u \ell'_\wedge \in T$, then in $\bool(T\rest_u)$, 
the root is an $\AND$-gate for which one of the inputs (namely $u\ell'_\wedge$) is a 
$\FALSE$-gate. Hence, $\bool(T\rest_u)$ evaluates to $\FALSE$.
Moreover, Lemma~\ref{lemma_inf:3} implies that 
$[T\rest_u]_1 \not\cong [T\rest_u]_2$.
On the other hand, if $u \ell'_\vee \in T$, then in $\bool(T\rest_u)$, 
the root is an $\OR$-gate for which one of the inputs (namely
$u\ell'_\vee$) is a 
$\TRUE$-gate. Hence, $\bool(T\rest_u)$ evaluates to $\TRUE$.
Moreover, Lemma~\ref{lemma_inf:2} implies that 
$[T\rest_u]_1 \cong [T\rest_u]_2$.
This concludes the induction base.

Next, let $u \in \pref(\cut(T))$ be a proper prefix of a node from
$\cut(T)$. In particular $u \not\in \cut(T)$.
We can distinguish 4 different cases:

\medskip
\noindent
{\em Case 1.} $\child(u,T) = \{ u\ell_\wedge, ur_\wedge \}$.
We must have $\{ u\ell_\wedge,  u r_\wedge\} \subseteq \pref(\cut(T))$.
Hence, the induction hypothesis (IH) holds for 
$u\ell_\wedge$ and $ur_\wedge$.
We get:
\begin{eqnarray*}
\bool(T\rest_u) \text{ evaluates to } \TRUE   & \Longleftrightarrow & 
     \bool(T\rest_{u\ell_\wedge}) \text{ evaluates to } \TRUE \text{
       and } \\
& &  \bool(T\rest_{ur_\wedge}) \text{ evaluates to } \TRUE \\
& \stackrel{\text{(IH)}}{\Longleftrightarrow} &  
     [T\rest_{u\ell_\wedge}]_1 \cong [T\rest_{u\ell_\wedge}]_2
       \text{ and } \\
& &  [T\rest_{ur_\wedge}]_1 \cong [T\rest_{ur_\wedge}]_2 \\
& \stackrel{\text{Lemma~\ref{lemma_inf:5}}}{\Longleftrightarrow} & 
      [T\rest_u]_1 \cong [T\rest_u]_2
\end{eqnarray*}
{\em Case 2.} $\child(u,T) = \{ u\ell_\vee, ur_\vee \}$.
This case is analogous to Case 1, using
Lemma~\ref{lemma_inf:4}.

\medskip
\noindent
{\em Case 3.} $\child(u,T) = \{ u\ell'_\wedge, ur_\wedge \}$.
Since $u \not\in \cut(T)$, we have $u a \not\in T$.
We must have $u r_\wedge \in \pref(\cut(T))$.
Moreover, in $\bool(T\rest_u)$, the root is an $\AND$-gate,
where one of the inputs is a $\TRUE$-gate and the other 
input is the root for the Boolean expression $\bool(T\rest_{ur_\wedge})$.
Hence, we get:
\begin{eqnarray*}
\bool(T\rest_u) \text{ evaluates to } \TRUE   & \Longleftrightarrow & 
    \bool(T\rest_{ur_\wedge}) \text{ evaluates to } \TRUE \\
& \stackrel{\text{(IH)}}{\Longleftrightarrow} & 
     [T\rest_{ur_\wedge}]_1 \cong [T\rest_{ur_\wedge}]_2 \\
& \stackrel{\text{Lemma~\ref{lemma_inf:1}}}{\Longleftrightarrow} & 
      [T\rest_u]_1 \cong [T\rest_u]_2
\end{eqnarray*}
{\em Case 4.} $\child(u,T) = \{ u\ell'_\vee, ur_\vee \}$.
This case is analogous to Case 3.
\qed
\end{proof}
Our last auxiliary lemma states that an NFA for the tree
$[L]_i$ can be easily computed from an NFA for $L$.

\begin{lemma} \label{prop:transducer}
There is a logspace machine that computes
from a given prefix-closed NFA $\mcA$ with terminal alphabet $\Omega$
a prefix-closed NFA $\mcB$ such that $L(\mcB) =
[L(\mcA)]_i$ for $i \in \{1,2\}$.
\end{lemma}

\begin{proof}
Let $\mcA = (Q, \Omega, \delta, p_0,Q)$.
Recall that all states of $\mcT_i$ and $\mcA$ are final.
The prefix-closed NFA $\mcB$ is obtained from the direct product
of $\mcA$ and $\mcT_i$ by adding further states so
that every transition is labeled with a single symbol.
Thus, the set of states of $\mcB$ contains
$Q \times \{q_1,q_2,s\}$ 
and the initial state of $\mcB$ is $(p_0,q_i)$.
If $q \xrightarrow{b} q'$ in $\mcA$ and 
$t \xrightarrow{b|w} t'$ in $\mcT_i$ for 
$w \in \{\ell,r\}^+$, then we add $|w|-1$ many new states
to $\mcB$, which built up a $w$-labeled path from from $(q,t)$ to $(q',t')$. 
\qed
\end{proof}

\subsubsection{$\EXPTIME$-hardness.} \label{sec:lower-bound-exptime-tree}

We are now in the position to prove the main result of this section.

\begin{theorem}\label{thm:EXPTIME-hard}
The following problem is $\EXPTIME$-hard (and hence $\EXPTIME$-com\-plete):

\medskip
\noindent
INPUT: Two prefix-closed NFAs $\mcA_1$ and $\mcA_2$.

\noindent
QUESTION: $(L(\mcA_1); \leq_{\pref}) \cong (L(\mcA_2); \leq_{\pref})$?
\end{theorem}

\begin{proof}
The upper bound is stated in Corollary~\ref{coro-NFA-prefix}.
For the lower bound we use the fact that $\EXPTIME$ equals the
class of all sets that can be accepted in polynomial space on an
alternating Turing machine \cite{CKS81}.
Hence, let $M$ be a polynomial space bounded alternating Turing
machine such that the accepted language $L(M) \subseteq \{0,1\}^*$
is $\EXPTIME$-complete. 
We can assume that $M$ has no infinite computation paths.
By padding inputs, we can moreover assume that 
$M$ works in space $n$ for an input of length $n$.
Let $Q = Q_\exists \cup Q_\forall$ be the set
of states of $M$ and let $\Gamma \supseteq \{0,1\}$ be the tape
alphabet. W.l.o.g. we can assume that in every computation step,
$M$ moves from an existential state to a universal state or vice
versa, and that the initial state $q_0$ is universal.

Let us now fix an input $w \in \{0,1\}^*$ of length $n$.
We will construct two prefix-closed NFAs $\mcA_1$ and $\mcA_2$ such that
$w \in L(M)$ if and only if $(L(\mcA_1); \leq_{\pref}) \cong (L(\mcA_2);
\leq_{\pref})$. Let $\Theta = \Gamma \cup Q$.
As usual, a configuration of $M$ can be represented
by a string from the language $\Theta^{n+1}$ (more precisely,
from $\bigcup_{j=0}^{n-1} \Gamma^j Q \Gamma^{n-j}$).
A word $u \in \Theta^*$ is a \emph{valid computation of $M$ on input $w$}
if $u$ is of the form $c_1 \cdots c_m$ for
some  $m \geq 0$ such that
the following holds:
\begin{itemize}
\item $c_i \in \bigcup_{j=0}^{n-1} \Gamma^j Q \Gamma^{n-j}$ 
for all $1 \leq i \leq m$
\item $c_i \vdash_M c_{i+1}$ (i.e., $c_{i+1}$ is a successor
  configuration of $c_i$)  for all $1 \leq i \leq m-1$
\item $q_0 w \vdash_M c_1$
\end{itemize}
Note that $\varepsilon$ is a valid computation in this sense.
It is well known that from $w$ one can construct in logspace
a coaccessible NFA $\mcA_w$ such that $\mcA_w$ accepts all words over $\Theta$
that are {\em not} a valid computation of $M$ on $w$ \cite{StMe73}.

Next, we will define a regular well-formed tree 
$T_w \subseteq \Omega^*$
(depending only on $w$)
such that 
$\bool(T_w)$ evaluates to $\TRUE$ if and only if 
$w \in L(M)$.
In the following, we identify the symbols in $\Theta$
with the integers $0, \ldots, |\Theta|-1$ in an arbitrary way.
We can assume that $|\Theta|\geq 2$.
We define two morphisms
\begin{align*}
\phi_\wedge : & \;\Theta^* \to \{\ell_\wedge,r_\wedge\}^* \\
\phi_\vee  :  & \;\Theta^* \to \{\ell_\vee,r_\vee\}^* 
\end{align*}
as follows ($\circ \in \{\wedge,\vee\}$):
\begin{eqnarray*}
\phi_\circ(a) & = & 
  \begin{cases}
  r_\circ^a \ell_\circ & \text{if } 0 \leq a < |\Theta|-1 \\
  r_\circ^a   & \text{if } a = |\Theta|-1
  \end{cases}
\end{eqnarray*}
For $i  \geq 1$, let $\phi_i$ be the mapping $\phi_\wedge$
(resp. $\phi_\vee$) if $i$ is odd (resp., even).
Similarly, for $x \in \{\ell,\ell',r\}$, let $x_i$ be $x_\wedge$
(resp. $x_\vee$) if $i$ is odd (resp., even).
Then, the tree $T_w \subseteq \Omega^*$ is $\pref(T'_w)$, where
\begin{align*}
T'_w = \; & \bigg\{ \bigg(\prod_{i=1}^{m} r_i \phi_i(c_i) \bigg) \ell'_{m+1} \mid
              m \geq 0, c_1, \ldots, c_m \in \Theta^{n+1} \bigg\} \;\cup \\
         & \bigg\{ \bigg(\prod_{i=1}^{m} r_i \phi_i(c_i) \bigg) a \mid
              m \geq 0, c_1, \ldots, c_m \in \Theta^{n+1}, c_1 \cdots c_m  \in L(\mcA_w) 
            \bigg\} 
\end{align*}
Clearly, $T_w$ is regular, and a prefix-closed NFA for $T_w$ can be computed in
logspace from $w$ (using the logspace computable coaccessible NFA $\mcA_w$).

\medskip
\noindent
{\em Claim 1:} $T_w$ is well-formed.

\medskip
\noindent
{\em Proof of Claim~1:} The first three conditions for well-formed
trees are easy to check. For the last condition, we have to consider
an arbitrary infinite path $P$ of $T_w$ and show that 
there exists $u \in T_w$ such that $u a \in T$. But this means that
$u$ is of the form 
$$
u = \prod_{i=1}^{m} r_i \phi_i(c_i) 
$$
with $m \geq 0$, $c_1, \ldots, c_m \in \Theta^{n+1}$, and
$c_1 \cdots c_m  \in L(\mcA_w)$. The latter condition means that
$c_1 \cdots c_m$ is not a valid computation of $M$ on input $w$.
Claim~1 now follows from the fact that for every infinite sequence
$c_1 c_2 c_3 \cdots$ with $c_i \in \Theta^{n+1}$ for $i \geq 1$
there exists $m \geq 1$ such that $c_1 \cdots c_m$ is not a valid
computation of $M$ on input $w$ 
(since $M$ does not have infinite computation paths).

\medskip
\noindent
{\em Claim 2:} $w \in L(M)$ if and only if $\bool(T_w)$ evaluates
to $\TRUE$.

\medskip
\noindent
{\em Proof of Claim~2:}  
Let us consider the {\em finite} tree $\pref(\cut(T_w))$.
For every node 
$$
g = r_\wedge \phi_\wedge(c_1) r_\vee \phi_\vee(c_2) r_\wedge \cdots \phi_{m-1}(c_{m-1}) r_m
\phi_m(c_m) \in \pref(\cut(T_w))
$$
with $m \geq 0$ and  $c_1, \ldots, c_m \in \Theta^{n+1}$ we will prove
(by induction on the height of $g$) the following: If $c_1 \cdots c_m$ is a valid computation of $M$ on
input $w$, then $c_m$ is an accepting configuration if and only if 
$g$ evaluates to true in $\bool(T_w)$. Here, for $m = 0$, we define
$c_0$ as the initial configuration $q_0 w$.

So, assume that $g \in \pref(\cut(T_w))$ is of the above form
and that $c_1 \cdots c_m$ is a valid computation of $M$ on input $w$.
W.l.o.g. assume that 
$m$ is odd (the case that $m$ is even can be dealt analogously).
Thus,
$$
g = r_\wedge \phi_\wedge(c_1) r_\vee \phi_\vee(c_2) r_\wedge \cdots \phi_\vee(c_{m-1}) r_\wedge
\phi_\wedge(c_m) .
$$
Then, in $\bool(T_w)$, the input gates for the $\OR$-gate $g$ are
$g \ell'_\vee$ and $g r_\vee$.
Since $c_1 \cdots c_m$ is a valid computation of $M$ on
input $w$, $g a$ does not belong to the tree $T_w$.
Hence, in $\bool(T_w)$, $g \ell'_\vee$ is a $\FALSE$-gate.
Thus, $g$ evaluates to $\TRUE$ if and only if $g r_\vee$ evaluates
to $\TRUE$. 
From the structure of $T_w$ we see that the latter holds if and only
if there exists $c_{m+1} \in \Theta^{n+1}$ such that
$g r_\vee \phi_\vee(c_{m+1})$ evaluates to $\TRUE$.
First assume that $c_{m+1}$ is such that
$c_1 \cdots c_m c_{m+1}$ is not a valid computation.
The inputs for the $\AND$-gate $g r_\vee \phi_\vee(c_{m+1})$
are $g r_\vee \phi_\vee(c_{m+1}) \ell'_\wedge$ and 
$g r_\vee \phi_\vee(c_{m+1}) r_\wedge$. Since $c_1 \cdots c_m c_{m+1}$ 
is not a valid computation, $g r_\vee \phi_\vee(c_{m+1}) a$
belongs to the tree $T_w$. Thus, in $\bool(T_w)$,
$g r_\vee \phi_\vee(c_{m+1}) \ell'_\wedge$ is a $\FALSE$-gate
and $g r_\vee \phi_\vee(c_{m+1})$ evaluates to $\FALSE$.
This holds for all $c_{m+1}$  such that
$c_1 \cdots c_m c_{m+1}$ is not a valid computation.
Hence, $g r_\vee$ evaluates
to $\TRUE$ if and only if there exists a configuration
$c_{m+1} \in \Theta^{n+1}$ such that $c_1 \cdots c_m c_{m+1}$
is a valid computation (which means that $c_{m+1}$ is a 
successor configuration of $c_m$) and $g r_\vee \phi_\vee(c_{m+1})$
evaluates to $\TRUE$ in $\bool(T_w)$. 
Now, if $c_1 \cdots c_m c_{m+1}$ is a valid computation,
then by induction, $g r_\vee \phi_\vee(c_{m+1})$
(which belongs to $\pref(\cut(T_w))$ as well)
evaluates to $\TRUE$ in $\bool(T_w)$ if and only if 
$c_{m+1}$ is an accepting configuration of $M$.

We have shown that $g$ evaluates to $\TRUE$
if and only if $c_m$ has an accepting successor 
configuration. Finally, since $m$ is odd, $c_m$ 
is an existential configuration (recall  that the initial
configuration $c_0 = q_0w$
is universal). Thus, indeed,
$g$ evaluates to $\TRUE$ if and only if $c_m$ is
accepting. This proves Claim~2.

\medskip
\noindent
Let $\mcT_1$ and $\mcT_2$ be the rational transducers from
Section~\ref{sec:inf-formulas}. Using Lemma~\ref{prop:transducer}
we can compute in logspace from a prefix-closed NFA for $T_w$ two
prefix-closed NFAs $\mcA_1$ and $\mcA_2$ such that 
$L(\mcA_i) = [T_w]_i$ for $i \in \{1,2\}$.
By Lemma~\ref{lemma_inf:toran} and Claim~2, we have
$$
w \in L(M) \ \Longleftrightarrow \ \bool(T_w) 
\text{ evaluates to } \TRUE \ \Longleftrightarrow \
(L(\mcA_1); \leq_{\pref}) \cong (L(\mcA_2); \leq_{\pref}) .
$$
This concludes the proof of the $\EXPTIME$ lower bound.
\qed
\end{proof}

\subsubsection{$\PSPACE$-hardness}

\begin{theorem} \label{thm:PSPACE-hard}
The following problem is $\PSPACE$-hard (and therefore
$\PSPACE$-com\-plete):

\medskip
\noindent
INPUT: Two prefix-closed acyclic NFAs $\mcA_1$ and $\mcA_2$.

\noindent
QUESTION: $(L(\mcA_1); \leq_{\pref}) \cong (L(\mcA_2); \leq_{\pref})$?
\end{theorem}

\begin{proof}
The upper bound is stated in Theorem~\ref{thm:PSPACE-acyclic}.
For the lower bound, we use the same idea as in the proof of
Theorem~\ref{thm:EXPTIME-hard}. In fact, we will use most 
of the notations from that proof; some of them will be 
slightly modified.
This time, we use the fact that $\PSPACE$ equals the
class of all sets that can be accepted in polynomial time on an
alternating Turing machine. 
Hence, let $M$ be a polynomial time bounded alternating Turing
machine such that the accepted language $L(M) \subseteq \{0,1\}^*$
is $\PSPACE$-complete. Let $p(n)$ (a polynomial) be the time
bound and let $q(n)=p(n)+1$. We can assume that $q(n)$ is odd for all $n \geq 0$.
W.l.o.g. we can assume again that $M$ works 
in space $n$ for an input of length $n$. Let $w \in \{0,1\}^*$
be an input for $M$ of length $n$. 

Let us add to the alphabet $\Omega$ in \eqref{eq:Omega} an additional
symbol $r'_\vee$. The notions from Section~\ref{sec:inf-formulas} have to 
be extended to this new alphabet $\Omega$. In condition (a) for the
definition of a well-formed tree $T$, we also allow the set
$\{ua, u\ell'_\vee, u r'_\vee\}$ for $\child(u,T)$. Moreover, every node 
$u r'_\vee \in T$  is a leaf of $T$.
The new definition for the set $\cut(T)$ can be overtaken from
\eqref{eq:def_C(T)}. Also the Boolean expression
$\bool(T)$ can be defined as in Section~\ref{sec:inf-formulas};
the truth value of a leaf ending with $r'_\vee$ is set arbitrarily (say $\TRUE$).
Finally, let us extend the two transducers $\mcT_1$ and $\mcT_2$ such
that, from $q_1$ and $q_2$ they can read the new symbol $r'_\vee$ 
and output $\ell$ and then terminate in a sink state $s$.

We now define the well-formed tree $U_w \subseteq \Omega^*$ as 
$U_w = \pref(U'_w)$, where:
\begin{align*}
U'_w = \; & \bigg\{ \bigg(\prod_{i=1}^{m} r_i \phi_i(c_i) \bigg) \ell'_{m+1} \mid
              0 \leq m \leq q(n), c_1, \ldots, c_m \in \Theta^{n+1} \bigg\} \;\cup \\
         & \bigg\{ \bigg(\prod_{i=1}^{m} r_i \phi_i(c_i) \bigg)  a \mid
              0 \leq m \leq q(n), c_1, \ldots, c_m \in \Theta^{n+1}, c_1 \cdots c_m  \in L(\mcA_w) 
            \bigg\} \;\cup \\
   & \bigg\{ \bigg(\prod_{i=1}^{q(n)} r_i \phi_i(c_i)\bigg) r'_\vee   \mid
     c_1, \ldots, c_{q(n)} \in \Theta^{n+1} \bigg\} .
\end{align*}
Note that $U_w$ is finite. An acyclic prefix-closed NFA for $U_w$ can 
be produced in logspace from $w$. Moreover,
since every  word from $\Theta^{(n+1)q(n)}$
is not a valid computation (since $M$ terminates after $\leq p(n)=q(n)-1$ 
steps), the Boolean expression
$\bool(U_w)$ and $\bool(T_w)$ (where 
$T_w$ was defined in the proof of Theorem~\ref{thm:EXPTIME-hard})
evaluate to the same truth value.
Hence, using Claim~2 from the proof
of  Theorem~\ref{thm:EXPTIME-hard}, it follows 
that $w \in L(M)$ if and only if $\bool(U_w)$
evaluates to $\TRUE$. Using an analogon of
Lemma~\ref{lemma_inf:toran},
this holds if and only if $[U_w]_1 \cong [U_w]_2$.
Acyclic NFAs for $[U_w]_1$ and $[U_w]_2$ can be easily
constructed in logspace from $w$ (using an acyclic NFA
for $U_w$). This concludes the proof of the theorem.
\qed
\end{proof}

\subsubsection{$\Ptime$-hardness}

\begin{theorem} \label{thm:P-hard}
The following problem is $\Ptime$-hard
(and hence $\Ptime$-complete):

\medskip
\noindent
INPUT: Two prefix-closed acyclic DFAs $\mcA_1$ and $\mcA_2$.

\noindent
QUESTION: $(L(\mcA_1); \leq_{\pref}) \cong (L(\mcA_2); \leq_{\pref})$?
\end{theorem}

\begin{proof}
The upper bound is stated in Theorem~\ref{thm:upper-bound-P}.
For the lower bound,
we reduce the $\Ptime$-complete monotone circuit value problem \cite{Gol77}
to the problem from the theorem. Note that the tree $(L(\mcA); \leq_{\pref})$, where
$\mcA$ is  a prefix-closed acyclic DFA, is just the unfolding of the
underlying dag (directed acyclic graph) in the initial of $\mcA$. Vice versa, from a dag $D$ with
a root node $r$ one can construct a prefix-closed acyclic DFA $\mcA$ 
such that $(L(\mcA); \leq_{\pref})$ is isomorphic to the unfolding
of $D$ in $r$ (let us denote the latter tree by  $\unfold(D,r)$).
One only has to associate labels to the edges of $D$.
Hence, it suffices to construct from a given monotone
circuit $C$ a dag $D$ which contains for every gate $g$
of $C$ two nodes $g_1, g_2$ such that $g$ evaluates to $\TRUE$ if and only
if $\unfold(D, g_1) \cong \unfold(D, g_2)$. This is straightforward
for the input gates of $C$. For $\AND$- and $\OR$-gates of $C$, we can
use again the construction of \cite{JKMT03}. Take the constructions from
Figure~\ref{fig:lemma_inf:4} and \ref{fig:lemma_inf:5}, where in Figure~\ref{fig:lemma_inf:4}
each of the  subtrees $[U]_1$, $[U]_2$, $[V]_1$, and $[V]_2$ is represented only once.
The construction for $\OR$-gates is shown in Figure~\ref{fig:P-hard}. Assume that the dag $D$
below the nodes $u_1$, $u_2$, $v_1$, and $v_2$ is already constructed. Here $u_1$ and $u_2$
correspond to a gate $u$ and $v_1$ and $v_2$ correspond to a gate $v$. 
Hence, $u$ (resp., $v$) evaluates to $\TRUE$ if and only if 
$\unfold(D, u_1) \cong \unfold(D, u_2)$ (resp., $\unfold(D, u_1) \cong \unfold(D, u_2)$).
Let $t$ be an $\OR$-gate with inputs $u$ and $v$. We add the nodes and edges as shown in 
Figure~\ref{fig:P-hard}. Then the arguments from the proof of Lemma~\ref{lemma_inf:4}
show that $u$ or $v$ evaluates to $\TRUE$ if and only if 
$\unfold(D, t_1) \cong \unfold(D, t_2)$.
\qed
\end{proof}
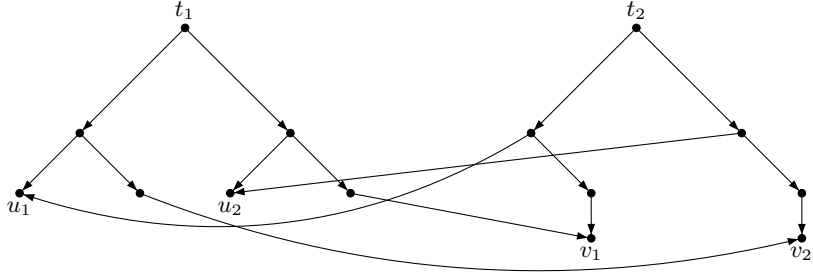
\begin{figure}[t]
\begin{center}                                                        
\begin{picture}(115,40)(-5,15)                                           
\unitlength=1mm 
\gasset{Nframe=y,Nfill=y,Nw=1,Nh=1,ELdist=.7,ExtNL=y,NLangle=270,NLdist=0.7}
\node[NLangle=90](root)(22,50){$t_1$}
\node(l)(8,36){}
\node(r)(36,36){}
\drawedge[ELside=r](root,l){}
\drawedge(root,r){}
\node(ll)(0,28){$u_1$}
\node(lr)(16,28){}
\node(rl)(28,28){$u_2$}
\node(rr)(44,28){}
\drawedge[ELside=r](l,ll){}
\drawedge[ELside=l](l,lr){}
\drawedge[ELside=r](r,rl){}
\drawedge[ELside=l](r,rr){}
\node[NLangle=90](root')(82,50){$t_2$}
\node(l')(68,36){}
\node(r')(96,36){}
\drawedge[ELside=r](root',l'){}
\drawedge(root',r'){}
\node(lr')(76,28){}
\node(rr')(104,28){}
\drawedge[ELside=r,curvedepth=8](l',ll){}
\drawedge[ELside=l](l',lr'){}
\drawedge[ELside=r,curvedepth=0](r',rl){}
\drawedge[ELside=l](r',rr'){}
\node(lrl')(76,22){$v_1$}
\node(rrl')(104,22){$v_2$}
\drawedge[ELside=r,curvedepth=0](lr',lrl'){}
\drawedge[ELside=r](rr',rrl'){}
\drawedge[ELside=r,curvedepth=-7](lr,rrl'){}
\drawedge[ELside=l,curvedepth=0](rr,lrl'){}
\end{picture}
\end{center}
\caption{\label{fig:P-hard}The $\OR$-construction in the proof of Theorem~\ref{thm:P-hard}}
\end{figure}

\section{Isomorphism problem for regular words}

In this section we study the isomorphism problem for regular 
words that are represented by partitioned DFAs. We prove that
this problem as well as the isomorphism problem for regular
linear orders that are represented by DFAs 
are $\Ptime$-complete.  It follows that the isomorphism problem for regular
linear orders that are represented by NFAs can be solved in 
exponential time. We show that this problem is $\PSPACE$-hard.
For the case of acyclic DFAs and NFAs, respectively, we obtain
completeness results for counting classes ($\mathsf{C}_=\mathsf{L}$-completeness
for acyclic DFAs and $\mathsf{C}_=\mathsf{P}$-completeness for acyclic NFAs).

\subsection{Upper bounds}

The main result of this section is:
 
\begin{theorem} \label{thm:Ptime-reg-words}
The following problem can be solved in polynomial time:

\medskip
\noindent
INPUT: Two partitioned DFAs $\mcA_1$ and $\mcA_2$.

\noindent
QUESTION: $w(\mcA_1) \cong w(\mcA_2)$?
\end{theorem}
In Section~\ref{sec:blocks}--\ref{sec:main-alg} we prove Theorem~\ref{thm:poly_ST}.
Section~\ref{sec:blocks} will introduce some of the machinery from
\cite{BloomE05} concerning blocks. Blocks allow to condensate a
generalized word to a coarser word (whose elements are the blocks of
the original word). In Section~\ref{sec:expressions} we will formally
introduce succinct regular expressions (expressions in form of dags)
and in Section~\ref{sec:heilbrunner} we will
argue that Heilbrunner's algorithm from \cite{Heilbrunner80}
allows to transform a given partitioned DFA in polynomial time into an equivalent
succinct (regular) expression. Hence, the remaining goal
is to develop a polynomial time algorithm for checking whether
two given succinct expressions represent isomorphic regular words.
For the special case that these regular words 
consist of only one block (so called primitive regular words),
this will be accomplished in Section~\ref{sec:equiv-primitive}.
In this step, we will make use of algorithms for straight-line programs
(succinctly represented finite words) \cite{Ryt04}. Finally, in
Section~\ref{sec:main-alg} we will present a polynomial time algorithm
or checking whether two given succinct expressions represent isomorphic regular words.

\subsection{Blocks and their combinatorics} \label{sec:blocks}

In this section, we will introduce the crucial notion of a block,
and we recall some of the results from \cite{BloomE05} that we
are using later. 

Let $u = (L; \leq, \tau)$ be a generalized word.
An \emph{interval} of $u$ is an interval of the underlying linear order $(L; \leq)$.
A \emph{subword} of $u$ is an interval $I$ of $u$ together with the
coloring $\tau$ restricted to $I$. Let $\Gamma\subseteq \Sigma$ be
finite. A \emph{$\Gamma$-uniform} subword of $u$ is a subword that is
isomorphic to $\Gamma^\eta$.  A subword is \emph{uniform} if it is
$\Gamma$-uniform for some $\Gamma\subseteq \Sigma$. A uniform subword is a
{\em maximal uniform subword} if it is not properly contained in another
uniform subword. Now let $v$ be a subword such that no point of $v$ is
contained in a uniform subword of $u$. Then $v$ is
\emph{successor-closed} if for each point $p$ of $v$, whenever the
successor and the predecessor of $p$ exist, they are contained in
$v$ as well. A successor-closed subword is {\em minimal} if it does
not strictly contain another successor-closed subword.
Following \cite{BloomE05} we define:

\begin{definition}[blocks]
  Let $u$ be a regular word.  A \emph{block} of $u$ is either a
  maximal uniform subword of $u$ or a minimal successor-closed
  subword of $u$.
\end{definition}
A regular word which consists of a single block is called
\emph{primitive.}\footnote{In combinatorics on words, a finite word is
  called primitive, if it is not a proper power of a non-empty
  word. Our notion of a primitive word should not be confused with
  this definition.}  
By \cite{BloomE05} a word $u$ is primitive if and only if it
is of one of the following forms (where $x,z \in \Sigma^+$, $y \in
\Sigma^*$):
A finite non-empty word, a scattered word of
the form $x^{\omegaop}y$, a scattered
word of the form $yz^{\omega}$, a
scattered word of the form $x^{\omegaop} y z^{\omega}$, or a 
uniform word ($\Gamma^\eta$ for some $\Gamma \subseteq \Sigma$).
Let $D(\Sigma)$ be the set of all
primitive words over $\Sigma$.

Let $u$ be a regular word. Each point $p$ of $u$ belongs to some
unique block $\Bl(p)$, which induces a regular (and hence primitive)
word. Moreover we can order the blocks of $u$ linearly by setting
$\Bl(p)< \Bl(q)$ if and only if $p<q$.  The order obtained that way is denoted
$(\Bl(u); \leq)$.  Then we extend the order $(\Bl(u); \leq)$ to a
generalized word $\widehat{u}$ over $D(\Sigma)$ (here it is useful to allow infinite
alphabets, since $D(\Sigma)$ is infinite), called the {\em skeleton} of $u$, by
labeling each block with the corresponding isomorphic word in
$D(\Sigma)$. Implicitly, it is shown in \cite{BloomE05} that 
for every regular word $u$ there exists a {\em finite} subset of 
$D(\Sigma)$ such that every block of $u$ is isomorphic to a primitive word
from that finite subset. Moreover, $\widehat{u}$ is again a regular
word. Later it will be convenient to have the following
renaming notion available. Let $V$ be a finite alphabet, let
$\phi:V\to D(\Sigma)$ be an injective mapping and suppose that all
blocks of a regular word $u$ belong to the image of $\phi$. The word $v$ that
has $(\Bl(u);\leq)$ as underlying order and each block $B$ of $u$
labeled with $\phi^{-1}(B)$ is called the $\phi$-skeleton of $u$.  We
will need the following result from \cite{BloomE05}:

\begin{proposition}[see {\cite[Corollary 73]{BloomE05}}]\label{prop:skel}
  Let $u,v \in \Reg(\Sigma)$. Let $V$ be a finite
  alphabet and let $\phi:V\to D(\Sigma)$ be injective such that all
  blocks of $u$ and $v$ are in the image of $\phi$. Then $u$ and $v$
  are isomorphic if and only if the $\phi$-skeletons of $u$ and $v$ are
  isomorphic.
\end{proposition}
We will consider finite and infinite sequences, whose symbols
are regular words and where the underlying order type 
is either finite, $\omega$ or $\omegaop$.
In the following, when writing $(u_i)_{i \in I}$, we assume that
either 
$I = \{1,\ldots,n\} \neq \emptyset$ (i.e., $(u_i)_{i \in I}$ is the finite
sequence $(u_1,\ldots,u_n)$) or 
$I = \{1,2,3,\ldots\}$ (i.e., $(u_i)_{i \in I}$ is the infinite
sequence $(u_1,u_2,u_2,\ldots)$) or
$I = \{\ldots,-2,-1,0\}$ (i.e., $(u_i)_{i \in I}$ is the infinite
sequence $(\ldots,u_{-2},u_{-1},u_0)$).
The corresponding generalized word is $\prod_{i\in I} u_i$
(either $u_1 \cdots u_n$, or $u_1 u_2 u_3 \cdots$ or $\cdots u_{-2}
u_{-1} u_0$).  
We say that two sequences $(u_i)_{i \in I}$ and $(v_j)_{j \in J}$
are {\em equivalent}, if the generalized words $\prod_{i\in I} u_i$ and $\prod_{j\in J} v_j$
are isomorphic.
We use commas to separate the successive $u_i$ 
in the sequence  $(u_i)_{i \in I}$ in order to avoid misinterpretations. 
For instance $(a,a)$ viewed as a sequence over
regular words has length two whereas $(aa)$ has length 1. Of course,
$(a,a)$ and $(aa)$ are equivalent sequences.

\begin{definition}
  Let $\bar{u}=(u_i)_{i\in I}$ be a sequence of regular words. We say
  that $\bar{u}$ does not merge if the set of blocks of $\prod_{i\in I} u_i$
  is the union of the set of blocks of the $u_i$. If this is not the 
  case, then we say that $\bar{u}$ \emph{merges.}
\end{definition}
In other words, $\bar{u}$ merges if there exists a block that contains elements from
two different $u_i$.
In \cite[Corollary 32]{BloomE05} it is shown that a sequence $\bar{u}$ merges,
if and only if there exists a factor $(u_i, u_{i+1})$ or $(u_i, u_{i+1}, u_{i+2})$ that merges.

\begin{example}
Clearly if $u$ and $v$ are finite words, then $(u,v)$ merges.
Also, $(\Gamma^\eta, \Gamma^\eta)$ and $(\Gamma^\eta, a, \Gamma^\eta)$
merge for every $\Gamma \subseteq \Sigma$ and $a \in \Gamma$
(in both cases, the sequence is equivalent to $\Gamma^\eta$).
On the other hand, $( [ab]^\eta, [ab]^\eta )$ does not merge.
The reason is that the blocks of $[ab]^\eta$ are the copies of $ab$.
More generally,
if $u$ is not primitive and 
$X$ is a finite subset of regular words,
then $( (X\cup\{u\})^\eta, (X\cup\{u\})^\eta )$
does not merge. 
\end{example}
For the case of a sequence of primitive words, a complete 
description of merging sequences was given in \cite{BloomE05}.
Moreover, if a sequence of primitive words merges, then it 
can be simplified to a non-merging sequence of primitive
words. To make this more precise,
let $u,v,w$ be primitive words. If $(u,v)$ merges, 
then by \cite[Lemma 24]{BloomE05} either 
$u$ and $v$ are $\Gamma$-uniform for some $\Gamma \subseteq \Sigma$
or $u$ is right-closed and $v$ is left-closed.
Then,  the regular word $uv$ has a
single block. If $(u,v,w)$ merges, then by \cite[Lemma 24]{BloomE05}
either $(u,v)$ merges, or $(v,w)$ merges, or 
$u,w$ are $\Gamma$-uniform and $v$ is a singleton from $\Gamma$. 
This motivates the definition of the following rewriting 
system $R$ over finite sequences over $D(\Sigma)$.
\begin{definition}[rewriting system $R$]\label{def:systemR}
The rewriting system $R$ over the set $D(\Sigma)$ consists of the following rules:
\begin{itemize}
\item  $(u_1,u_2,u_3) \to u$ if  $u_1=u_3=u=\Gamma^\eta$ for some
    $\Gamma\subseteq \Sigma$ and $u_2\in \Gamma$
\item $(u_1,u_2)\to u$ if one of the following holds:
  \begin{itemize}
  \item $u_1$ is right-closed, $u_2$ is left-closed and $u=u_1u_2$
  \item $u_1=u_2=u=\Gamma^\eta$ for some $\Gamma\subseteq \Sigma$.
  \end{itemize}
\end{itemize}
\end{definition}
In the following, we will use some basic facts from rewriting theory,
see e.g. \cite{BoOt93} for further details.
For sequences $\bar{x}$ and $\bar{y}$ over $\Reg(\Sigma)$, we write
$\bar{x} \to_R \bar{y}$ if there exist a rewrite rule
$\bar{u} \to u$ and an occurrence of the sequence $\bar{u}$ in $\bar{x}$
such that replacing that occurrence by $u$ gives the sequence $\bar{y}$.
Here, $\bar{x}$ and $\bar{y}$ may be infinite sequences.
Moreover, those $x_i$ of $\bar{x} = (x_i)_{i \in I}$ that 
are not primitive are left untouched in the rewrite step 
$\bar{x} \to_R \bar{y}$. Clearly, $\bar{x} \to_R \bar{y}$ implies
that the sequences $\bar{x}$ and $\bar{y}$ are equivalent.
A (possibly infinite) sequence $\bar{u}$ is irreducible w.r.t.~$R$ 
if there does not exist a sequence $\bar{v}$ with $\bar{u} \to_R \bar{v}$.
Clearly, on infinite sequences, $R$ cannot be {\em terminating} 
(e.g., $(a^\eta,a^\eta,a^\eta \ldots) \to_R (a^\eta,a^\eta,a^\eta\ldots)$ is a loop).
On the other hand, $R$ is trivially terminating on finite sequences, since it is length-reducing.
Moreover, by analyzing overlapping left-hand sides of $R$, 
one can easily show:

\begin{lemma}
 The rewriting system $R$ is 
 strongly confluent (on finite and infinite sequences), i.e.,
 for all $\bar{u}$, $\bar{v}$, $\bar{w}$ such that  $\bar{u} \to_R \bar{v}$ and 
 $\bar{u} \to_R \bar{w}$ there exists $\bar{x}$ such that  ($\bar{v} = \bar{x}$ or 
 $\bar{v} \to_R \bar{x}$) and  ($\bar{w} = \bar{x}$ or 
 $\bar{w} \to_R \bar{x}$). 
\end{lemma}
By a simple fact from rewriting theory, it follows that 
$R$ is also {\em confluent}, i.e., 
for all $\bar{u}$, $\bar{v}$, $\bar{w}$ such that  $\bar{u} \to_R^* \bar{v}$ and 
$\bar{u} \to_R^* \bar{w}$ there exists $\bar{x}$ such that  
$\bar{v} \to_R^* \bar{x}$ and  
$\bar{w} \to_R^* \bar{x}$. Termination (on finite sequences) and confluence imply that $R$ produces unique
 normal forms for finite sequences, i.e., for every finite sequence 
 $\bar{u}$ there exists a unique finite sequence $\bar{v}$ such that
 $\bar{u} \to_R^* \bar{v}$ and $\bar{v}$ is irreducible w.r.t. $R$.
 This $\bar{v}$ is called the {\em irreducible normal form} of $\bar{u}$. 

The following is a direct consequence of \cite[Lemma 24 \& Corollary 32]{BloomE05}.
\begin{lemma}\label{lem:merge}
  Let $\bar{u}$ be a sequence of primitive words. Then
  $\bar{u}$ does not merge if and only if $\bar{u}$ is irreducible w.r.t. $R$.
\end{lemma}
We also have to verify that a sequence $\bar{u}$ over
$\Reg(\Sigma)$ containing non-primitive words does not merge. 
We use the definition below. Note that a regular word need not have
a first or last block. For instance, $(a^\omega)^\omega$ has a first
block but no last block, whereas $(a^\omega)^{\omegaop}
(a^\omega)^\omega$ and $[aa]^\eta$ neither have a first block nor a
last block.

\begin{definition}[good and semi-good sequences]\label{def:good}
The sequence $\bar{u} = (u_i)_{i \in I}$ is {\em good} if 
the following conditions hold:
\begin{enumerate}[(1)]
\item $\bar{u}$ is irreducible with respect to $R$.
\item For all $i\in I$ we have:
  \begin{enumerate}[(a)]
  \item If $u_i$ is not primitive and has a first block, then either
    ($i-1 \in I$, $u_{i-1}$ is uniform, and $(u_{i-1},u_i)$ does not merge) or
    ($i-1, i-2 \in I$, $u_{i-1}$ and $u_{i-2}$ are primitive, and $(u_{i-2},u_{i-1},u_i)$
    does not merge).
  \item If $u_i$ is not primitive and has a last block, then either
    ($i+1 \in I$, $u_{i+1}$ is uniform, and $(u_{i},u_{i+1})$ does not merge) or
    ($i+1, i+2 \in I$, $u_{i+1}$ and $u_{i+2}$ are primitive, and $(u_{i},u_{i+1},u_{i+2})$
    does not merge).
  \end{enumerate}
\end{enumerate}
If only (2) holds, then $\bar{u}$ is said to be \emph{semi-good}.
\end{definition}

\begin{lemma}\label{lem:goodtorporper}
  If $\bar{u}$ is good, then $\bar{u}$ does not merge.
\end{lemma}

\begin{proof}
Assume that $\bar{u}$ is good but merges. By \cite[Corollary 32]{BloomE05},
one of the following cases holds:

\medskip
\noindent
{\em Case 1.} $\bar{u}$ contains a factor $(u_i, u_{i+1})$ that merges.
If $u_i$ and $u_{i+1}$ would be both primitive, then $\bar{u}$ would be
not irreducible, which is a contradiction ($\bar{u}$ is good).
Hence, $u_i$ or $u_{i+1}$ must be not primitive. W.l.o.g. assume that
$u_i$ is not primitive (the other case is symmetric).
If $u_i$ has no last block, then \cite[Corollary 30(1)]{BloomE05}
implies that $(u_i, u_{i+1})$ does not merge, which is a
contradiction. Hence, we can assume that $u_i$ has a last block.
But then, since $\bar{u}$ is good, $(u_i, u_{i+1})$ does not merge,
which is again a contradiction.

\medskip
\noindent
{\em Case 2.} $\bar{u}$ contains a factor $(u_i,u_{i+1},u_{i+2})$ that merges
but neither $(u_i,u_{i+1})$ nor $(u_{i+1},u_{i+2})$ merges.
Since $\bar{u}$ is irreducible w.r.t. $R$, it follows that
$u_i, u_{i+1}$, or $u_{i+2}$ is not primitive. The case that $u_{i+2}$ is not primitive
is symmetric to the case that $u_i$ is not primitive. Hence, it
suffices to consider the following two subcases:

\medskip
\noindent
{\em Case 2a.} $u_i$ is not primitive. 
If $u_i$ has no last block, then \cite[Corollary 31(1)]{BloomE05}
implies that $(u_i, u_{i+1}, u_{i+2})$ does not merge, which is a
contradiction. Hence, we can assume that $u_i$ has a last block, call
it $b_i$. Since $\bar{u}$ is good and $(u_i,u_{i+1},u_{i+2})$ merges,
$u_{i+1}$ must be uniform. If $u_{i+2}$ has no first block, then 
again \cite[Corollary 31(1)]{BloomE05} implies that 
$(u_i, u_{i+1}, u_{i+2})$ does not merge, which is a
contradiction. Let $b_{i+2}$ be the first block of $u_{i+2}$.
Moreover, \cite[Corollary 31(2)]{BloomE05} implies that
$(b_i, u_{i+1}, b_{i+2})$ merges. 
Since $(u_i, u_{i+1})$ and $(u_{i+1}, u_{i+2})$ do not merge,
also $(b_i, u_{i+1})$ and $(u_{i+1}, b_{i+2})$ do not merge.
It follows (from the form of our rewriting system $R$) that
$b_i = b_{i+2}$ is uniform and $u_{i+1}$ is a singleton word.
But we have already shown that $u_{i+1}$ is uniform, which is 
a contradiction.

\medskip
\noindent
{\em Case 2b.} $u_{i+1}$ is not primitive.
Then $u_{i+1}$ has more than one block and 
\cite[Corollary~31(1)]{BloomE05} directly
implies that $(u_i,u_{i+1},u_{i+2})$ does not merge,
which is again a contradiction.
\qed
\end{proof}

\begin{lemma} \label{lemma:semi-good-preserved}
  If $\bar{u}$ is semi-good and $\bar{u}\to_R\bar{v}$, then $\bar{v}$ is
  semi-good as well.
\end{lemma}

\begin{proof}
Assume that $\bar{u} = (u_i)_{i\in I}$ is semi-good and $\bar{u}\to_R\bar{v}$.
We have to show that $\bar{v} = (v_j)_{j\in J}$ is semi-good.
For this, consider an $j \in J$ such that
$v_j$ is not primitive.
Since the system $R$ does not introduce non-primitive
words, $v_j$ must have been already present in $\bar{u}$.
Let $i \in I$ be the position in $\bar{u}$ that corresponds
to position $j$ in $\bar{v}$. Hence, $u_i = v_j$.
By symmetry it suffices to show that
condition (2a) from Definition~\ref{def:good} holds for
$j \in J$.
The case that $u_i = v_j$ has no first block is clear. So, assume that
$u_i$ has a first block. Since $\bar{u}$ is semi-good,
we can distinguish the following two cases.

\medskip
\noindent
{\em Case 1.} $i-1 \in I$, $u_{i-1}$ is uniform, and $(u_{i-1},u_i)$
does not merge. 
From the form of the rewrite rules, it follows that
$v_{j-1} = u_{i-1}$. Hence, $v_{j-1}$ is uniform, and $(v_{j-1},v_j) =
(u_{i-1},u_i)$ does not merge. Thus, we have shown condition (2a) 
from Definition~\ref{def:good} for $j$.

\medskip
\noindent
{\em Case 2.} $i-1, i-2 \in I$, $u_{i-2}, u_{i-1}$ are primitive,  
and $(u_{i-2}, u_{i-1},u_i)$ does not merge. 
We make a case distinction on the position, where the rewrite rule 
is applied.

\medskip
\noindent
{\em Case 2a.} 
$i-3 \in I$ and 
in the rewrite step $\bar{u}\to_R\bar{v}$, $(u_{i-3}, u_{i-2}, u_{i-1})$
is replaced by $u  \in D(\Sigma)$. Thus, 
$u_{i-3} = u_{i-1} = u$ is uniform. Hence, $v_{j-1} = u$ is uniform.
Moreover, $(v_{j-1}, v_j) = (u_{i-1}, u_i)$ does not merge.

\medskip
\noindent
{\em Case 2b.} 
$i-4 \in I$ and 
in the rewrite step $\bar{u}\to_R\bar{v}$, $(u_{i-4}, u_{i-3}, u_{i-2})$
is replaced by $u  \in D(\Sigma)$. Thus,
$u_{i-4} = u_{i-2} = u$ is uniform, $v_{j-2} = u = u_{i-2}$,
and $u_{i-1} = v_{j-1}$. It follows that
$v_{j-2}$ and $v_{j-1}$ are primitive, and that
$(v_{j-2}, v_{j-1}, v_j) = (u_{i-2}, u_{i-1}, u_i)$ does not merge.

\medskip
\noindent
{\em Case 2c.}  
In the rewrite step $\bar{u}\to_R\bar{v}$, $(u_{i-2}, u_{i-1})$
is replaced by $u  \in D(\Sigma)$. Then, $(u_{i-2}, u_{i-1})$
merges. But this contradicts the assumption that 
$(u_{i-2}, u_{i-1},u_i)$ does not merge.

\medskip
\noindent
{\em Case 2d.}  
$i-3 \in I$ and 
in the rewrite step $\bar{u}\to_R\bar{v}$, $(u_{i-3}, u_{i-2})$
is replaced by $u  \in D(\Sigma)$.
If $u_{i-3} = u_{i-2} = u$ is uniform, then
$v_{j-2} = u_{i-2}$ and $v_{j-1} = u_{i-1}$
are primitive and 
$(v_{j-2}, v_{j-1}, v_j) = (u_{i-2}, u_{i-1}, u_i)$ does not merge.
Finally, assume that $u_{i-3}$ is right-closed, $u_{i-2}$ is 
left-closed and $v_{j-2} = u = u_{i-3}u_{i-2}$.
We have $v_{j-1} = u_{i-1}$. Thus  $v_{j-1}$ and $v_{j-2}$ are 
primitive. It remains to show that 
$(v_{j-2}, v_{j-1}, v_j) = (u_{i-3}u_{i-2}, u_{i-1}, u_i)$
does not merge. 
We know that $(u_{i-1}, u_i)$ does not merge (since 
$(u_{i-2}, u_{i-1}, u_i)$ does not merge). 
Assume that $(u_{i-3}u_{i-2}, u_{i-1})$ merges.
Then (since $u_{i-3}u_{i-2}$ is primitive and scattered and $u_{i-1}$
is primitive) $u_{i-3}u_{i-2}$ must be right-closed and $u_{i-1}$
must be left-closed. But then, $u_{i-2} \neq \varepsilon$ is
right-closed  as well and
$(u_{i-2}, u_{i-1})$  merges. This is a contradiction.
Hence, $(u_{i-3}u_{i-2}, u_{i-1})$ does not merge.
Let $b_i$ be the first block of $u_i$. If 
$(u_{i-3}u_{i-2}, u_{i-1}, u_i)$ merges, then
by  \cite[Corollary~31(2)]{BloomE05}, $(u_{i-3}u_{i-2}, u_{i-1}, b_i)$
merges. Since neither $(u_{i-3}u_{i-2}, u_{i-1})$ nor 
$(u_{i-1}, b_i)$ merges, $u_{i-3}u_{i-2}$ and $b_i$ must be uniform.
But we know that  $u_{i-3}u_{i-2}$ is scattered, which leads
again to a contradiction. Thus, indeed 
$(u_{i-3}u_{i-2}, u_{i-1}, u_i)$ does not merge.

\medskip
\noindent
If the rewrite rule is applied at a position different from
those considered in Case 2a--2d, then 
$(v_{j-2}, v_{j-1}, v_j) = (u_{i-2}, u_{i-1}, u_i)$.
Since $(u_{i-2}, u_{i-1}, u_i)$ fulfills condition (2a) 
from Definition~\ref{def:good}, so does $(v_{j-2}, v_{j-1}, v_j)$.
This concludes the proof of the lemma.
\qed
\end{proof}
Lemma~\ref{lemma:semi-good-preserved} implies that 
from a given finite semi-good sequence $\bar{u}$ 
we can compute an equivalent good sequence, by computing the (unique)
irreducible normal form of $\bar{u}$.

\subsection{Expressions and succinct expressions} \label{sec:expressions}

Regular words can be naturally described by expressions using
the operations of concatenation, $\omega$-power,  $\omegaop$-power, and shuffle.
Formally,  the set $T(V,\Sigma)$ of {\em expressions} over $V$ and $\Sigma$ is
inductively defined as follows:
\begin{enumerate}[(a)]
\item $V \cup \Sigma \subseteq T(V,\Sigma)$
\item If $\alpha_1, \ldots, \alpha_n \in T(V,\Sigma)$ ($n \geq 1$), then
  $\alpha_1 \cdots \alpha_n \in T(V,\Sigma)$.
\item If $\alpha \in T(V,\Sigma)$, then $\alpha^\omega \in
  T(V,\Sigma)$ and $\alpha^{\omegaop} \in T(V,\Sigma)$.
\item If $\alpha_1,\ldots,\alpha_n \in T(V,\Sigma)$ ($n \geq 1$), then
  $[\alpha_1,\ldots,\alpha_n]^\eta\in T(V,\Sigma)$.
\end{enumerate}
A mapping $f : V \to \Reg(\Sigma)$ will be extended homomorphically to a mapping $f :
T(V,\Sigma) \to \Reg(\Sigma)$ inductively as follows, where $\alpha,
\alpha_1, \ldots, \alpha_n \in T(V,\Sigma)$:
\begin{itemize}
\item $f(a) = a$ for $a \in \Sigma$
\item $f(\alpha_1 \cdots \alpha_n) = f(\alpha_1) \cdots f(\alpha_n)$
\item $f(\alpha^{\omega})=f(\alpha)^{\omega}$
\item $f(\alpha^{\omegaop})=f(\alpha)^{\omegaop}$
\item
  $f([\alpha_1,\ldots,\alpha_n]^\eta)=([f(\alpha_1),\ldots,f(\alpha_n)]^\eta$
\end{itemize}
For $\alpha \in T(V,\Sigma)$ we define the size $|\alpha|\in \mathbb{N}$ 
inductively as follows:
\begin{itemize}
\item $|\alpha|=1$
for $\alpha \in V \cup \Sigma$
\item $|\alpha_1 \cdots \alpha_n| = |\alpha_1|+\cdots+|\alpha_n|$
\item $|\alpha^{\omega}|=|\alpha^{\omegaop}|=|\alpha|+1$
\item
  $|[\alpha_1,\ldots,\alpha_n]^\eta|=|\alpha_1|+\cdots+|\alpha_n|+1$
\end{itemize}
A {\em succinct expression system (SES)} is a tuple $\dA = (V,\Sigma, \rhs)$
such that:
\begin{itemize}
\item $V$ (the set of variables) and $\Sigma$ (the terminal alphabet)
  are disjoint finite alphabets.
\item $\rhs$ (for right-hand side) is a mapping from $V$ to
  $T(V,\Sigma)$ such that the relation
  $\{ (Y,X) \in V \times V \mid Y \text{ occurs in } \rhs(X)\}$
  is acyclic. The reflex transitive closure of this relation is called the {\em hierarchical order} of 
  $\dA$ and denoted by $\preceq_{\dA}$.
 \end{itemize}
The property for $\rhs$ ensures that there exists a unique mapping
$\val_{\dA} : V \to \Reg(\Sigma)$ such that $\val_{\dA}(X) =
\val_{\dA}(\rhs(X))$ for all $X \in V$.  If $\dA$ is clear from the context,
we will simply write $\val(X)$.

In the following a quadruple
$\dA=(V,\Sigma,\rhs,S)$ where $(V,\Sigma,\rhs)$ is as above and $S \in V$ (i.e., an
SES with a distinguished start variable $S$) we will be called a
\emph{succinct expression}. In this case let us set $\val(\dA) =
\val_{\dA}(S)$.  
A succinct expression may be also seen as a dag (directed acyclic
graph), whose unfolding is an expression in the above sense.
\begin{example}
Consider the succinct expression 
$$\dA = (\{X_1, X_2, X_3, X_4, X_5\}, \{a,b\},
\rhs, X_1)$$ 
with
\begin{alignat*}{3}
\rhs(X_1) & = [X_2,X_3]^\eta \qquad &  
\rhs(X_2) & = X_3 X_3 \qquad &  
\rhs(X_3) & = X_4 X_4 \\
\rhs(X_4) & = X_5 X_6 & 
\rhs(X_5) & = ab & 
\rhs(X_6) & = ba . 
\end{alignat*}
We have $\val(\dA) = [abbaabba, abbaabbaabbaabba]^\eta$. The
corresponding
dag looks as follows:
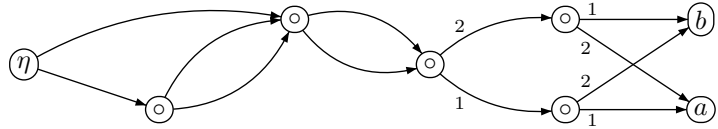
\begin{center}
\setlength{\unitlength}{1.2mm}
\begin{picture}(75,20)
\gasset{AHnb=1,ELdist=0.5,linewidth=.15,Nfill=n,Nframe=y,Nw=1,Nh=1,Nadjust=wh,Nadjustdist=0.8}
\node(x1)(0,10){$\eta$}
\node(x2)(15,5){$\circ$}
\node(x3)(30,15){$\circ$}
\node(x4)(45,10){$\circ$}
\node(x5)(60,5){$\circ$}
\node(x6)(60,15){$\circ$}
\node(a)(75,5){$a$}
\node(b)(75,15){$b$}
\drawedge(x1,x2){}
\drawedge[curvedepth=3](x1,x3){}
\drawedge[curvedepth=-3](x2,x3){}
\drawedge[curvedepth=3](x2,x3){}
\drawedge[curvedepth=-3](x3,x4){}
\drawedge[curvedepth=3](x3,x4){}
\drawedge[ELpos=30,curvedepth=-2,ELside=r](x4,x5){{\scriptsize $1$}}
\drawedge[ELpos=30,curvedepth=2](x4,x6){{\scriptsize $2$}}
\drawedge[ELpos=20,ELside=r](x5,a){{\scriptsize $1$}}
\drawedge[ELpos=20,ELside=l](x5,b){{\scriptsize $2$}}
\drawedge[ELpos=20,ELside=r](x6,a){{\scriptsize $2$}}
\drawedge[ELpos=20](x6,b){{\scriptsize $1$}}
\end{picture}
\end{center}
Nodes labelled with $\circ$ compute the concatenation of their 
successor nodes. In case the order of the successor nodes matters,
we specify it by edge labels.
\end{example}
For an SES $\dA$ we define $$|\dA|=\sum_{X\in V}|\rhs(X)|.$$ 
An SES $\dA = (V,\Sigma, \rhs)$ is in \emph{normal form} if all
right-hand sides are in $(V\cup\Sigma)^+$ or of the form $Y^\omega,
Y^{\omegaop},[Y_1,\ldots,Y_n]^\eta$ for some $Y,Y_1,\ldots,Y_n\in
V\cup\Sigma$.  
For such an SES $\dA$, we define
$\depth_\dA(X)$ and $\height_\dA(X)$ for $X\in V$ inductively as
follows (below, we set $\depth_\dA(a) = \height_\dA(a) = 0$
for $a \in \Sigma$):
\begin{itemize}
\item  If $\rhs(X) = Y_1 \cdots Y_n$ ($n \geq 1$, $Y_1, \ldots, Y_n
  \in \Sigma\cup V$), then
\begin{eqnarray*}
\depth_\dA(X) &=& \max(\depth_\dA(Y_1),\ldots,\depth_\dA(Y_n))+1, \\
\height_\dA(X) &=& \max(\height_\dA(Y_1),\ldots,\height_\dA(Y_n)) .
\end{eqnarray*}
\item 
If $\rhs(X)=Y^{\omega}$ or $\rhs(X)=Y^{\omegaop}$, then
\begin{eqnarray*} 
\depth_\dA(X) &=& \depth_\dA(Y)+1, \\
\height_\dA(X) &=& \height_\dA(Y)+1 .
\end{eqnarray*}
\item If $\rhs(X)=[Y_1,\ldots,Y_n]^\eta$, then
\begin{eqnarray*} 
  \depth_\dA(X) &=& \max(\depth_\dA(Y_1),\ldots,\depth_\dA(Y_n))+1, \\
 \height_\dA(X) &=& \max(\height_\dA(Y_1),\ldots,\height_\dA(Y_n))+1 . 
\end{eqnarray*}  
\end{itemize}

\paragraph{\bf Straight-line programs.}

A succinct expression, where all right-hand sides belong to
$(V \cup \Sigma)^+$ is called a {\em straight-line program (SLP)} \cite{PlRy99}. 
In this case, $\val(\dA)$ is a finite non-empty word.
An SLP $\dA$ can be viewed as a succinct representation of the word
$\val(\dA)$. More precisely, the length of $\val(\dA)$ may be exponential in $|\dA|$.
We will make heavy use of the fact that certain algorithmic problems on SLP-encoded finite
words can be solved in polynomial time. More precisely, we use the
following  results:

\begin{remark} \label{remark:SLP-alg}
There exist polynomial time algorithms for the following problems:
\begin{enumerate}[(a)]
\item Given an SLP $\dA$, calculate $|\val(\dA)|$.
\item Given an SLP $\dA$ and a number $k\in \N$ (coded in binary) we
  can produce an SLP $\dB$ of size $|\dA|+O(\log k)$ such that
  $\val(\dB)=\val(\dA)^k$.
\item Given an SLP $\dA$ and numbers $i\leq j\leq|\val(\dA)|$,
  compute an SLP $\dB$ with $\val(\dB)=\val(\dA)[i:j]$. 
  Here $w[i:j] = a_i\ldots a_j$ for a finite word
   $w=a_1\ldots a_n$.
 \item Given SLPs $\dA$ and $\dB$ decide whether $\val(\dA) =
   \val(\dB)$ \cite{Pla94}.
\item Given SLPs $\dA$ and $\dB$ decide whether $\val(\dA)$ is a
  factor of $\val(\dB)$ \cite{GaKaPlRy96,Lif07,MiShTa97}.
\end{enumerate}
The proofs for (a), (b), and (c) are straightforward.
\end{remark}

\paragraph{\bf 2-level systems.}

A \emph{2-level system} is a tuple $\dA=(\Up,\Lo,\Sigma,\rhs)$ such
that the following holds ($f \rest_A$ denotes the restriction of a
function $f$ to the set $A$):
\begin{itemize}
\item The tuple $(\Up, \Lo, \rhs\rest_{\Up})$ is an SES (w.l.o.g. in
  normal form) over the
  terminal alphabet $\Lo$.
\item The tuple $(\Lo, \Sigma, \rhs\rest_{\Lo})$ is an SES over the
  terminal alphabet $\Sigma$.
\end{itemize}
The set $\Up$ (resp. $\Lo$) is called the set of {\em upper level
  variables} ({\em lower level variables}) of $\dA$.  Moreover, we set
$V = \Up \cup \Lo$ and call it the set of variables of $\dA$.  The SES
$(\Up, \Lo, \rhs\rest_{\Up})$ is called the {\em upper part of $\dA$},
briefly $\up(\dA)$, and the SES $(\Lo, \Sigma, \rhs\rest_{\Lo})$ is
the {\em lower part of $\dA$}, briefly, $\lo(\dA)$.  The upper level
evaluation mapping $\uval_{\dA} : \Up \to \Reg(\Lo)$ of $\dA$ is
defined as $\uval_{\dA} = \val_{\up(\dA)}$.  The evaluation mapping
$\val_{\dA}$ is defined by $\val_{\dA}(X) =
\val_{\lo(\dA)}(\val_{\up(\dA)}(X))$ for $X \in \Up$ and
$\val_{\dA}(X) = \val_{\lo(\dA)}(X) $ for $X \in
\Lo$. 

\subsection{Heilbrunner's algorithm} \label{sec:heilbrunner}

\begin{theorem} \label{thm:heilb} 
From a given partitioned DFA $\mcA$, we can compute in polynomial
time a succinct expression $\dA$  such that $w(\mcA) \cong \val(\dA)$. 
\end{theorem}

\begin{proof}
There is  nothing new about the proof. 
We just have to follow \mbox{Heilbrunner's} algorithm carefully.
Let $\mcA = (Q,\Gamma,\delta,q_0,(F_a)_{a \in\Sigma})$
be a partitioned DFA and let $F = \bigcup_{a \in\Sigma} F_a$.
 We can assume that every state in $F$ is a dead end, i.e.,
 does not have outgoing transitions. For this, take
a new symbol $\$$, as well as a copy $q'$
together with the transition $(q,\$,q')$
for every final state $q \in F$. We set 
$F'_a = \{q' \mid q \in F_a\}$ and
let $\$$ be the smallest symbol in $\Gamma \cup \{\$\}$.
The resulting partitioned DFA produces the same generalized word as
$\mcA$.

So, assume that every state in $F$ is a dead end.
W.l.o.g. we can also assume that $\mcA$ is coaccessible.
The variables of the succinct expression $\dA$ will be the states of $\mcA$.
Consider a state $p \in Q$ and let
$(p, a_i, q_i)$ ($1  \leq i \leq k$) be all outgoing
transitions for $p$, where $a_1 < a_2 < \cdots < a_k$.
Let us define $\out(p) = q_1 q_2 \cdots q_k$.
Next, consider the graph with node set $Q$ and an edge from $p \in Q$ 
to $q \in Q$ if there is a transition from $p$ to $q$.
We partition this graph into its strongly connected components (SCCs).
An SCC $C$ is smaller than an SCC $D$  if
there exists a path from a state in $C$ to a state in $D$; this defines
a partial order on the set of SCCs.
We eliminate all SCCs starting with the maximal ones.
When eliminating an SCC $C$, we define $\rhs_{\dA}(p)$ 
for each state $p \in C$.  If the SCC $C$ is a singleton set $\{p\}$ with
$p \in F_a$, then we set $\rhs_{\dA}(p) = a$. 
If the SCC $C = \{p\}$ is a singleton set with
$p \not\in F$, then we set 
$\rhs_{\dA}(p) = \out(p)$. Note that $\out(p) \neq \varepsilon$, since
$p \not\in F$ and $\mcA$ is coaccessible.
Now, consider an SCC $C$ of size $|C| \geq 2$.
Then every word $\out(p)$ ($p \in C$) contains at least one occurrence
of a state from $C$. Hence $\out(p)$ can be factored as $\out(p) = u_p x_p v_p$,
where $u_p$ and $v_p$ do not contain occurrences of states from the
SCC $C$ (i.e., all states occurring in $u_p$ and $v_p$ belong to larger
SCCs), and $x_p$ starts and ends with a state from $C$ ($x_p$ might 
consist of a single state from $C$). 
Define functions $\ell : C \to C$
and $r : C \to C$ as follows: $\ell(p)$ (resp. $r(p)$) is the first
(resp. last) state of the word $x_p$.
Then, for every $p \in C$, the sequences $p, \ell(p), \ell^2(p), \ldots$
and $p, r(p), r^2(p), \ldots$ become periodic after at most $|C|$
steps. 
We now define regular expressions $\ell_p$ and $r_p$ 
as follows:
Let $p_0, p_1, \ldots, p_a$ and $q_0, q_1, \ldots, q_c$
be shortest sequences such that 
$p_0 = q_0 = p$, $p_{i+1} = \ell(p_i)$, $q_{i+1} = r(q_i)$,
and $\ell(p_a) \in \{p_0, p_1, \ldots, p_a\}$,
$r(q_c) \in \{q_0, q_1, \ldots, q_c\}$. 
Assume that $\ell(p_a) = p_b$ and $r(q_c) = q_d$
for $0 \leq b \leq a$, $0 \leq d \leq c$.
Then, we define 
\begin{eqnarray*}
\ell_p & = & (u_{p_0}  \cdots u_{p_{b-1}}) (u_{p_b} \cdots u_{p_a})^\omega, \\
r_p & = & (v_{q_c} \cdots v_{q_d})^{\overline{\omega}} (v_{q_{d-1}}
\cdots v_{q_0}) .
\end{eqnarray*}
Next, let $T$ be the set of all regular expressions
of the form $\ell_s y r_t$ ($s, t \in C$) such that some word 
$\out(p)$ ($p \in C$) contains a factor $s y t$, where the word $y$  
does not contain a state from $C$.
Then we finally set $\rhs_{\dA}(p) = \ell_p [T]^\eta r_p$ for all $p \in C$.
This concludes the elimination step for the SCC $C$.
By \cite{Heilbrunner80}, for every state $p \in Q$ we have
$w(Q,\Gamma,\delta,p,(F_a)_{a \in\Sigma}) \cong \val_{\dA}(p)$.
\qed
\end{proof}
By Theorem~\ref{thm:heilb}, it suffices to prove the following 
result in order to prove Theorem~\ref{thm:Ptime-reg-words}. 

\begin{theorem} \label{thm:poly_ST}
The following problem can be solved in polynomial time:

\medskip
\noindent
INPUT: Two succinct expressions $\dA_1$ and $\dA_2$.

\noindent
QUESTION: $\val(\dA_1) \cong \val(\dA_2)$?
\end{theorem}
In the next section, we will prove this result for the special
case that both $\val(\dA_1)$ and $\val(\dA_2)$ are primitive.

\subsection{A polynomial time equivalence test for succinct primitive expressions}
\label{sec:equiv-primitive}

By Theorem~\ref{thm:heilb}, the remaining goal is  
to test in polynomial time, whether two succinct expressions 
represent isomorphic regular words. In a first step, we accomplish this
for succinct expressions that represent primitive words. 
In the following, $\Sigma$ will always refer to a {\em finite} alphabet.
Let us first show that we can decide in polynomial time whether
a succinct expression represents a primitive word.

\begin{lemma}\label{lem:oneblock}
  Given a succinct expression $\dA$, we can decide in polynomial time whether
  $\val(\dA)$ is a primitive word, and in case it is we can compute in polynomial time
  a representation, which has one of the following forms, where
  $\dB, \dC, \dD$ are SLPs and $\Gamma \subseteq \Sigma$ (here, we should
  allow also the empty word for $\val(\dC)$):
  $\val(\dB)$, $\val(\dC) \val(\dD)^\omega$, $\val(\dB)^{\omegaop} \val(\dC)$, 
  $\val(\dB)^{\omegaop} \val(\dC) \val(\dD)^\omega$, $\Gamma^\eta$. 
\end{lemma}

\begin{proof}
  We proceed along the hierarchical order of $\dA$ and compute for each
  variable $A$ of $\dA$ whether $\val(A)$ is of one of the following
  forms ($u,w \in \Sigma^+,  v \in \Sigma^*$, $\Gamma \subseteq \Sigma$, $a,b \in \Gamma$): 
  $v$,  $u^{\omegaop} v$,  $v w^{\omega}$,  $u^{\omegaop} v w^{\omega}$, 
  $\Gamma^\eta$, $a \Gamma^\eta$, $\Gamma^\eta b$, $a \Gamma^\eta b$. Moreover, SLPs for the 
  finite words $u$, $v$, and $w$ can computed simultaneously. 
  Observe that from $\rhs(A)$ and the
  information already computed we can easily obtain whether $\val(A)$
  is of such a form and in this case of which form. 
  The following identities have to be used for shuffles ($\Gamma \subseteq \Sigma$, 
  $n \geq 0$, $m \geq 1$,
  $a, a_1,\ldots, a_n \in \Gamma$, and every $u_i$ ($1 \leq i \leq m$) has one of the forms $\Gamma^\eta$, $c \Gamma^\eta$, $\Gamma^\eta c$, $c \Gamma^\eta d$
  with $c,d \in \Gamma$)
  \begin{gather*}
  [a_1, \ldots, a_n, u_1, \ldots, u_m]^\eta \cong \Gamma^\eta   \\
  \Gamma^\eta \Gamma^\eta \cong \Gamma^\eta a \Gamma^\eta \cong (\Gamma^\eta)^\omega \cong (\Gamma^\eta)^{\omegaop} \cong (\Gamma^\eta a)^\omega \cong (a \Gamma^\eta)^{\omegaop} \cong \Gamma^\eta \\
  (a \Gamma^\eta)^\omega  \cong  a \Gamma^\eta \\
  (\Gamma^\eta a)^{\omegaop}  \cong  \Gamma^\eta a 
\end{gather*}
All these identities can be deduced from the axioms for regular expressions in \cite{BloomE05}.
  Now $\val(\dA)$ is primitive if and only if $\val(S)$  is of one of the following
  forms ($u,w \in \Sigma^+,  v \in \Sigma^*$, $\Gamma \subseteq \Sigma$): 
  $v$,  $u^{\omegaop} v$,  $v w^{\omega}$,  $u^{\omegaop} v w^{\omega}$, 
  $\Gamma^\eta$.
 \qed
\end{proof}
For our polynomial  time equivalence test for succinct expressions that represent primitive
words, we need the following
technical lemma.

\begin{lemma}\label{lem:factor}
  Let $u_i,v_i,w_i$ $(i\in \{1,2\}$) be finite words such that
  $|u_1|=|u_2|=|v_1|=|v_2|=|w_1|=|w_2|>0$.  Then $u_1^{\omegaop} v_1
  w_1^\omega = u_2^{\omegaop} v_2 w_2^\omega$ if and only if one of
  the following conditions hold:
  \begin{itemize}
  \item $u_2 v_2 w_2^2$ is a factor of $u_1^2v_1w_1^2$.
  \item $u_1 v_1 w_1^2$ is a factor of $u_2^2v_2w_2^2$.
  \item $v_1 = w_1$, $u_2 = v_2$, and $u_2 w_2^2$ is a factor of
    $u_1^2 w_1^2$.
  \item $u_1 = v_1$, $v_2 = w_2$, and $u_1 w_1^2$ is a factor of
    $u_2^2 w_2^2$.
  \end{itemize}
\end{lemma}

\begin{figure}[t]
\begin{center}
\setlength{\unitlength}{1mm}
\begin{picture}(110,10)
\put(5,0){\framebox(10,5){$u_2$}}
\put(15,0){\framebox(10,5){$v_2$}}
\put(25,0){\framebox(10,5){$w_2$}}
\put(35,0){\framebox(10,5){$w_2$}}
\put(0,5){\framebox(10,5){$u_1$}}
\put(10,5){\framebox(10,5){$u_1$}}
\put(20,5){\framebox(10,5){$v_1$}}
\put(30,5){\framebox(10,5){$w_1$}}
\put(40,5){\framebox(10,5){$w_1$}}
\put(65,0){\framebox(10,5){$u_1$}}
\put(75,0){\framebox(10,5){$v_1$}}
\put(85,0){\framebox(10,5){$w_1$}}
\put(95,0){\framebox(10,5){$w_1$}}
\put(60,5){\framebox(10,5){$u_2$}}
\put(70,5){\framebox(10,5){$u_2$}}
\put(80,5){\framebox(10,5){$v_2$}}
\put(90,5){\framebox(10,5){$w_2$}}
\put(100,5){\framebox(10,5){$w_2$}}
\end{picture}
\end{center}
\caption{\label{fig:overlap1}}
\end{figure}

\begin{figure}[t]
\begin{center}
\setlength{\unitlength}{1mm}
\begin{picture}(90,10)
\put(5,0){\framebox(10,5){$u_2$}}
\put(15,0){\framebox(10,5){$w_2$}}
\put(25,0){\framebox(10,5){$w_2$}}
\put(0,5){\framebox(10,5){$u_1$}}
\put(10,5){\framebox(10,5){$u_1$}}
\put(20,5){\framebox(10,5){$w_1$}}
\put(30,5){\framebox(10,5){$w_1$}}
\put(50,0){\framebox(10,5){$u_2$}}
\put(60,0){\framebox(10,5){$u_2$}}
\put(70,0){\framebox(10,5){$w_2$}}
\put(80,0){\framebox(10,5){$w_2$}}
\put(55,5){\framebox(10,5){$u_1$}}
\put(65,5){\framebox(10,5){$w_1$}}
\put(75,5){\framebox(10,5){$w_1$}}
\end{picture}
\end{center}
\caption{\label{fig:overlap2}}
\end{figure}

\begin{proof}
  The four conditions from the lemma are shown in
  Figure~\ref{fig:overlap1} and Figure~\ref{fig:overlap2}.  It is
  straightforward to show that any of these four situations implies
  $u_1^{\omegaop} v_1 w_1^\omega = u_2^{\omegaop} v_2 w_2^\omega$.
  For instance, if the left situation in Figure~\ref{fig:overlap1}
  occurs, then there exist words $x,y,x',y'$ such that $u_1 = xy$,
  $u_2 = yx$, $w_1 = x'y'$, $w_2 = y'x'$ and $v_2 w_2 = y v_1 x'$.
  Hence,
  $$
  u_1^{\omegaop} v_1 w_1^\omega = (xy)^{\omegaop} v_1 (x'y')^{\omegaop}
  = (yx)^{\omegaop} y v_1 x' (y'x')^{\omegaop} =
  u_2^{\omegaop} v_2 w_2 w_2^{\omegaop} =  u_2^{\omegaop} v_2 w_2^\omega .
  $$
  Let us now assume that $u_1^{\omegaop} v_1 w_1^\omega =
  u_2^{\omegaop} v_2 w_2^\omega$.  We distinguish the following cases:
 
  \medskip
  \noindent
  {\em Case 1.} The occurrence of $v_1$ in $u_1^{\omegaop} v_1
  w_1^\omega$ overlaps the occurrence of $v_2$ in $u_2^{\omegaop} v_2
  w_2^\omega$. Then, either $u_2 v_2 w_2^2$ is a factor of
  $u_1^2v_1w_1^2$ (if $v_2$ starts before $v_1$) or $u_1 v_1 w_1^2$ is
  a factor of $u_2^2v_2w_2^2$ (if $v_1$ starts before $v_2$), see
  Figure~\ref{fig:overlap1}.

  \medskip
  \noindent
  {\em Case 2.} The occurrence of $v_1$ in $u_1^{\omegaop} v_1
  w_1^\omega$ does not overlap the occurrence of $v_2$ in
  $u_2^{\omegaop} v_2 w_2^\omega$.

  \medskip
  \noindent
  {\em Case 2.1.} The occurrence of $u_1 v_1 w_1$ in $u_1^{\omegaop}
  v_1 w_1^\omega$ overlaps the occurrence of $v_2$ in $u_2^{\omegaop}
  v_2 w_2^\omega$. Then, one of the following two situations occurs:
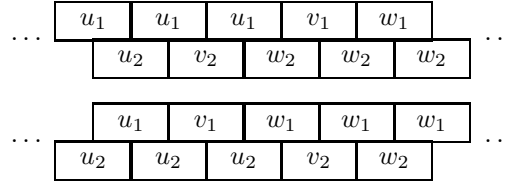
\begin{center}
\setlength{\unitlength}{1mm}
\begin{picture}(60,10)
\put(-6,5){$\ldots$}
\put(5,0){\framebox(10,5){$u_2$}}
\put(15,0){\framebox(10,5){$v_2$}}
\put(25,0){\framebox(10,5){$w_2$}}
\put(35,0){\framebox(10,5){$w_2$}}
\put(45,0){\framebox(10,5){$w_2$}}
\put(0,5){\framebox(10,5){$u_1$}}
\put(10,5){\framebox(10,5){$u_1$}}
\put(20,5){\framebox(10,5){$u_1$}}
\put(30,5){\framebox(10,5){$v_1$}}
\put(40,5){\framebox(10,5){$w_1$}}
\put(57,5){$\ldots$}
\end{picture}
\end{center}
\begin{center}
\setlength{\unitlength}{1mm}
\begin{picture}(60,10)
\put(-6,5){$\ldots$}
\put(0,0){\framebox(10,5){$u_2$}}
\put(10,0){\framebox(10,5){$u_2$}}
\put(20,0){\framebox(10,5){$u_2$}}
\put(30,0){\framebox(10,5){$v_2$}}
\put(40,0){\framebox(10,5){$w_2$}}
\put(5,5){\framebox(10,5){$u_1$}}
\put(15,5){\framebox(10,5){$v_1$}}
\put(25,5){\framebox(10,5){$w_1$}}
\put(35,5){\framebox(10,5){$w_1$}}
\put(45,5){\framebox(10,5){$w_1$}}
\put(57,5){$\ldots$}
\end{picture}
\end{center}
  In the first situation, we obtain $v_1 = w_1$ (since $v_1 w_1$ is
  a factor of $w_2^3$) and $u_2 = v_2$ (since $u_2v_2$ is a factor of 
  $u_1^3$). Hence, we get the left situation shown in
  Figure~\ref{fig:overlap2}, i.e., $u_2w_2^2$ is a factor of
  $u_1^2w_1^2$. In the second situation, we obtain $u_1 = v_1$ 
  (since $u_1v_1$ is a factor of $u_2^3$) and $v_2=w_2$ (since 
  $v_2w_2$ is a factor of $w_1^3$). Hence, we get the right situation shown in
  Figure~\ref{fig:overlap2}, i.e., $u_1w_1^2$ is a factor of
  $u_2^2w_2^2$.

  \medskip
  \noindent
  {\em Case 2.2.} The occurrence of $u_1 v_1 w_1$ in $u_1^{\omegaop} v_1
  w_1^\omega$ does not overlap the occurrence of $v_2$ in $u_2^{\omegaop} v_2
  w_2^\omega$. Then $u_1 v_1 w_1$ either occurs in
  $u_2^{\omegaop}$ or $w_2^{\omega}$. Hence, $u_1 = v_1 =
  w_1$ and similarly $u_2 = v_2 = w_2$. But
  $u_1^{\omegaop}u_1^{\omega} = u_2^{\omegaop}u_2^{\omega}$ implies
  that $u_2^3$ is a factor of $u_1^4$. Hence, the third condition
  from the lemma holds.
  \qed
\end{proof}

\begin{lemma}\label{lem:equalitytest}
  Given two succinct expressions $\dA_1,\dA_2$ over $\Sigma$ such that $\val(\dA_1)$
  and $\val(\dA_2)$ are primitive words, we can decide in polynomial time
  whether $\val(\dA_1)=\val(\dA_2)$.
\end{lemma}

\begin{proof}
  We have to distinguish the following cases:
  
  \medskip
  \noindent
  \emph{Case 1.} 
  $\val(\dA_i)$ ($i \in \{1,2\}$) is finite. Then $\val(\dA_1)=\val(\dA_2)$
  can be checked in polynomial time by Remark~\ref{remark:SLP-alg}(d).

  \medskip
  \noindent 
  \emph{Case 2.} $\val(\dA_i)$ is $\Gamma_i$-uniform ($i \in \{1,2\}$).
  Then $\val(\dA_1)=\val(\dA_2)$ if and only if  $\Gamma_1=\Gamma_2$ 
  which can be checked in polynomial time.

  \medskip
  \noindent
  \emph{Case 3.}  $\val(\dA_i)=u_iv_i^\omega$ ($i \in \{1,2\}$).  By Lemma~\ref{lem:oneblock}
  we can produce SLPs for $u_i$ and $v_i$ ($i \in \{1,2\}$) from
  $\dA_1$ and $\dA_2$, respectively, in polynomial time.  Let $k_i = |u_i|$ and
  $\ell_i = |v_i|$.  Let $\text{gcm}(\ell_1,\ell_2)$ denote the greatest common
  multiple of $\ell_1$ and $\ell_2$.
  By replacing $v_i$ by $v_i^{\max(k_1,k_2) \cdot
    \text{gcm}(\ell_1, \ell_2)/\ell_i}$ (for which we can compute an
  SLP in polynomial time by Remark~\ref{remark:SLP-alg}(b)), 
  we can assume that $|v_1| = |v_2| \geq
  k_1,k_2$.  Let $\ell = |v_1| = |v_2|$.  W.l.o.g assume that $k_1
  \leq k_2$ and let $k = k_2-k_1 \leq \ell$.  Then, we can replace
  $u_1$ and $v_1$ by $u_1 v_1[1:k]$ and $v_1[k+1:\ell] v_1[1:k]$,
  respectively (we can compute SLPs for these words in
  polynomial time by Remark~\ref{remark:SLP-alg}(c)).  
  Hence, we can also assume that $|u_1| = |u_2|$.
  But then, $u_1v_1^\omega = u_2v_2^\omega$ if and only if $u_1 = u_2$
  and $v_1 = v_2$, which can be checked in polynomial time by
   Remark~\ref{remark:SLP-alg}(d).

  \medskip
  \noindent
  \emph{Case 4.}  $\val(\dA_i)=u_i^{\omegaop} v_i$ ($i \in \{1,2\}$). 
  This case can be dealt with analogously to Case~3.

  \medskip
  \noindent
  \emph{Case 5.}  $\val(\dA_i)=u_i^{\omegaop} v_i w_i^{\omega}$ ($i
  \in \{1,2\}$). By Lemma~\ref{lem:oneblock}
  we can produce SLPs for $u_i$, $v_i$, and $w_i$ in
  polynomial time.  As in Case~3, by replacing the words $u_i, w_i$ by
  appropriate powers, we can enforce the condition $|u_1| = |u_2| =
  |w_1| = |w_2| = \ell \geq |v_1|, |v_2|$.  In addition, we can
  enforce the condition $|v_1| = |v_2| = \ell$ as follows: Let $k_i =
  |v_i| \leq \ell$. Then we can replace $v_i$ and $w_i$ by $v_i
  w_i[1:\ell-k_i]$ and $w_i[\ell-k_i+1:\ell]w_i[1:\ell-k_i]$, respectively.  Now,
  that we have $|u_1|=|u_2|=|v_1|=|v_2|=|w_1|=|w_2|$, we can check
  $u_1^{\omegaop} v_1 w_1^{\omega} = u_2^{\omegaop} v_2 w_2^{\omega}$
  in polynomial time using Lemma~\ref{lem:factor} and 
  Remark~\ref{remark:SLP-alg}(e).
  \qed
\end{proof}

\subsection{A polynomial time equivalence test for succinct expressions} \label{sec:main-alg}

In this section, we will finally prove Theorem~\ref{thm:poly_ST}.
The general strategy is very similar to \cite{BloomE05}.
We will incrementally reduce the $\height$ of the two 
given succinct expressions, until one of them (or both) describe
primitive words. This allows to use the results from the previous
section. We have to analyze carefully the size of 
the intermediate succinct expressions.
In the following, $\Sigma$ will always refer to a {\em finite} alphabet.
We will need certain nice properties of SESs.

\begin{definition}[primitive]
  A primitive SES is an SES $\dA = (V,\Sigma, \rhs)$ such that
  $\val_{\dA}(X)$ is primitive for all $X\in V$.  A 2-level system
  $\dB$ is primitive if $\lo(\dB)$ is
  primitive.
\end{definition}

\begin{definition}[irredundant]
  An irredundant SES is an SES $\dA = (V,\Sigma, \rhs)$
  such that $\val_{\dA}(X) \neq \val_{\dA}(Y)$ 
  for all  $X,Y\in V$ with 
  $X \neq Y$. Again we say that a 2-level system $\dB$ is
  irredundant if $\lo(\dB)$ is irredundant.
\end{definition}
One can think of a primitive and irredundant SES as a succinct
representation of a finite subset of $D(\Sigma)$ where $\val_{\dA} : V
\to D(\Sigma)$ defines an injective mapping from $V$ to this finite
subset. Hence, for a regular word $u$ such that all blocks belong to
the image of $\val_{\dA}$, we can define the $\val_{\dA}$-skeleton of
$u$. In the following, we will simply call it the $\dA$-skeleton of
$u$.  A primitive and irredundant 2-level system intuitively is a
system, where the terminal alphabet is a finite subset of $D(\Sigma)$
(namely the valuations of the variables of the lower part $\lo(\dB)$).

\begin{remark}\label{remark:making-irredundant}
If a primitive 2-level system $\dB$ is not irredundant then, using
Lemma~\ref{lem:equalitytest}, one can
produce in polynomial time an irredundant 2-level system $\dC$ such
that $\val(\dB) = \val(\dC)$. Indeed, if there are two different
variables $X,Y\in \Lo$ such that $\val_\dB(X) = \val_\dA(Y)$, then one has
to replace $X$ in all right-hand sides by $Y$. Thereafter $X$ can be
removed from $\Lo$.  Note that this process does not change the set of
upper level variables of $\dB$.
\end{remark}
Assume that $\dB$ is an SES or 2-level system and let
$u = (A_i)_{i\in I}$ be a (possibly infinite) sequence of variables of $\dB$.
We say that $u$ does not merge (is good,  semi-good, irreducible), if the sequence 
$(\val(A_i))_{i \in I}$ does not merge (is good, semi-good, irreducible).  Moreover, 
two sequences $u = (A_i)_{i\in I}$ and $v = (B_j)_{j\in J}$ of variables
(possibly from two different SESs or 2-level systems) are equivalent
if the sequences $(\val(A_i))_{i \in I}$ and $(\val(B_j))_{j \in J}$
are equivalent (i.e., $\prod_{i\in I} \val(A_i)$ and 
$\prod_{j\in J} \val(B_j)$ are isomorphic generalized words).
The following definition is an adaption of the definition of a proper
expression in \cite{BloomE05}.

\begin{definition}[proper]\label{def:proper}
  Let $\dB= (\Up,\Lo,\Sigma,\rhs)$ be a primitive 2-level system. A variable $X\in \Lo \cup \Up$ is proper if
  one of the following cases holds:
  \begin{enumerate}[(1)]
  \item $X \in \Lo$
  \item $\rhs(X)=Y_1\cdots Y_n$, where $Y_1 \cdots Y_n$ does not merge and
    $Y_1,\ldots,Y_n$ are proper.
  \item $\rhs(X)=Y^\omega$ or $\rhs(X)=Y^{\omegaop}$,
  where $Y$ is proper and $YYY$ does not merge.
   \item $\rhs(X)=[Y_1,\ldots,Y_n]^\eta$ where $Y_1,\ldots,Y_n$ are
    proper and $\val(X)$ is not primitive.
  \end{enumerate}
  The 2-level system $\dB$ is proper if $\dB$ is irredundant,
  primitive, and all variables are proper.
\end{definition}
Note that the condition that $YYY$ does not
merge in Definition~\ref{def:proper}(3) implies that
$YYY\cdots$ and
$\cdots YYY$ both do not merge
by~\cite[Corollary 32]{BloomE05}.
Moreover, condition (4) from Definition~\ref{def:proper}
means that $Y_1,\ldots,Y_n$ are
proper and at least on $\val(Y_i)$ is not a single symbol.

\begin{lemma}[see {\cite[Corollary 75]{BloomE05}}]\label{lem:skel}
  Let $\dB$ be a proper 2-level system and $X$ an upper level
  variable. Then $\uval(X)$ is the $\lo(\dB)$-skeleton of
  $\val(X)$.
\end{lemma}
The next two lemmas will be used to make a given 2-level system proper.

\begin{lemma}\label{cor:makegood}
  Given a primitive 2-level system $\dB$ and a finite semi-good sequence $A_1 \cdots A_m$
  of variables of $\dB$, we can produce in polynomial time 
  a primitive 2-level system $\dC$ and a sequence 
  $B_1 \cdots B_n$ of variables of $\dC$ such that the following holds:
  \begin{itemize}
  \item The upper parts of $\dB$ and $\dC$ are the same,
    and the lower part of $\dC$ extends the lower part of $\dB$ by at
    most $m-1$ many new lower level variables, whose right-hand sides
     have length 2.
  \item The sequence $B_1 \cdots B_n$ is good.
  \item $A_1 \cdots A_m$ and  $B_1 \cdots B_n$ are equivalent sequences.
  \item The subsequence of upper level variables in $A_1 \cdots A_m$
     is the same as the subsequence of  upper level variables in $B_1 \cdots B_n$.
    \item $n\leq m$.
  \end{itemize}
\end{lemma}

\begin{proof}
As long as the sequence $A_1 \cdots A_m$ contains a factor
$A_i A_{i+1}$ or $A_i A_{i+1} A_{i+2}$, whose evaluation is a left-hand
side of our rewriting system $R$, we do the following:

If $\val(A_i)$ is right-closed and $\val(A_{i+1})$ is left-closed, then 
we introduce a new lower level variable
$A$, set $\rhs(A) = A_i A_{i+1}$, and  replace the sequence 
$A_1 \cdots A_m$ by the sequence
$A_1 \cdots A_{i-1} A A_{i+2} \cdots A_m$.
If $\val(A_i) = \val(A_{i+1}) = \Gamma^\eta$ for some $\Gamma \subseteq \Sigma$,
we continue with the sequence 
$A_1 \cdots A_{i-1} A_{i+1} \cdots A_m$.
Finally, if $\val(A_i) = \val(A_{i+2}) = \Gamma^\eta$ for some $\Gamma \subseteq \Sigma$
and $\val(A_{i+1}) = a \in \Gamma$,
we continue with the sequence 
$A_1 \cdots A_{i-1} A_{i+2} \cdots A_m$.
We iterate this process as long as possible.
\qed
\end{proof}

\begin{lemma}\label{cor:makegood2}
  Given a primitive 2-level system $\dB$ and a finite
  irreducible sequence $A_1 \cdots A_k$ ($k \geq 3$), where 
  every $A_i$ is a lower level variable of $\dB$, we can produce in polynomial time 
  a primitive 2-level SES $\dC$ and sequences 
  $B_1 \cdots B_m$, $C_1 \cdots C_n$ ($m \geq 0$, $n \geq 1$)
  of lower level variables of $\dC$ such that the following holds:
  \begin{itemize}
  \item The upper parts of $\dB$ and $\dC$ are the same,
    and the lower part of $\dC$ extends the lower part of $\dB$ by at
    most one new lower level variable, whose right-hand side
     has length 2.
   \item The infinite sequence $B_1 \cdots B_m (C_1 \ldots C_n)^\omega$ is irreducible.
  \item $(A_1 \cdots A_k)^\omega$ and $B_1 \cdots B_m (C_1 \cdots C_n)^\omega$
    are equivalent sequences.
  \item $m,n\leq k$.
  \end{itemize}
\end{lemma}

\begin{proof}
W.l.o.g.  assume that $(A_1 \cdots A_k)^\omega$ is not irreducible.
Since $A_1 \cdots A_k$ is irreducible, an $R$-reduction in the infinite sequence
$A_1 \cdots A_k A_1 \cdots A_k A_1 \cdots A_k \cdots$ can only occur at a border
between $A_k$ and $A_1$.
There are the following cases, according to the left-hand sides of the
system $R$.

\medskip
\noindent
{\em Case 1.} $\val(A_k) = \val(A_1) = \Gamma^\eta$ for some $\Gamma
\subseteq \Sigma$. Then, the infinite sequence 
$A_1 A_2 \cdots A_k (A_2 \cdots A_k)^\omega$ is irreducible and 
equivalent to our original sequence (recall that $k \geq 3$).

\medskip
\noindent
{\em Case 2.} $\val(A_k)$ is scattered and right-closed, 
$\val(A_1)$ is scattered and left-closed.
Then, we introduce a new lower level variable $A$ with $\rhs(A) = A_k A_1$.
It follows that the infinite sequence 
$A_1 A_2 \cdots A_{k-1} (A A_2 \cdots A_{k-1})^\omega$ is irreducible
and equivalent to our original sequence.

\medskip
\noindent
{\em Case 3.} $\val(A_k) = \Gamma^\eta$, $\val(A_1) = a$, 
$\val(A_2) = \Gamma^\eta$ for some $\Gamma \subseteq \Sigma$ and $a
\in \Gamma$. If $k = 3$, then $A_1 A_2 \cdots A_k = A_1 A_2 A_3$
would not be irreducible (since $\val(A_2) = \val(A_3)  =
\Gamma^\eta$), which contradicts our assumptions.
Hence, assume that $k \geq 4$. 
Then, the sequence 
$A_1 A_2 \cdots A_k (A_3 \cdots A_k)^\omega$ is again 
irreducible and equivalent to our original sequence.

\medskip
\noindent
{\em Case 4.} $\val(A_{k-1}) = \Gamma^\eta$, $\val(A_k) = a$, 
$\val(A_1) = \Gamma^\eta$ for some $\Gamma \subseteq \Sigma$ and $a
\in \Gamma$. This case is similar to Case~3.
\qed
\end{proof}
Let $\dB$ be an SES and $X$ a variable with $\height(X)=h \geq
1$. Then there is a sequence of variables $X_1,\ldots,X_h$ such
that $X_h=X$, $X_i \preceq_{\dB} X_{i+1}$,  and $\height(X_i)=i$.
Note that $\val(X_1)$ is either primitive or a shuffle of finite words.
If $\val(X_1) = [u_1,\ldots,u_k]^\eta$ where at least one of the $u_i$ is in
$\Sigma^{\geq 2}$ (thus, $\val(X_1)$ is not primitive), 
then this sequence is called a {\em bad sequence}. If a
variable $X$ has a bad sequence, then we say it is of {\em bad
shape}. Otherwise it is of {\em good shape}.
For instance, if $\rhs(X) = [Y]^\eta$ and $\rhs(Y) = ab$, then
$X$ is of bad shape.

\begin{proposition}\label{prop:mainsstep}
   Let $\dB=(V,\Sigma, \rhs)$  be an SES such that for every variable 
  $X \in V$, either $\rhs(X) \in \Sigma^+ \cup \Sigma^*V\Sigma^* \cup VV$
  or $\rhs(X)$ is of the form $Y^\omega$, $Y^{\omegaop}$, or
  $[Y_1,\ldots,Y_n]^\eta$ for $Y, Y_1,\ldots,Y_n \in V \cup \Sigma$.
  Given $\dB$ 
  we can produce in polynomial
  time a proper 2-level system $\dC=(\Up,\Lo,\Gamma,\rhs)$ such that
  every variable $X\in V$, where $\val_\dB(X)$ is not
  primitive, belongs to $\Up$ and for each of these variables $X$ we have:
 \begin{enumerate}[(a)]
  \item $\val_\dB(X)=\val_\dC(X)$
  \item If $X$ is of good shape in $\dB$, then $\height_{\dB}(X)>\height_{\up(\dC)}(X)$.
  \item If $X$ is of bad shape in $\dB$, then 
    $\height_{\dB}(X)=\height_{\up(\dC)}(X)$ and $X$ is of good
    shape in $\up(\dC)$.
  \end{enumerate}
\end{proposition}

\begin{proof}
  W.l.o.g. we can assume that
  $\val(\dB)$ is not primitive.  We start with some preprocessing.

  \paragraph{\bf Preprocessing.}
  First we transform our succinct expression $\dB$ into a 2-level system $\dC$ by
  collecting in $\Lo$ all variables $X$ such that $\val(X)$ is primitive.
  This can be done in polynomial time using Lemma~\ref{lem:oneblock}.
  Note that if $\val(X)$ is primitive and scattered, then for every $Y$ in
  $\rhs(X)$, $\val(Y)$ is primitive too. But if $\val(X)$ is primitive
  and dense (i.e., of the form $\Gamma^\eta$ for some
  $\Gamma \subseteq \Sigma$), then this is not necessarily
  true.\footnote{Let, for instance, $\rhs(X) = [Y]^\eta$ 
  with $\val(Y) = a [a]^\eta$. Then $\val(X) = [a]^\eta$ is primitive but 
  $\val(Y)$ is not primitive.}
  Hence, in this case we have to redefine $\rhs(Y) = \Gamma^\eta$.
  After this process the 2-level system $\dC$
  is already primitive, satisfies conditions (a), (b), and  (c) in 
  our proposition, and for all $X\in \Up$ the word $\val(X)$ is not
  primitive. All these properties will stay invariant throughout the
  remaining proof where we manipulate the system $\dC$ in order to
  make it proper.

  Before we come to the actual algorithm we transform $\dC$ for
  technical convenience such that for all $X\in \Up$ one of the
  following holds:
  \begin{enumerate}[(1)]
  \item $\rhs(X)\in \Lo^{\geq 2}\cup\Lo^*\Up\Lo^*$,
  \item $\rhs(X)=[Y_1,\ldots Y_n]^\eta$ for some $Y_1,\ldots,Y_n\in \Up\cup\Lo$,
  \item $\rhs(X) \in \Up\Up$,
  \item $\rhs(X)=Y^\omega$ for $Y \in \Up\cup\Lo$,
  \item $\rhs(X)=Y^{\omegaop}$ for $Y \in \Up\cup\Lo$.
  \end{enumerate}
  In order to achieve this form we simply introduce for each upper
  level variable $X$ with $\rhs(X)=uYv$ where $u,v\in\Sigma^*$ and
  $Y\in V$ two variables $X_u,X_v\in\Lo$ and set 
  $\rhs(X) = X_u Y X_v$,
  $\rhs(X_u)=u$, and
  $\rhs(X_v)=v$ (if e.g. $u = \varepsilon$, then $X_u$ is not present).  
  Moreover, if a symbol $a \in \Sigma$ occurs in a right-hand
  side of the form $Y^\omega$, $Y^{\omegaop}$, or
  $[Y_1,\ldots,Y_n]^\eta$, then we replace that occurrence by 
  a new $\Lo$-variable with right-hand side $a$.
  
  In fact, by this preprocessing all right-hand sides
  of the form (1) have length at most $3$.  This fact will be important
  when we estimate the size of the final system.
  From now on variables in $\Up$
  that have a right-hand side of form (1) or (2) are said to be of 
  type (1, 2), all other variables are said to be of type (3-5).

  Following \cite[proof of Theorem 65 \& 66]{BloomE05} we will now
  give an algorithm that produces a proper 2-level system.  We will proceed along the
  hierarchical order of the variables in $\Up$ where in each step we
  possibly add a constant number of new variables and change the
  right-hand sides of the old variables such that all variables are
  proper and of the form (1)--(5) and, moreover, all old variables $X$
  are of type (1, 2) and fulfill the following technical condition (TEC):
  \begin{framed}
  \begin{enumerate}[(a)]
  \item If $\val(X)$ has a first block, then $\rhs(X)\in \Lo^{\geq
      2}\cup\Lo^+\Up\Lo^*$ and the first variable of $\rhs(X)$
    evaluates to the first block of $\val(X)$.
  \item If $\val(X)$ has a second block and the first block is
    scattered, then $\rhs(X)\in \Lo^{\geq 2}\cup\Lo^{\geq 2}\Up\Lo^*$
    and the second variable of $\rhs(X)$ evaluates to the second block
    of $\val(X)$.
  \item If $\val(X)$ has a last block then $\rhs(X)\in \Lo^{\geq
      2}\cup\Lo^*\Up\Lo^+$ and the last variable of $\rhs(X)$
    evaluates to the last block of $\val(X)$.
  \item If $\val(X)$ has a second last block and the last block is
    scattered, then $\rhs(X)\in \Lo^{\geq 2}\cup\Lo^*\Up\Lo^{\geq 2}$
    and the second last variable of $\rhs(X)$ evaluates to the second
    last block of $\val(X)$.
  \end{enumerate}
  \end{framed}
  We need the following claim about this property (TEC):
  \begin{claim} \label{Claim}
    If $\rhs(X)\in\Lo^+\cup\Lo^*\Up\Lo^*$ and $\rhs(X)$ is good, then
    $X$ satisfies (TEC).
  \end{claim}
  \begin{proof}
    By symmetry let us only consider conditions (a) and (b) of
    (TEC). Assume that $\rhs(X)$ is a good sequence.
    If $\rhs(X) \in \Lo^*$, then 
    Lemma~\ref{lem:goodtorporper} implies that the variables
    in $\rhs(X)$ evaluate to the blocks of $\val(X)$ (recall that $\rhs(X)$ is good).
    Hence (a) and (b) hold. Next, assume that $\rhs(X) \in \Lo^{\geq
      2}\Up\Lo^*$.  Again, since $\rhs(X)$ is good, Lemma~\ref{lem:goodtorporper}
      implies that the first two variables in $\rhs(X)$ evaluate to the first
      two blocks of $\val(X)$. Thus, (a) and (b) hold again.
     If  $\rhs(X)\in\Up\Lo^*$, then the first variable of $\rhs(X)$
     evaluates to a non-primitive word. Since $\rhs(X)$ is good,
     it follows that $\val(X)$ does not have a first block and (a) and (b) hold.
     Finally assume that $\rhs(X)\in\Lo\Up\Lo^*$ and
    the first two variables of $\rhs(X)$ are $A\in\Lo$ and
    $Z\in\Up$. Then, $\val(A)$ is the first block of $\val(X)$.
    Since $\rhs(X)$ is good either $\val(Z)$ does not have
    a first block or $\val(Z)$ has a first block, $\val(A)$ is uniform,
    and $(\val(A),\val(Z))$ does not merge. In both cases (a) and (b) are
    obviously satisfied.
    This proves the claim.
  \end{proof}

  \paragraph{\bf Actual algorithm.}
  We can now outline our procedure. 
  Consider a variable $X \in \Up$ such that 
  every variables in $\rhs(X)$ is either in $\Lo$ or 
  was already processed and is therefore now proper, satisfies (TEC), 
  and is  of type (1, 2).
  We need to distinguish 
  on the form of the right-hand side of $X$. 
  In all of the following cases, we reset $\rhs(X)$ 
  either 
  \begin{enumerate}[(i)]
  \item to a shuffle of variables that are already proper or
  \item to a good sequence from
   $\Lo^+ \cup \Lo^* \Up \Lo^*$ (and all variables in that sequence
  are already proper).
  \end{enumerate}
  In (i), $X$ is proper by Definition~\ref{def:proper}(4) (note that $\val(X)$ is not primitive since $X \in \Up$).
  In (ii) it follows from
  Lemma~\ref{lem:goodtorporper} and Claim~\ref{Claim},
  that $X$ is proper and satisfies (TEC).   
  For every other new upper level variables $Y$ that is  introduced, the right-hand side
  is  either 
  \begin{enumerate}[(i)]
  \item
  a non-merging sequence of (already proper) variables or
  \item
  $Z^\omega$ or $Z^{\omegaop}$, where $Z$ is already proper and $ZZZ$ does not merge.
  \end{enumerate}
  In both cases it follows from Definition~\ref{def:proper} that  $Y$ is proper too.
  
  \medskip
  \noindent\emph{Case 1.} $\rhs(X)\in
  \Lo^2\cup\Lo^3$ (hence $\rhs(X)$ is semi-good).  
  By applying Lemma~\ref{cor:makegood} to $\rhs(X)$, we can compute in polynomial
  time an equivalent good sequence of at most three possibly new $\Lo$-variables
  (and their corresponding right-hand sides). This sequence 
  becomes the new right-hand side of $X$. 
  
  \medskip
  \noindent\emph{Case 2.} $\rhs(X)\in \Lo^{\leq 1}\Up\Lo^{\leq 1}$.  
  Let $Z$ be the unique $\Up$-variable in $\rhs(X)$.
  Note that $Z$ is one of the old variables, which has already been processed
  and hence is proper, of type (1, 2), and satisfies (TEC).
  If $\rhs(Z) \in\Lo^{\geq 2}\cup\Lo^*\Up\Lo^*$, then we replace
  $Z$ in $\rhs(X)$ by $\rhs(Z)$ (if $\rhs(Z)$ is a shuffle, then 
  we leave $Z$ in $\rhs(X)$). Recall that $Z$ is proper and satisfies
  (TEC). It follows easily that the resulting new right-hand side of $X$ is semi-good and in
  $\Lo^{\geq 2}\cup\Lo^*\Up\Lo^*$. Thus, we can apply
  Lemma~\ref{cor:makegood} and obtain an equivalent good sequence in
  $\Lo^+\cup\Lo^*\Up\Lo^*$ (as in 
  Case~1, we will introduce new $\Lo$-variables thereby).
  This good sequence will be the
  new right-hand side of $X$. 
 
  \medskip

  \noindent\emph{Case 3.} $\rhs(X)=[Y_1,\ldots,Y_k]^\eta$. Then
  there is nothing to do. Recall that we assumed that $\val(X)$ is not
  primitive and hence $X$ is proper and satisfies the technical
  condition (TEC) as $\val(X)$ neither has a first nor a last block.

  \medskip
  
  \noindent\emph{Case 4.} $\rhs(X)=YZ$ for some $Y,Z\in \Up$.
   Here $Y$ and $Z$ are old variables, which 
   have already been processed and therefore are proper, of type (1, 2), and satisfy (TEC).
   If $\rhs(Y)\in\Lo^{\geq 2}\cup\Lo^*\Up\Lo^*$ then we replace $Y$ in
   $YZ$ by $\rhs(Y)$ (if $\rhs(Y)$ is a shuffle, we leave $Y$ in
   $YZ$). We proceed analogously with $Z$ in $YZ$.
   Since $Y$ and $Z$ are proper and satisfy (TEC), it follows (as in
   Case 2) that the resulting new right-hand side of $X$ is semi-good
   and contains at most two variables from $\Up$. Thus we can apply
   Lemma~\ref{cor:makegood} and obtain an equivalent good sequence $u$
   of variables with at most
   two variables from $\Up$ (again, we introduce new $\Lo$-variables thereby).
  
   Now, we replace parts in the sequence $u$ in order to get $\rhs(X)$.
   First, assume that $u = A_1\cdots A_k\in \Lo^+$. 
   If $k \leq 5$, then $\rhs(X)$ simply becomes $u$ (which is good).  
   If $k \geq 6$, then we introduce a new $\Up$-variable $U$ and set
   $$
    \rhs(X) = A_1 A_2 U A_{k-1}A_k, \quad
    \rhs(U) = A_3 \cdots A_{k-2} .
   $$
   Since $u$ is good, both right-hand sides are good as well.
    Second, assume that $u = A_1\cdots A_k U B_1\cdots
    B_\ell\in\Lo^*\Up\Lo^*$ with $U\in\Up$. If $k \leq 2$ and $\ell
    \leq 2$ then we we simply set $\rhs(X) = u$.
     On the other hand, if $k > 2$ or $\ell > 2$, then we
    introduce a new $\Up$-variable $V$ and set 
    $$
    \rhs(X) =  A_1 A_2 V B_{\ell-1} B_\ell, \quad
    \rhs(V) = A_3\cdots A_k U B_1\cdots B_{\ell-2}
    $$
    (if e.g. $k > 2$ but $\ell = 1$, then $B_1\cdots B_{\ell-2}$ and
    $B_{\ell-1}$ disappear).  
    Since $u$ is good, $\rhs(X)$ will be good too. Moreover,
    since $u$ does not merge (by Lemma~\ref{lem:goodtorporper}),
    $\rhs(V)$ does not merge as well ($\rhs(V)$ is not necessarily good).
    Third, assume that  
    $u = A_1\cdots A_k U B_1\cdots B_\ell V C_1\cdots C_n\in\Lo^*\Up\Lo^*\Up\Lo^*$ with
    $U,V\in\Up$. In this case we introduce two new $\Up$-variables
    $W_1$ and $W_2$ and set 
    $$
    \rhs(X) = A_1 A_2 W_1 C_1\cdots C_n, \;\;
    \rhs(W_1) = W_2 V, \;\;
    \rhs(W_2) = A_3\cdots A_k U B_1\ldots B_\ell .
    $$
    Again, since $u$ is good, $\rhs(X)$ is good as well. 
    Moreover, since $u$ does not merge, neither $\rhs(W_1)$ nor
    $\rhs(W_2)$ merges. Note that the number $n$ in the right-hand
    side of $X$ above is bounded by $|\rhs(Z)|$. This will be
    important for estimating the length of right-hands. 
    
    \medskip
  \noindent\emph{Case 5.} $\rhs(X)=Y^\omega$. 
   Note that $Y$ is either a $\Lo$-variable, or it is 
   an old $\Up$-variable, which has already been processed
   and hence is proper, of type (1, 2), and satisfies (TEC).
   We can therefore distinguish the following subcases.

   \medskip

  \noindent
  \emph{Case 5(a).} $\rhs(Y)=[Z_1,\ldots,Z_n]^\eta$ for some
  $Z_1,\ldots,Z_n\in\Lo\cup\Up$. Then by the general identity
  $(\Gamma^\eta)^\omega \cong \Gamma^\eta$ (which follows from Cantor's theorem), we have
  $\val(X)=\val(Y)$ and we set $\rhs(X)=Y$. Then $X$ is obviously
  proper. Since we assumed that $\val(X)$ is not primitive $\val(X)$
  does not have a first or a last block and (TEC) is satisfied.
 
  \medskip

  \noindent
  \emph{Case 5(b).} $\rhs(Y) \in \Lo^* \Up \Lo^*$.
   Let $\rhs(Y) = u Z v$ with $Z \in \Up$ and $u, v \in \Lo^*$.
   Since $Y$ is proper and satisfies (TEC),
   the infinite sequence $u Z v u Z v \cdots = 
   u (Z v u)^\omega$ is semi-good. 
   By applying Lemma~\ref{cor:makegood} to the sequence
   $vu$ of $\Lo$-variables,
   we obtain an
   equivalent good sequence
   $u (Z w)^\omega$. Here $w$ is a sequence of (possibly
   new) $\Lo$-variables such that $w$ represents the 
   irreducible normal form w.r.t.~$R$ of the sequence
   represented by $vu$. Note that $|w| \leq  |uv|$.
   We set
   $$
   \rhs(X) = u V, \quad \rhs(V) = U^\omega, \quad \rhs(U) = Z w .
   $$  
   Since the sequence $u(Zw)^\omega$ is good, also the sequence $uV$ is good. 
   Moreover, since $u(Zw)^\omega$ does not merge 
   (by Lemma~\ref{lem:goodtorporper}), the same holds for
   $\rhs(U)$ and $UUU$ (so $U$ and $V$ are proper by definition).

   \medskip

  \noindent
  \emph{Case 5(c).} $Y\in \Lo$ and hence $\val(Y)$ is primitive. Then the infinite sequence
  $YYY\cdots$ must be irreducible, because otherwise
  $\val(Y)$ would be either finite or uniform and 
  $\val(X) = \val(Y^\omega)$ would be primitive.
  We introduce a new $\Up$-variable $Z$
  and set 
  $$
  \rhs(X)=YYZ, \quad \rhs(Z)=Y^\omega .
  $$
  Then $\rhs(X)$ is
  good and $YYY$ does not merge.
   
  \medskip 

  \noindent
  \emph{Case 5(d).} $\rhs(Y) \in \Lo^2$. Let $\rhs(Y)=A_1A_2$ for 
  $A_1,A_2\in\Lo$. Since $Y$ is already proper, we know that $A_1 A_2$
  is irreducible.
  If the infinite sequence $A_1 A_2 A_1 A_2 \cdots$ is irreducible
  too, then we introduce a new $\Up$-variables $Z$ 
  and set 
  $$
  \rhs(X)= A_1 A_2 Z, \quad \rhs(Z) = Y^\omega .
  $$ 
  Clearly, $\rhs(X)$ is good and $YYY$ does not merge.
  On the other hand, if $A_1 A_2 A_1 A_2 \cdots$ is not irreducible,
  then (since $A_1 A_2$ is irreducible), an $R$-reduction can only 
  occur at a border between $A_2$ and $A_1$. 
  The case that $\val(A_1) = \val(A_2) = \Gamma^\eta$ for some 
  $\Gamma \subseteq \Sigma$ cannot occur (since $A_1 A_2$ is
  irreducible). If $\val(A_2)$ is scattered and right-closed 
  and $\val(A_1)$ is scattered and left-closed, then we introduce a 
  new $\Lo$-variable $B$ and a new $\Up$-variable $Z$ 
  and set
  $$
  \rhs(X)= A_1 B Z, \quad \rhs(Z) = B^\omega, \quad \rhs(B) = A_2 A_1 .
  $$
  It is straightforward to show that the infinite sequence $A_1 B B B
  \cdots$ is irreducible. Hence $\rhs(X)$ is good and $BBB$ does not merge.
  Next, if $\val(A_1) = \Gamma^\eta$ and $\val(A_2) = a$
  for some  $\Gamma \subseteq \Sigma$ and $a \in \Gamma$, then 
  $A_1 A_2 A_1 A_2 \cdots$ evaluates to 
  $\Gamma^\eta$. Hence, $\val(X)$ is primitive, which is a contradiction.
  Finally, if $\val(A_2) = \Gamma^\eta$ and $\val(A_1) = a \in \Gamma$,
  then $A_1 A_2 A_1 A_2 \cdots$ evaluates to 
  $a \Gamma^\eta = \val(Y)$ and we set $\rhs(X) = Y$.

  \medskip

  \noindent 
  \emph{Case 5(d).}  $\val(Y)\in \Lo^{\geq 3}$.
  We apply Lemma~\ref{cor:makegood2} to the irreducible sequence $\rhs(Y)$ and compute
  sequences $u$, $v$ of (possibly
  new) $\Lo$-variables with their corresponding right-hand sides. 
  The infinite sequence $u v^\omega$ of $\Lo$-variables is irreducible and
  evaluates to $\val(Y)$. W.l.o.g. we can assume $|u| \geq 2$
  (otherwise, we can replace $u$ by $uvv$). We introduce new $\Up$-variables $U$ and $V$
  and set
  $$
  \rhs(X) = u V, \quad \rhs(V) = U^\omega, \quad \rhs(U) = v .
  $$
  (if $|v|=1$, i.e., $v$ consists of a single $\Lo$-variable, 
  then we do not need $U$).
  
  \medskip

  \noindent\emph{Case 6.} $\rhs(X)=Y^{\omegaop}$. This case is symmetric
  to Case 4.

  \medskip
  \noindent
  The resulting system $\dC$ is primitive and all $\Up$-variables are proper.
  On the other hand, $\dC$ is not 
  necessarily irredundant. But this can be easily achieved as described in 
  Remark~\ref{remark:making-irredundant}.
  \qed
\end{proof}
We are now in the position to prove Theorem~\ref{thm:poly_ST}.

\medskip
\noindent
{\em Proof of Theorem~\ref{thm:poly_ST}.}
It suffices to show that
the following problem can be solved in polynomial time:

\medskip
\noindent
INPUT: An SES $\dA$ and two variables $X,Y$ of $\dA$.

\noindent
QUESTION: $\val(X) \cong \val(Y)$?

\medskip
\noindent
  If both variables $X$ and $Y$ evaluate to primitive words, then we just need to
  apply Lemma~\ref{lem:equalitytest}. If only one of the two 
  evaluates to a primitive word, then $\val(X) \not\cong \val(Y)$. 
  Hence, we may assume that
  both $\val(X)$ and $\val(Y)$ are not primitive. In particular, we 
  have $\height(X), \height(Y) >0$. It is easy to bring $\dA$ into
  the normal form required in Proposition~\ref{prop:mainsstep}.
  Applying Proposition~\ref{prop:mainsstep} 
  to $\dA$ gives a proper 2-level system $\dA_0$. The variables
  $X$ and $Y$ belong to the upper level part of $\dA_0$. 
  Starting with $\dA_0$ we construct a sequence of proper 2-level systems
  $\dA_j=(\Up_j,\Lo_j,\Lo_{j-1},\rhs_j)$ (with $\Lo_{-1}=\Sigma$).
  In order to obtain $\dA_j$ we apply the procedure of
  Proposition~\ref{prop:mainsstep} to $\up(\dA_{j-1})$.  
  Let $k$ be maximal such that $X$ and $Y$ belong to 
  the upper level part of $\dA_k$.  
  Since by Proposition~\ref{prop:mainsstep} in every
  second step the $\height$ of $X$ and $Y$ strictly decreases we
  have $k \leq 2\cdot |\dA|$. 
  
  Let $0 \leq j \leq k$. By
  Lemma~\ref{lem:skel} $\uval_j(X)$ is the $\lo(\dA_j)$-skeleton of
  $\val_j(X)$ 
  and similarly for
  $Y$. Hence $\val_j(X) \cong \val_j(Y)$ if and only if $\uval_j(X) \cong \uval_j(Y)$
  by Proposition~\ref{prop:skel}. 
  Recall that $\dA_{j+1}$ is obtained by applying the procedure of
  Proposition~\ref{prop:mainsstep} to $\up(\dA_{j})$.  
  We obtain 
  $\val_j(X) \cong \val_j(Y)$ if and only if $\val_{j+1}(X) \cong \val_{j+1}(Y)$
  for all $0 \leq j < k$. Hence, $\val(X) \cong \val(Y)$ if and only if 
  $\val_k(X) \cong \val_k(Y)$ if and only if $\uval_k(X) \cong \uval_k(Y)$.
  Now, by the maximality of $k$, $\uval_k(X)$ or $\uval_k(Y)$ must be primitive.
  Hence, using Lemma~\ref{lem:equalitytest},
  we can check in polynomial time whether $\uval_k(X) \cong \uval_k(Y)$.

  \paragraph{\bf Runtime.}
  Let us analyze the system $\up(\dA_j)$ for $1\leq j\leq k$. The
  2-level system $\dA_j$ is obtained by applying
  Proposition~\ref{prop:mainsstep} to $\up(\dA_{j-1})$.  Observe that
  by the construction in the proof, 
  the system $\up(\dA_j)$ already has the normal form that we require
  in Proposition~\ref{prop:mainsstep}.
  Let $\NotNice_j$ be the set of variables in $\Up_j$ that are 
  of type (3-5).

  Now let us estimate the number $|\Up_j|$ for $1\leq j \leq k$. Observe
  that in the proof of Proposition~\ref{prop:mainsstep} in each of the
  Cases (1)--(3) only new lower level variables are introduced. In
  each of the Cases (4)--(6) the old variable is turned into a 
  variable of type (1, 2) and at most one new variable of type (3-5) is added to
  $\Up_j$. Moreover, additionally at most one new  variables of type (1, 2) is
  added to $\Up_j$. We conclude that $|\NotNice_j|\leq
  |\NotNice_{j-1}|$ and the total number of variables in $\Up_j$ is
  bounded by $|\Up_{j-1}|+2\cdot |\NotNice_{j-1}|$.  Recall that $j
  \leq k \leq 2|\dA|$. Hence
  $|\Up_j|\leq |\Up_0|+2j\cdot |\NotNice_0|\leq |\dA_0|\cdot 
  (4\cdot|\dA|+1)$ for all $0\leq j\leq k$.

  Let us now estimate the maximal length of a right-hand side in
  $\dA_j$.  
  Let us first bound the length of the right-hand side of a variable $X \in
  \Up_j\cap\Up_{j-1}$ (i.e., an old variable). 
  By reanalyzing all cases from the proof of 
  Proposition~\ref{prop:mainsstep}, we see that 
  for such 
  a variable $X$, $|\rhs_j(X)|$ is either at most 5 or it is
  bounded by $3+|\rhs_j(Y)|$, where $Y \in
  \Up_j\cap\Up_{j-1}$ is an old variable, which was processed before.
  We therefore obtain
  $|\rhs_j(X)| \leq 3 \cdot |\Up_j \cap \Up_{j-1}| + 5$.
  Hence, $|\rhs_j(X)| \leq 3\cdot |\dA_0| \cdot (4\cdot|\dA|+1) + 5$.
  For the newly added variables, $X \in \Up_j\setminus\Up_{j-1}$ 
  the size of the right-hand
  side is bounded by twice the maximal size of a
  right-hand side of an old variable in $\Up_j\cap\Up_{j-1}$
  (the factor 2 comes from Case~4). Hence
  $|\rhs_j(X)|\leq 6 \cdot |\dA_0| \cdot (4\cdot|\dA|+1) + 10$
  for all $X\in\Up_j$. Finally, note that $|\dA_0|$ is 
  bounded polynomially bounded in $|\dA|$.

  Concerning lower level variables of $\dA_j$, 
  note that the length $|\rhs_j(A)|$ for a lower level
  variable of $\dA_j$ is bounded by 2 (if $A$ is introduced in one
  of the Cases~1--6) or by the maximal length of the right-hand side
  of a variable from $\dA_{j-1}$ (if $A$ is introduced in the 
  preprocessing step). Moreover, in each of the Cases~1--6,
  the number of new lower level variables that are introduced
  is bounded by twice the maximal size of a
  right-hand side of an old variable in $\Up_j\cap\Up_{j-1}$
  (the factor 2 comes again from Case~4). 
  Hence the number of lower level variables is also
  bounded polynomially in $|\dA|$.
  
  We have shown that the total size of very 2-level system $\dA_j$
  ($1 \leq j \leq k$)
  is bounded polynomially in $|\dA|$.
  As the time needed to construct $\dA_{j+1}$ from $\dA_j$ is 
  polynomially bounded by Proposition~\ref{prop:mainsstep},
  we conclude that the overall running time of our algorithm
  is polynomially bounded as well.
\qed

\subsection{Lower bounds for regular linear orders}

In this section we prove lower bounds for the isomorphism
problem for regular words. In fact, all these lower bounds 
only need a unary alphabet, i.e., they hold for 
regular linear orders. The results in this section nicely
contrast the results from Section~\ref{sec:reg-trees}, where
we studied the isomorphism problem for the prefix order
trees on regular languages. In this section, we replace
the prefix order by the lexicographical order.

\begin{theorem} \label{thm:Ptime-complete-ST}
The following problem is $\Ptime$-hard (and hence $\Ptime$-complete)
for every finite alphabet $\Sigma$:

\medskip
\noindent
INPUT: Two succinct expressions $\dA_1$ and $\dA_2$ over the alphabet $\Sigma$. 

\noindent
QUESTION: $\val(\dA_1) \cong \val(\dA_2)$?
\end{theorem}

\begin{proof}
Note that the problem can be solved in 
polynomial time
by Theorem~\ref{thm:poly_ST}. 
$\Ptime$-hardness will be shown by 
a reduction from the monotone circuit value problem.
So, let $C$ be a monotone Boolean circuit.
We can assume that the gates of $C$ are 
partitioned into layers $L_1, \ldots, L_n$, where
$L_1$ contains all input gates, $L_n$ only contains
the output gate, and all inputs for a gate from $L_{i+1}$
belong to $L_i$. Moreover, $L_i$ ($i > 1$) either contains only
$\AND$-gates or $\OR$-gates.
We construct an SES $\dA$ (over a unary terminal alphabet $\{a\}$), 
which contains for each
gate $v$ of $C$ a variable $\test_v$ and for each layer
$d \in \{1,\ldots,n\}$ two variables 
$\good_d$, and $\bad_d$
such that the following holds for all gates $v \in L_d$:
\begin{enumerate}[(a)]
\item Either $\val_{\dA}(\test_v) \cong  \val_{\dA}(\bad_d)$ 
or $\val_{\dA}(\test_v) \cong  \val_{\dA}(\good_d)$.
\item $\val_{\dA}(\test_v) \cong  \val_{\dA}(\good_d)$
if and only if gate $v$ evaluates to $\TRUE$.
\item The linear orders
$\val_{\dA}(\good_d)$ and $\val_{\dA}(\bad_d)$
do not contain an interval isomorphic to $\omega \cdot d$ (recall that
$\omega \cdot d$ denotes the linear order $\underbrace{\omega + \cdots + \omega}_{\text{$d$ times}}$).
\end{enumerate} 
The base case for the first layer is trivial.
Set $\rhs_{\dA}(\good_1) = a$ and $\rhs_{\dA}(\bad_1) = aa$.
In other words, $\val_{\dA}(\good_1) \cong \mathbf{1}$ and
$\val_{\dA}(\bad_1) \cong \mathbf{2}$.
Moreover, $\rhs_{\dA}(\test_v) = a$ if $v \in L_1$ is a $\TRUE$-gate
and $\rhs_{\dA}(\test_v) = aa$ if $v \in L_1$ is a $\FALSE$-gate.

Now assume that $v \in L_{d+1}$ is a gate with inputs $v_1, v_2 \in L_d$.
For $n \in \mathbb{N}$ we use the abbreviation
$$
\omega \cdot n = \underbrace{a^\omega a^\omega \cdots
  a^\omega}_{\text{$n$ times}} .
$$
Moreover, we write $\alpha+\beta$ for the concatenation $\alpha\beta$
of the regular expression $\alpha$ and $\beta$ (which denote regular
linear orders since the alphabet is unary).
There are two cases:

\medskip
\noindent
{\em Case 1.} $L_{d+1}$ consists of $\AND$-gates. 
Then we set
\begin{eqnarray*}
\rhs_{\dA}(\test_v) & = &  [\omega\cdot d + \test_{v_1},
\omega\cdot d + \test_{v_2}, \omega\cdot d + \good_d]^\eta \\
\rhs_{\dA}(\good_{d+1}) & = &  [\omega\cdot d + \good_d]^\eta \\
\rhs_{\dA}(\bad_{d+1}) & = &   [\omega\cdot d + \good_d, \omega\cdot d + \bad_d]^\eta .
\end{eqnarray*}

\medskip
\noindent
{\em Case 2.} $L_{d+1}$ consists of $\OR$-gates.
\begin{eqnarray*}
\rhs_{\dA}(\test_v) & = & [\omega\cdot d + \test_{v_1},
\omega\cdot d + \test_{v_2}, \omega\cdot d + \bad_d]^\eta \\
\rhs_{\dA}(\good_{d+1}) & = &  [\omega\cdot d + \good_d, \omega\cdot d + \bad_d]^\eta\\
\rhs_{\dA}(\bad_{d+1}) & = &  [\omega\cdot d + \bad_d]^\eta .
\end{eqnarray*}
The above three properties (a), (b), and (c) can be shown by
induction on the layer. For layer $L_1$ all three properties 
are trivially true.
Now, consider layer $L_{d+1}$. 
Property (a) follows directly from the induction hypothesis
for layer $L_d$.
Since  the linear orders
$\val_{\dA}(\good_d)$ and $\val_{\dA}(\bad_d)$ are shuffles, 
(c) holds for layer $L_{d+1}$ too.
Finally, for (b) we consider two cases:

\medskip
\noindent
{\em Case 1.} $v \in L_{d+1}$ is an $\AND$-gate. Let $v_1, v_2 \in L_d$ 
be the inputs for $v$. First, assume that $v$ evaluates to
$\TRUE$.
Then, $v_1$ and $v_2$ both evaluate to $\TRUE$. Hence, by induction,
we get $\val_{\dA}(\test_{v_1}) \cong \val_{\dA}(\test_{v_2})
\cong \val_{\dA}(\good_d)$. Thus,
\begin{eqnarray*}
\val_{\dA}(\test_{v}) & = & 
[\omega\cdot d + \val_{\dA}(\test_{v_1}),\omega\cdot d +
\val_{\dA}(\test_{v_2}), 
\omega\cdot d + \val_{\dA}(\good_d)]^\eta \\
& \cong & [\omega\cdot d + \val_{\dA}(\good_d)]^\eta \\
& = & \val_{\dA}(\good_{d+1} ) .
\end{eqnarray*}
For the other direction assume that 
\begin{eqnarray*}
\val_{\dA}(\test_{v}) & = & 
[\omega\cdot d + \val_{\dA}(\test_{v_1}),\omega\cdot d +
\val_{\dA}(\test_{v_2}), 
\omega\cdot d + \val_{\dA}(\good_d)]^\eta \\
& \cong & [\omega\cdot d + \val_{\dA}(\good_d)]^\eta .
\end{eqnarray*}
Since neither $\val_{\dA}(\test_{v_1})$ nor 
$\val_{\dA}(\test_{v_2})$ nor $\val_{\dA}(\good_d)$ 
contains an interval isomorphic to $\omega \cdot d$, 
\cite[Lemma~23]{KuLiLo10} implies that 
$$
\omega\cdot d + \val_{\dA}(\test_{v_1}) \cong \omega\cdot d + \val_{\dA}(\test_{v_2})
\cong \omega\cdot d + \val_{\dA}(\good_d) .
$$
This implies
$$
\val_{\dA}(\test_{v_1}) \cong \val_{\dA}(\test_{v_2}) \cong \val_{\dA}(\good_d).
$$
Finally, the induction hypothesis yields that both $v_1$ and $v_2$,
and hence also $v$ evaluate to $\TRUE$.

\medskip
\noindent
{\em Case 2.} $v \in L_{d+1}$ is an $\OR$-gate. We can use similar
arguments as for Case~1.
\qed
\end{proof}
We do not know, whether the lower bound from Theorem~\ref{thm:Ptime-complete-ST}
holds for ordinary expressions too (instead of succinct expressions).

\begin{theorem} \label{thm:lower-bound-DFA-lex}
The following problem is $\Ptime$-hard
(and hence $\Ptime$-complete):

\medskip
\noindent
INPUT: Two DFAs $\mcA_1$ and $\mcA_2$.

\noindent
QUESTION: $(L(\mcA_1); \leq_{\lex}) \cong (L(\mcA_2); \leq_{\lex})$?
\end{theorem}

\begin{proof}
Note that by Theorem~\ref{thm:Ptime-reg-words} 
the problem belongs to $\Ptime$. For $\Ptime$-hardness, 
it suffices by Theorem~\ref{thm:Ptime-complete-ST} to construct in logspace 
from a given succinct expression $\dA$ (over a unary terminal 
alphabet) a DFA $\mcA$ such that 
the linear order $\val(\dA)$ is isomorphic to 
$(L(\mcA); \leq_{\lex})$. But this is accomplished
by the construction in the proof of \cite[Proposition~2]{Thomas86}.
\qed
\end{proof}
Theorem~\ref{thm:Ptime-reg-words} implies that 
it can be checked in $\EXPTIME$ whether the lexicographical
orderings on two regular languages, given by NFAs, are isomorphic.
We do not know whether this upper bound is sharp. Currently,
we can only prove a lower bound of $\PSPACE$:

\begin{theorem} \label{thm:PSPACE-NFA-lex}
The following problem is $\PSPACE$-hard:

\medskip
\noindent
INPUT: Two NFAs $\mcA_1$ and $\mcA_2$.

\noindent
QUESTION: $(L(\mcA_1); \leq_{\lex}) \cong (L(\mcA_2); \leq_{\lex})$?
\end{theorem}

\begin{proof}
We prove $\PSPACE$-hardness by a reduction from the $\PSPACE$-complete
problem whether
a given NFA $\mcA$ (over the terminal alphabet $\{a,b\}$) accepts
$\{a,b\}^*$ \cite{StMe73}.
So let $\mcA$ be an NFA  over the terminal alphabet $\{a,b\}$ and let
$K = L(\mcA)$.
Let $\Sigma = \{0,1,a,b,\$_1, \$_2\}$ and fix the following order on $\Sigma$:
$$
\$_1 < 0 <  1 < \$_2 < a < b .
$$
Under this order, 
$( \{0,1\}^* 1; \leq_{\lex}) \cong ( \{a,b\}^* b; \leq_{\lex}) \cong \eta$.

It is straightforward to construct from $\mcA$ in logspace NFAs for the following languages:
\begin{eqnarray}
L_1 & = & \{a,b\}^* b \, \$_1 \nonumber \\
L_2 & = & K  \,b \, \{0,1\}^* 1 \nonumber \\
L_3 & = & \{a,b\}^* b \, \$_2 \nonumber \\
L & = & L_1 \cup L_2 \cup L_3 \label{eq:language-L}
\end{eqnarray}
It follows that 
$$
(L; \leq_{\lex}) \cong \sum_{w \in \{a,b\}^*b} \mathcal{L}(w) ,
$$
(the sum is taken over all words from $\{a,b\}^*b$ in lexicographic order), where 
$$
\mathcal{L}(w)  \cong \begin{cases}
 \mathbf{1} + \eta + \mathbf{1} & \text{if } w \in K \\
  \mathbf{2}   & \text{else.} 
\end{cases}
$$
Hence, if $K \neq \{a,b\}^*$, then $(L; \leq_{\lex})$ contains an interval isomorphic to 
$\mathbf{2}$ and therefore is not dense. Hence 
$(L; \leq_{\lex}) \not\cong \eta$. On the other hand, if 
$K = \{a,b\}^*$, then $(L; \leq_{\lex}) \cong (\mathbf{1} + \eta + \mathbf{1}) \cdot \eta \cong \eta$.
This proves the theorem.
\qed
\end{proof}

\begin{remark}
The proof of Theorem~\ref{thm:PSPACE-NFA-lex} shows that it is $\PSPACE$-hard
to check for a given NFA $\mcA$, whether $(L(\mcA);  \leq_{\lex}) \cong \eta$. In fact,
this problem is $\PSPACE$-complete, since we can check in polynomial space
whether $(L(\mcA);  \leq_{\lex}) \cong \eta$: 
In polynomial time, we can construct 
an NFA $\mcB$ that accepts a convolution of 
two words\footnote{The convolution of the words 
$a_1 a_2 \cdots a_m$ and $b_1 b_2 \cdots b_n$
is the word $(a_1,b_1) (a_2,b_2) \cdots (a_k,b_k)$,
where $k = \max\{m,n\}$, 
$a_i = \#$ (a dummy symbol) for $m < i \leq k$ and 
$b_i = \#$ for $n < i \leq k$.}
$u \otimes v$ if and only if $u, v \in L(\mcA)$ and there exist words
$w_1, w_2, w_3 \in L(\mcA)$ such that
$w_1 <_{\lex} u <_{\lex} w_2$ and ($v \leq_{\lex} u$ or 
$u <_{\lex} w_3 <_{\lex} v$). Then, $(L(\mcA);  \leq_{\lex}) \cong \eta$
if and only if $\mcB$ accepts the set of all convolutions 
$u \otimes v$ with $u, v \in L(\mcA)$. The latter can be checked
in polynomial space.
\end{remark}

\begin{remark}
In \cite{Esik11} it is shown that the problem, whether for a given context-free language
$L$ the linear order $(L; \leq_{\lex})$ is isomorphic to $\eta$, is undecidable. This result is shown
by a reduction from Post's correspondence problem. Note that this result can be also easily 
deduced using the technique from the above proof: If we start with a pushdown automaton for $\mcA$
instead of an NFA, then the language $L$ from \eqref{eq:language-L} is context-free. 
Hence, $(L; \leq_{\lex}) \cong \eta$ if and only if $L(\mcA) = \{a,b\}^*$. The latter property
is  a well-known undecidable problem. 
\end{remark}
In Section~\ref{sec:reg-trees} we also studied the isomorphism 
problem for finite trees that are succinctly given by the prefix
order on the finite language accepted by a DFA (resp., NFA).
To complete the picture, we will finally consider the isomorphism
problem for linear orders that consist of a lexicographically ordered
finite language, where the latter is represented by a DFA (resp., NFA). 
Of course, this problem is somehow trivial, since 
two finite linear orders are isomorphic if and only if they
have the same cardinality. Hence, we have to consider the problem
whether two given acyclic DFAs (resp. NFAs) accept languages of the 
same cardinality.

\begin{proposition}
It is $\mathsf{C}_=\mathsf{L}$-complete
(resp. $\mathsf{C}_=\mathsf{P}$-complete) to check
whether two given acyclic DFAs (resp., acyclic NFAs) accept languages
of the same size.
\end{proposition}

\begin{proof}
The upper bounds are easy: There exists a nondeterministic 
polynomial time (resp., logspace) machine, which gets an
NFA (resp. a DFA) $\mcA$  over an alphabet $\Sigma$ 
as input, and has precisely
$|L(\mcA)|$ many accepting paths. Let $n$ be the number
of states of $n$. The machine first branches nondeterministically
for at most $n \cdot \log(|\Sigma|)$ steps  and thereby produces 
a word $w \in \Sigma^{\leq n}$. Then it checks whether $w \in L(\mcA)$
and only accepts it this holds. The checking step can be done
in deterministic polynomial time for an NFA and in 
deterministic logspace for a DFA.

For the lower bound, we first consider the DFA-case.
Given two nondeterministic logspace machines $M_1, M_2$
(over the same input alphabet) together with an input $w$
we can produce in logspace the configuration graphs $G_1$ and $G_2$
of $M_1$ and $M_2$, respectively, on input $w$.
W.l.o.g. we can assume that $G_1$ and $G_2$ are acyclic (one 
can add a step counter to $M_i$). 
Now, from $G_i$ it is straightforward to construct 
an acyclic DFA $\mcA_i$ such that $|L(\mcA_i)|$ is the number of
paths in $G_i$ from the initial configuration to the (w.l.o.g. unique)
accepting configuration. The latter number is the number of accepting 
computations of $M_i$ on input $w$.

Finally, $\mathsf{C}_=\mathsf{P}$-hardness for NFAs follows from
\cite[Theorem~2.1]{KannanSM95}, where it was shown that counting the 
number of words accepted by an NFA is $\#\mathsf{P}$-complete.
\qed
\end{proof}

\subsection{Ordered trees} \label{sec:order-trees}

Let us briefly discuss the isomorphism problem for 
ordered regular trees, i.e., regular trees,
where the children of a node are linearly ordered.
An ordered tree can be viewed as a triple $(A; \leq, R)$,
where $(A; \leq)$ is a tree as defined in Section~\ref{sec:trees}
and the binary relation $R$ is the disjoint union of relations
$R_a$ ($a \in A$), where $R_a$ is a linear order 
on the children of $a$.
Now, assume that
$\mcA$ is a (deterministic or 
nondeterministic) finite automaton with input alphabet
$\Sigma$ and let $\leq_\Sigma$ be a linear order
on $\Sigma$.   Assume that $\varepsilon \in L(\mcA)$.
Then, we can define a finitely 
branching ordered regular tree $\oT(\mcA,\leq_\Sigma)$ with 
$\mcA$ as follows:
$$
\oT(\mcA,\leq_\Sigma) = (L(\mcA);\, \leq_{\pref},\, \textstyle\bigcup_{u \in
  L(\mcA)} R_u),
$$
where $R_u$ is the relation
$$
R_u = \{ (v,w) \mid v,w \text{ are children of $u$ in $(L(\mcA);
  \leq_{\pref})$}, v \leq_{\lex} w \}.
$$
This means that we order the children of a node $u \in L(\mcA)$
lexicographically.  
In the following, we will omit the order $\leq_\Sigma$ on the alphabet.
The proof of the following result 
combines ideas from the proof of Theorem~\ref{thm:upper-bound-P}
with Theorem~\ref{thm:Ptime-reg-words}.

\begin{proposition} \label{prop:ordered-trees-P}
The following problem is $\Ptime$-complete:

\medskip
\noindent
INPUT: Two DFAs $\mcA_1$ and $\mcA_2$ with 
$\varepsilon \in L(\mcA_1) \cap L(\mcA_2)$.

\noindent
QUESTION: $\oT(\mcA_1) \cong \oT(\mcA_2)$?
\end{proposition}

\begin{proof}
Similarly to the proof of
Theorem~\ref{thm:upper-bound-P}, it suffices  
to take a DFA $\mcA = (Q, \Sigma, \delta, F)$ 
without initial state and two
states $p,q \in F$, and to check in polynomial time, whether
$\oT(\mcA,p) \cong \oT(\mcA,q)$, where
$\oT(\mcA,r) = \oT(Q, \Sigma, \delta, r, F)$ for $r \in F$. 
Define the following equivalence relation on $F$:
$$
\iso = \{ (p,q) \in F \times F \mid \oT(\mcA,p) \cong \oT(\mcA,q) \} .
$$
We show that $\iso$ can be computed in polynomial time. As in the proof of 
Theorem~\ref{thm:upper-bound-P}, this will be done with a partition refinement algorithm.
We need a few definitions.

Recall from the proof of Theorem~\ref{thm:upper-bound-P}
the definition of the languages $L(\mcA,p,C)$
and $K(\mcA,p,C) \subseteq L(\mcA,p,C)$
 for $p \in F$ and $C \subseteq F$.
Assume that $R$ is an equivalence relation on $F$ and let $m$ be the number of 
equivalence classes of $R$. Fix an arbitrary bijection $f$ between the 
the alphabet $\{1,\ldots,m\}$ and the set of 
equivalence classes of $R$.  With $R$ and $p \in F$ we associate
a partitioned DFA $\mcA(p,R)$ as follows: Take the DFA for the language
$L(\mcA,p,F)$ as defined in the proof of Theorem~\ref{thm:upper-bound-P} and set $F_i = f(i)$ ($1 \leq i \leq m$), which 
is the set of final states associated with symbol $i$.
Finally, define the regular word $w(p,R) = w(\mcA(p,R))$ over the alphabet $\{1,\ldots,m\}$.
We define the new equivalence relation $\widetilde{R}$ on $F$
as follows:
$$
\widetilde{R} = \{ (p,q) \in R \mid w(p,R)  \cong w(q,R) \} .
$$
Thus, $\widetilde{R}$ is a refinement of $R$ which, by Theorem~\ref{thm:Ptime-reg-words}, 
can be computed in polynomial time from $R$.
Let us define a sequence of equivalence relations $R_0, R_1,
\ldots$ on $F$ as follows: $R_0 = F \times F$, 
$R_{i+1} = \widetilde{R}_i$.
Then, there exists $k < |F|$ such that $R_k = R_{k+1}$.
We claim that  $R_k = \iso$.

For the inclusion $\iso \subseteq R_k$, one shows, by induction on $i$, that
$\iso \subseteq R_i$ for all $1 \leq i \leq k$. The point is that for every equivalence
relation $R$ on $F$ with $\iso \subseteq R$, we also have $\iso \subseteq \widetilde{R}$.
To see this, assume that $\iso \subseteq R$ but there is $(p,q) \in \iso$, which does
not belong to $\widetilde{R}$. Since $(p,q)$ belongs to $R$, we must have $w(p,R)  \not\cong w(q,R)$.
On the other hand, since $(p,q) \in \iso$,
it follows that the regular words $w(p,\iso)$ and $w(q,\iso)$ are isomorphic.
But since $\iso \subseteq R$, $w(p,R)$ is a homomorphic image of $w(p,\iso)$ and 
similarly for $w(q,R)$. Thus, also $w(p,R)$ and $w(q,R)$ are isomorphic, which is 
a contradiction.

For the inclusion $R_k \subseteq \iso$, we show that if $R$ is an equivalence
relation on $F$ such that $R = \widetilde{R}$ (this holds for $R_k$),
then $R \subseteq \iso$. 
For this, take a pair $(p_1,p_2) \in R$.
Take the tree $\oT(\mcA,p_i)$. 
We assign types in form of final states to the nodes of 
$\oT(\mcA,p_i)$ 
in the same way as in the proof of Theorem~\ref{thm:upper-bound-P}.
We now construct an isomorphism $f: \oT(\mcA,p_1) \to \oT(\mcA,p_2)$ as the
limit of isomorphisms $f_n$, $n \geq 1$. Here, $f_n$ is an isomorphism between the 
trees that result from $\oT(\mcA,p_1)$ and $\oT(\mcA,p_2)$ by cutting off
all nodes below level $n$. Let us call these trees 
$\oT(\mcA,p_i)\rest_n$ ($i \in \{1,2\}$).
Moreover, if an $f_n$ maps a node 
$u_1$ of type $q_1$ to a node $u_2$ of type $q_2$, then 
we will have $(q_1, q_2) \in R$. 
Assume that $f_n$ is already constructed and let $u_1$ of type $q_1$ be a 
leaf of $\oT(\mcA,p_1)\rest_n$. Let $u_2 = f(u_1)$ be of type $q_2$.
Then we have $(q_1, q_2) \in R$ and hence the regular
words $w(q_1,R)$ and $w(q_2,R)$ are isomorphic. Let $g$ be an 
isomorphism.
The elements of these regular words correspond to the children of 
$u_1$ and $u_2$, respectively. More precisely, if $v_i$ belongs to 
the domain of $w(q_i,R)$, then $u_i v_i$ is a child of $u_i$ and vice
versa. Clearly, $g$ can be also viewed as an isomorphism
between the lexicographical orderings on the children of $u_1$ and
$u_2$, respectively. Moreover, by definition of the regular words
$w(q_1,R)$ and $w(q_2,R)$, if $g$ maps some $u_1 v_1$ of type $r_1$
to $u_2 v_2$ of type $r_2$, then $(r_1, r_2) \in R$. By choosing
such an isomorphism $g$ for every pair $(u_1, f(u_1))$ of leaves
in $\oT(\mcA,p_1)\rest_n$ and $\oT(\mcA,p_2)\rest_n$, respectively,
we can extend $f_n$ to $f_{n+1}$.
\qed
\end{proof}
Let us now consider prefix-closed automata. Here, we can improve the upper bound from
Theorem~\ref{prop:ordered-trees-P} to {\sf NL}.

\begin{proposition} \label{prop:NL}
The following problem is {\sf NL}-complete:

\medskip
\noindent
INPUT: Two prefix-closed DFAs $\mcA_1$ and $\mcA_2$.

\noindent
QUESTION: $\oT(\mcA_1) \cong \oT(\mcA_2)$?
\end{proposition}

\begin{proof}
Again, it suffices to take a prefix-closed DFA $\mcA = (Q, \Sigma, \delta, Q)$ 
without initial state, and two
states $p,q \in Q$, and two check in {\sf NL}, whether
$\oT(Q, \Sigma, \delta, p, Q) \cong \oT(Q, \Sigma, \delta, p, Q)$.
By the complement closure of {\sf NL}, it suffices to check
nondeterministically in logarithmic space, whether 
$\oT(Q, \Sigma, \delta, p, Q) \not\cong \oT(Q, \Sigma, \delta, p, Q)$
This can be done as follows: Let $a_1 < a_2 \cdots < a_m$ and 
$b_1 < b_2 < \cdots < b_n$ the transition labels of the outgoing 
transitions of $p$ and $q$, respectively. If $m \neq n$ 
then clearly $\oT(Q, \Sigma, \delta, p, Q) \not\cong \oT(Q, \Sigma,
\delta, q, Q)$ and the algorithm can accept. 
If $n = m$, then $\oT(Q, \Sigma, \delta, p, Q) \not\cong \oT(Q, \Sigma, \delta, q, Q)$
if and only if there exists $1 \leq i \leq m$ such that
$\oT(Q, \Sigma, \delta, \delta(p,a_i), Q) \not\cong \oT(Q, \Sigma,
\delta, \delta(q,b_i), Q)$. Hence, the algorithm will simply guess
$1 \leq i \leq m$ and replace the state pair $(p,q)$ by
$(\delta(p,a_i), \delta(q,b_i))$. 
In this way, the algorithm only has to store two states of $\mcA$,
which is possible in logspace.

{\sf NL}-hardness can be shown by a reduction from the complement of 
the graph accessibility problem. Take a directed graph $G = (V,E)$ 
and two nodes $s, t \in V$. Add to each node of $V$ loops, so that
every node $v \in V \setminus \{t\}$ has outdegree $n$ (where $n$ can
be taken as the maximal outdegree of a node of $G$) and $t$ has
outdegree
$n+1$. Then label the edges of the resulting multigraph arbitrarily  
by symbols so that we obtain a DFA $\mcA$ (the initial state is $s$ and all
states are final). Then there is no path from $s$ to $t$ in $G$ if and
only if the tree $\oT(\mcA)$ is a full $n$-ary tree.
\qed
\end{proof}

\begin{corollary}
The following problem is {\sf PSPACE}-complete:

\medskip
\noindent
INPUT: Two prefix-closed NFAs $\mcA_1$ and $\mcA_2$.

\noindent
QUESTION: $\oT(\mcA_1) \cong \oT(\mcA_2)$?
\end{corollary}

\begin{proof}
The {\sf PSPACE} upper bound follows from Proposition~\ref{prop:NL},
using Lemma~\ref{PSPACE} and the obvious fact that the power set 
automaton of a given NFA can be produced by a {\sf PSPACE}-transducer.
For the {\sf PSPACE} lower bound, note that for an NFA $\mcA$ over
an alphabet $\Sigma$ we have $L(\mcA) = \Sigma^*$ 
if and only if $\oT(\mcA)$ is a full $|\Sigma|$-ary tree.
But universality for NFAs is {\sf PSPACE}-complete \cite{StMe73}.
\qed
\end{proof}

\section{Conclusion and open problems}

Table~\ref{tab-pref} (Table~\ref{tab-lex}) summarizes our complexity results 
for the isomorphism problem for regular trees (regular linear orders).
\setlength{\extrarowheight}{5pt}
\begin{table}[t]                                   
\begin{tabular}{>{\centering\arraybackslash}m{3cm}|>{\centering\arraybackslash}m{3cm}|>{\centering\arraybackslash}m{3cm}} 
        & DFA & NFA \\ \hline
acyclic &  & $\PSPACE$-complete  \\ \cline{1-1}\cline{3-3}
arbitrary & \raisebox{2.3ex}[-2.3ex]{$\Ptime$-complete} & $\EXPTIME$-complete  \\ \hline
\end{tabular} \medskip
\caption{\label{tab-pref} Main results for the isomorphism problem for regular trees}
\end{table}
\begin{table}[t]                                   
\begin{tabular}{>{\centering\arraybackslash}m{3cm}|>{\centering\arraybackslash}m{3cm}|>{\centering\arraybackslash}m{3cm}} 
        & DFA & NFA \\ \hline
acyclic &  $\mathsf{C}_=\mathsf{L}$-complete & $\mathsf{C}_=\mathsf{P}$-complete  \\ \hline
arbitrary & $\Ptime$-complete & $\PSPACE$-hard, \mbox{in $\EXPTIME$}  \\ \hline
\end{tabular} \medskip
\caption{\label{tab-lex} Main results for the isomorphism problem for regular linear orders}
\end{table}
Let us conclude with some open problems. As can be seen from Table~\ref{tab-lex}, there is 
a complexity gap for the isomorphism problem for regular linear orders that are represented
by NFAs. This problem belongs to $\EXPTIME$ and is $\PSPACE$-hard.
Another interesting problem concerns the equivalence problem for straight-line programs
(i.e., succinct expressions that generate finite words, or equivalently, acyclic partitioned DFAs, or equivalently,
context-free grammars that generate a single word).
Plandowski has shown that this problem can be solved in polynomial time. Recall that
this result is fundamental for our polynomial time algorithm for succinct expressions (Theorem~\ref{thm:poly_ST}).
In \cite{GaGiRy99}, it was conjectured that the equivalence problem for straight-line programs
is $\Ptime$-complete, but this is still open.


\def\cprime{$'$}

\end{document}